\documentclass[a4paper,11pt]{article}
\pdfoutput=1 

\usepackage{jheppub} 
\usepackage{lineno}
\usepackage{mathrsfs}
\usepackage{MnSymbol}
\usepackage{color}
\usepackage[dvipsnames]{xcolor}
\usepackage[many]{tcolorbox}
\usepackage{tikz}

\tikzstyle{Orange Dot}=[fill={rgb,255: red,241; green,143; blue,31}, draw=black, shape=circle]
\tikzstyle{Green Dot}=[fill={rgb,255: red,120; green,151; blue,95}, draw=black, shape=circle]
\tikzstyle{Cyan Dot}=[fill={rgb,255: red,0; green,159; blue,223}, draw=black, shape=circle]
\tikzstyle{none}=[]
\tikzstyle{Dashed}=[-, dashed]
\tikzstyle{Arrowright}=[->]
\tikzstyle{Average}=[-, dashed, draw = {rgb,255: red,120; green,151; blue,95}, thick]
\tikzstyle{Orange}=[-, draw = {rgb,255: red,241; green,143; blue,31}, thick]
\tikzstyle{Cyan}=[-, draw = {rgb,255: red,0; green,159; blue,223},thick]

\newtcolorbox{empheqboxed}{colback=gray!30, 
 colframe=white,
 width=\textwidth,
 sharpish corners,
 top=-2mm, 
 bottom=0mm
}

\def\tst{t}
\def\tws{\tau}
\def\psu{\mathfrak{psu}(1,1|2)_1}
\def\sl{\mathfrak{sl}(2,\mathbb{R})_1}
\def\su{\mathfrak{su}(2)_1}
\def\Fla{\mathcal{F}_\lambda}
\def\Flaw{\mathcal{F}_\lambda^w}
\def\therm{$AdS_3^T$}
\def\mcX{\mathcal{M}}

\def\SymNm{\textrm{Sym}^{N_-}(\mathcal{M})}
\def\SymNp{\textrm{Sym}^{N_+}(\mathcal{M})}
\def\SymN{\textrm{Sym}^{N}(\mathcal{M})}
\def\Ipa{\mathcal{I}^{(p)}_{|a_\pm\rangle}}

\arxivnumber{2504.00078} 
\preprint{YITP-25-32, DESY-25-041}

\title{Holographic Interfaces \\ in Symmetric Product Orbifolds}

\author[a,b]{Sebastian Harris,}
\author[c,d]{Yasuaki Hikida,}
\author[a,b,e]{Volker Schomerus,}
\author[c]{and Takashi Tsuda}

\affiliation[a]{Deutsches Elektronen-Synchrotron DESY, Notkestr. 85, D-22607 Hamburg, Germany}
\affiliation[b]{Zentrum f\"ur Mathematische Physik, Bundesstrasse 55, D-20146 Hamburg , Germany}
\affiliation[c]{Center for Gravitational Physics and Quantum Information, Yukawa Institute for Theoretical Physics, Kyoto University, Kitashirakawa Oiwakecho, Sakyo-ku, Kyoto 606-8502, Japan}
\affiliation[d]{Department of Information and Computer Science, Osaka Institute of Technology, Kitayama, Hirakata, Osaka 573-0196, Japan}
\affiliation[e]{II. Institut f\"ur Theoretische Physik, Universit\"at Hamburg, Luruper Chaussee 149,\\ D-22761 Hamburg, Germany}

\emailAdd{sebastian.harris@desy.de}
\emailAdd{yasuaki.hikida@oit.ac.jp}
\emailAdd{volker.schomerus@desy.de}
\emailAdd{takashi.tsuda@yukawa.kyoto-u.ac.jp}
 
\abstract{The study of non-local operators in gauge theory and holography, such as line-operators or interfaces, 
has attracted significant attention. Two-dimensional symmetric product orbifolds are close cousins of higher-dimensional 
gauge theory. In this work, we construct a novel family of interfaces in symmetric product orbifolds. These may be regarded 
as two-dimensional analogues of Wilson-line operators or Karch-Randall interfaces at the same time. The construction of the 
interfaces entails the choice of boundary conditions of the seed theory. For a generic seed theory, we construct the 
boundary states associated to the interfaces via the folding trick, compute their overlaps and extract the spectrum of 
interface changing  operators through modular transformation. Then, we specialise to the supersymmetric four-torus 
$\mathbb{T}^4$ and show that the corresponding interfaces of the symmetric product orbifold are dual to $AdS_2$ branes 
in the tensionless limit of type IIB superstring theory on $AdS_3 \times S^3 \times \mathbb{T}^4$. }

\begin{document}
\maketitle
\flushbottom
\section{Introduction}
\label{sec:Intro}

    Symmetric product orbifolds of two-dimensional conformal field theories (CFTs) have been studied extensively for more 
    than thirty years starting with \cite{Klemm:1990df}. They provide families of two-dimensional CFTs whose central charge 
    grows linearly with the number $N$ of copies of the underlying seed theory $\mcX$. Correlation functions of local operators 
    in these theories share many features with those of higher-dimensional gauge theories. In particular, they possess 
    a diagrammatic large $N$ expansion that is very similar to the 't Hooft expansion of non-Abelian gauge theories
    \cite{Pakman:2009zz}. Just as it is the case for their higher-dimensional cousins, these features of symmetric 
    product orbifolds are suggestive of their holographic correspondence with string theory in $AdS_3$ backgrounds, 
    as was first suggested in \cite{deBoer:1998gyt,Haehl:2014yla,Belin:2015hwa}. The work of Eberhardt, Gaberdiel 
    and Gopakumar \cite{Eberhardt:2018ouy} established that the symmetric product orbifold of a supersymmetric 
    four-torus is dual to string theory on $AdS_3$ with one unit of pure NSNS flux. This has become one of the key 
    examples for a holographic duality and it is one that could help to uncover the inner workings of the AdS/CFT 
    correspondence. 
        
    The study of symmetric product orbifolds has largely focused on the spectrum and correlation functions of 
    local fields. There are only relatively few papers that deal with boundary conditions, defects or interfaces, 
    see  \cite{Belin:2021nck,Gaberdiel:2021kkp,Martinec:2022ofs,Gutperle:2024vyp,Gutperle:2024rvo,Knighton:2024noc}. 
    This is in some contrast with higher-dimensional gauge theories, and in particular in ${\mathcal N}=4$ supersymmetric 
    Yang-Mills (SYM) theory, where the study of e.g.~Karch-Randall interfaces and Maldacena-Wilson lines has received 
    very considerable attention, both for their roles in gauge theory and for the insights they provide on 
    non-perturbative aspects of string theory in anti-de Sitter (AdS) space. In this work, we initiate a study of 
    interfaces in symmetric product orbifolds that can be considered the lower-dimensional cousins of both 
    Karch-Randall interfaces and of Wilson lines at the same time. 
    \smallskip 
        
    Before we go into the features of the interfaces between symmetric product orbifolds, it is instructive 
    to briefly recall basic facts about the higher-dimensional analogues\footnote{We would like to refer the reader to \cite{Harvey:2025ttz} for a recent overview of various top-down and bottom-up examples of holographic defects and boundaries.}. The Karch-Randall interface 
    connects two maximally SYM theories with gauge groups SU$(N_+)$ and SU$(N_-)$ on either side of the interface. 
    It arises from a D3-D5 brane system \cite{Karch:2000gx}, where D3 branes end on D5 branes.\footnote{More 
    precisely, one considers configurations in which $q$ of the $N_+$ D3 branes that realise the SU$(N_+)$ 
    SYM theory end on the D5 brane while $N_+-q=N = N_-$ run through it to realise the SU$(N_-)$ SYM theory on 
    the other side.}  Wilson lines in ${\mathcal N}=4$ SYM theory possess different holographic realisations 
    depending on a choice of representation of the gauge group. The most standard case is that of the fundamental 
    representation in which case the brane realisation is in terms of a single fundamental string in $AdS_5$. 
    For representations that arise through the tensor product of $k$ fundamental representations with $k$ of 
    order $N$, which is more relevant in our context, the brane realisation uses D-branes instead, either 
    D3 or D5 branes depending on whether the representation is symmetric or antisymmetric.\footnote{Other 
    representations of the gauge group whose Young diagram have more than one row or column require multiple 
    probe branes.} In both cases, the charge of $k$ fundamental strings is carried by the electric flux of 
    the D-brane, see e.g.~\cite{Imamura:2024pgp} and references therein, in particular \cite{Rey:1998ik,
    Maldacena:1998im,Drukker:2005kx,Yamaguchi:2006tq}. Such branes with non-vanishing electric flux 
    approach the boundary of AdS at an angle that depends on the amount of flux.  
    
    Let us now turn to the two-dimensional case. Certain symmetric product orbifolds can be thought of as describing 
    excitations of a system of $Q_1$ fundamental strings and $Q_5$ NS5 branes which wrap some four-dimensional
    compactification manifold, such as $\mathbb{T}^4$ or $K3$. We can create a defect in this theory by inserting 
    a probe D1 brane into the dual $AdS_3$ background. We shall focus on the case in which this probe brane is 
    localised along an $AdS_2$ inside $AdS_3$. More generally, we can consider $(n,1)$-strings, i.e.~bound 
    states of a D1 brane and $n$ fundamental strings, see e.g.~\cite{Bachas:2001vj}. As in the case of Wilson
    lines reviewed in the previous paragraph, one may think of these $(n,1)$ strings as a D1 brane with a 
    worldvolume electric field turned on with a strength that is determined by $n$. When we allow $n$ 
    fundamental strings to run along the worldvolume of the probe D1 brane, the number of fundamental 
    strings that appear in the brane realisation of the symmetric product orbifold can jump from one side 
    of the D1 brane junction to the other, just as in the case of Karch-Randall interfaces in gauge theory. 
    
    In contrast to the higher-dimensional theories, string theory on $AdS_3$ with a pure NSNS-background is 
    famously solvable using a description in terms (non-compact) Wess-Zumino-Novikov-Witten (WZNW) models. 
    The related WZNW model on the sphere with hyperbolic target space $H_3^+$ was first solved by Teschner 
    \cite{Teschner:1997ft,Teschner:1999ug}. With input from Teschner's CFT analysis, Maldacena and Ooguri
    were able to construct a theory of closed strings on $AdS_3$ background with pure NSNS flux 
    \cite{Maldacena:1997re,Maldacena:2000kv}. A central ingredient of their work was the inclusion of 
    spectrally flowed representations of the underlying $\mathfrak{sl}(2)$ affine Kac-Moody algebra. 
    These were needed in order to describe long strings. The worldsheet analysis of branes and open 
    strings in these models was initiated in \cite{Bachas:2001vj} with a classification of possible 
    brane geometries. Maximally symmetric branes, i.e.~branes that preserve the affine Kac-Moody algebra, 
    can be localised along several different submanifolds. Here, we shall mainly deal with $AdS_2$ branes.  
    In general, D-branes may be described by boundary states of the worldsheet CFT which encode
    all information about the open string spectrum and the couplings with closed strings. Boundary states 
    for $AdS_2$ branes (along with spherical branes) have been constructed in \cite{Ponsot:2001gt,Lee:2001gh} 
    building upon some preliminary studies in \cite{Giveon:2001uq,Petropoulos:2001qu,Lee:2001xe,Hikida:2001yi,
    Rajaraman:2001cr}. In agreement with the geometric intuition we recalled above, there is a continuous  
    family of such boundary states that is parametrized by one real parameter that is related to the angle 
    at which the $AdS_2$ brane approaches the boundary. Moreover, the branes support a family of `half-winding' long open strings with arbitrarily large winding number. The half-winding is related to the 
    fact that the $AdS_2$ brane approaches the asymptotic boundary of $AdS_3$ at two opposite sides which are 
    mapped onto each other by a half-rotation of $AdS_3$. While the spectrum of open strings on the 
    $AdS_2$ branes is rather rich, they only couple to a very small subset of the bulk states in the 
    WZNW model, namely to states of winding number $w=0$ only. 
    \smallskip 
    
\begin{figure}
        \centering
        \includegraphics[width=1\linewidth]{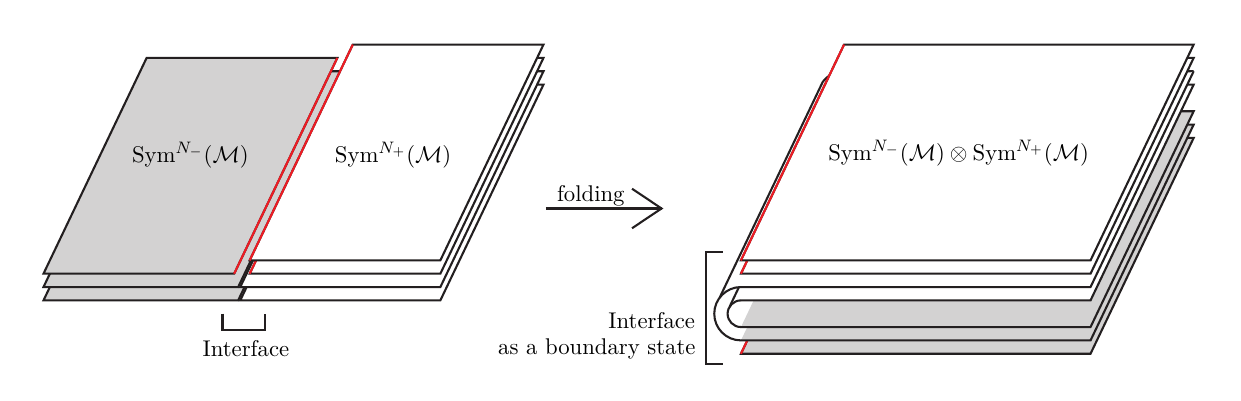}
        \caption{Interface between $\text{Sym}^{N_-}\mathcal{M}$ and $\text{Sym}^{N_+}\mathcal{M}$ as a boundary state of the folded theory. The red lines indicate reflective boundaries. This figure is for $N_-=3$, $N_+=4$, $p=2$. }
        \label{fig:InterfaceAsBdy}
    \end{figure}

\begin{figure}
        \centering
        \includegraphics[width=1\linewidth]{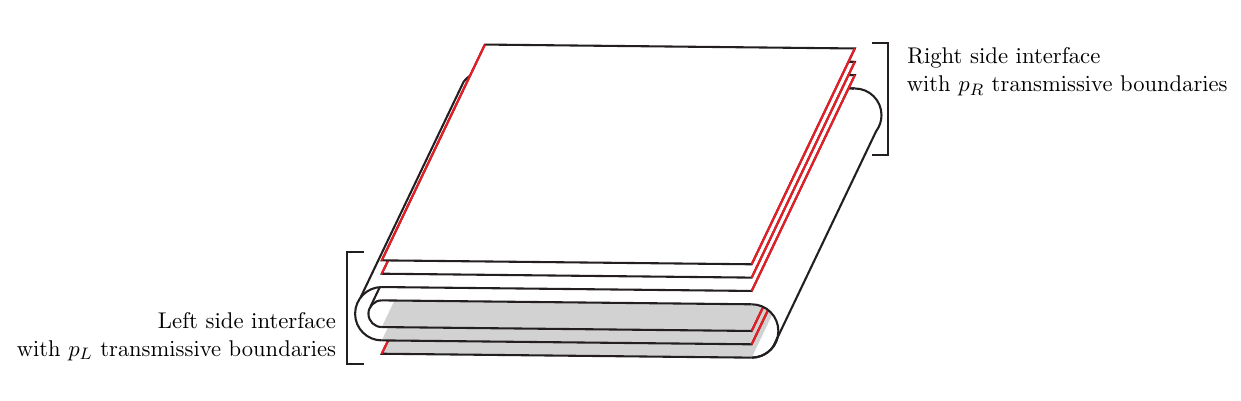}
        \caption{Contribution to the torus partition function in the presence of a pair of interfaces with parameters $N_-=3$, $N_+=4$, $p_L=2$, $p_R=1$. The red lines indicate reflective boundaries. As discussed in Section \ref{sec:interface_changing_partition_function}, the full partition function is a symmetrised sum over terms of the type illustrated in this figure.}
        \label{fig:InterfaceOverlap}
    \end{figure}

    The goal of our work is to address the holographic description of $AdS_2$ branes and to construct the 
    interface between symmetric orbifold theories that is created when an $AdS_2$ brane ends on the boundary 
    of $AdS_3$, see also \cite{Martinec:2022ofs} for a very insightful previous discussion. More precisely, 
    we construct a family of interfaces between any pair of symmetric product orbifolds of rank 
    $N_-$ and $N_+$ that are parametrised by some integer $p = 1, \dots, \textrm{min}(N_-,N_+)$. The 
    interfaces $\Ipa$ that we define are transmissive in $p$ components of the product theory and 
    reflective in the remaining ones with a reflective behaviour that depends on the choice of a 
    boundary conditions $a_\pm$ of the underlying seed conformal field theory $\mcX$. A precise definition 
    is given in section \ref{sec:def_sym_orbifold_boundary_states}, see eq.~\eqref{eq:def:pra} and 
    the explanation of notation below that formula. Our description uses the folding trick and hence it 
    describes the interface in terms of a boundary state of the folded theory, as shown in figure 
    \ref{fig:InterfaceAsBdy}. As usual, the boundary 
    state encodes the entire set of one-point couplings between the interface and bulk fields of the 
    CFT. From this description of the interfaces, we then determine the annulus amplitude and thereby 
    the exact spectrum of interface changing  operators between any pair of interfaces in section
    \ref{sec:interface_changing_partition_function}. This key result is stated in eqs.~(\ref{eq:GCZ} - \ref{eq:seedZo}). As is familiar in the context symmetric product orbifolds since the seminal paper 
    of Dijkgraaf-Moore-Verlinde-Verlinde \cite{Dijkgraaf:1996xw}, it is easiest to state the result in 
    terms of a grand canonical partition function. In the setup under consideration, this grand 
    canonical partition function involves four chemical potentials $\mu_\pm$ and $\rho_A$ with $ A\in \{L,R\}$. 
    The first two, i.e.~$\mu_-$ and $\mu_+$, correspond to the order $N_\pm$ of the symmetric product orbifold 
    on the two sides of the interfaces. The other two chemical potentials, $\rho_L$ and $\rho_R$, are associated 
    with the numbers $p_L$ and $p_R$ to the left and the right of the interface changing operator, see 
    figure \ref{fig:InterfaceOverlap}. 


    A key objective of this paper is to show that the interfaces constructed in section 
    \ref{sec:def_sym_orbifold_boundary_states} provide a holographic description of $AdS_2$ branes 
    (and the open strings thereon) in type IIB superstring theory on $AdS_3 \times S^3 \times 
    \mathbb{T}^4$. In the presence of $k$ units of NSNS-flux, the dual CFT is supposed to be given 
    by a marginal deformation of symmetric product orbifold of $\mathbb{T}^4$. When $k=1$, i.e.~for
    minimal amount of NSNS-flux, the dual CFT has been argued to coincide with the orbifold fixed 
    point \cite{Eberhardt:2018ouy}, without any need for further deformation. As 
    evidence, Gaberdiel et al.~were able to show that the spectrum of the string theory agreed with 
    that of the symmetric product orbifold CFT. Later, the matching was extended to string amplitudes 
    and correlation functions. In particular it was understood that the string worldsheet in 
    calculations of closed string amplitudes localises to the boundary of the $AdS_3$ target space. 
    This has been exploited to calculate string amplitudes and compare them with the corresponding 
    correlation functions in the dual CFT, see in particular \cite{Eberhardt:2019ywk,Dei:2020zui}. 
    The holographic description of spherical branes in the tensionless limit of type IIB superstring 
    theory on $AdS_3 \times S^3 \times \mathbb{T}^4$ was addressed in \cite{Gaberdiel:2021kkp}. More 
    concretely, it was found that (some) spherical branes are dual to special cases of the boundaries 
    in symmetric product orbifolds that were studied extensively by Belin et al.~in \cite{Belin:2021nck} 
    by applying standard constructions of boundary CFT, see in particular \cite{Recknagel:2002qq}. 
    Our work is complementary to these developments and focuses fully on $AdS_2$ branes.

    While for generic values of the level $k$ there exists a continuous family of $AdS_2$ 
    branes, the tensionless limit of IIB superstring theory has a unique $AdS_2$ brane in the $AdS_3$ factor of the target space.\footnote{In 
    \cite{Gaberdiel:2021kkp} the authors constructed two boundary states. Here we only admit one special 
    linear combination thereof in order to restore the rotational symmetry with respect to half-rotations 
    of $AdS_3$. This additional feature makes the boundary state unique.} 
    With our holographic proposal, we extend the match of bulk partition functions 
    \cite{Eberhardt:2018ouy} to this unique $AdS_2$ brane. Our matching of spectra is based on 
    formulas for the partition function of open strings on $AdS_2$ branes in (global and) thermal 
    $AdS_3$ backgrounds that we derive in section \ref{sec:tensionless_strings}, see in particular 
    eq.~\eqref{eq:STspectrum}. For a concise summary of the holographic relation we propose, the 
    reader might also find the box on page \pageref{central_box} useful.
    \smallskip     
    
    Let us now briefly outline the plan of this paper. In many respects, section \ref{sec:symmetric_product}
    contains the main new results of this work. After a short review \ref{sec:symmetric_product_notation} of 
    symmetric product orbifolds that we use to set up notations, section \ref{sec:def_sym_orbifold_boundary_states} 
    constructs the interfaces $\Ipa$. As a small interlude, we compute the transmissivity and reflectivity 
    of these interfaces in terms of the parameter $p$ in section \ref{sec:reflectivity_transmissitivty}. Then, 
    in section \ref{sec:interface_changing_partition_function}, we calculate the spectrum of interface changing  
    operators. While our construction of the interface may seem somewhat artificial, it is guided by the intention 
    to construct a holographic dual of $AdS_2$ branes in $AdS_3$. To show that this objective is indeed achieved, at least 
    in the tensionless limit of type IIB superstrings on $AdS_3 \times S^3 \times \mathbb{T}^4$, we discuss the 
    string theory of $AdS_2$ branes in section \ref{sec:boundary_states_for_ads2}. 
    The boundary states of the world sheet theory corresponding to $AdS_2$ branes have been constructed in \cite{Gaberdiel:2021kkp}. 
    We briefly review this construction and then, in section \ref{sec:global_ads3_partition_function}, use it to determine the spectrum of 
    worldsheet boundary excitations associated to open string states on the $AdS_2$ 
    branes in global $AdS_3$.
    The result serves as an input to section \ref{sec:thermal_ads3_partition_function}, where we perform an orbifold 
    construction to pass to thermal $AdS_3$. After 
    integration over the modular parameter of the toroidal worldsheet, we finally obtain the partition function 
    of open superstrings in thermal $AdS_3$. The two strands of the discussion are merged in section
    \ref{sec:matching_thermal_ads_with_sym_orbifold} where we compare the string partition function with the connected contribution to the torus partition function of our interface in the symmetric product orbifold of the supersymmetric four-torus
    to find complete agreement between these two quantities. The section also contains some qualitative discussion 
    of correlation functions in the symmetric product orbifold and the corresponding scattering amplitudes of open 
    and closed strings in the dual string theory. A more quantitative analysis will appear in forthcoming work. 
    This and other future directions are discussed in the concluding section \ref{sec:outlook}. The work 
    contains three appendices in which we provide some explicit calculations to prove formulas that are 
    stated in the main text. 
    
\section{Interfaces of symmetric product orbifolds}\label{sec:symmetric_product}

    The arguably simplest type of interfaces are those that are purely reflecting or purely transmitting. 
    Purely reflecting interfaces can be obtained by choosing boundary conditions for the two CFTs which are 
    separated by the interface, resulting simply in a product of two essentially decoupled boundary theories.
    In the specific context of symmetric product orbifolds, a detailed analysis of the spectrum of boundary 
    states has been provided by \cite{Belin:2021nck}.
    The most prominent example of a purely transmitting interface is the trivial interface between two identical CFTs.
    For symmetric product orbifolds, a broader class of purely transmitting defects was studied in \cite{Gutperle:2024vyp}.
    
    In this section, we introduce a more general family of interfaces between two different symmetric product 
    orbifolds Sym$^{N_\pm}(\mathcal{M})$ of the same seed theory $\mathcal{M}$ that interpolates between the purely 
    transmitting and purely reflecting extremes. For this purpose, section \ref{sec:symmetric_product_notation} 
    establishes the general notation which we use to describe symmetric product orbifolds.    
    Section \ref{sec:def_sym_orbifold_boundary_states} gives a precise definition of our interfaces.
    Section \ref{sec:reflectivity_transmissitivty} serves the purpose of making our definitions more accessible by 
    demonstrating how to compute the reflectivity and transmissivity coefficients of our interface. 
    It does not contain any information that is strictly necessary for the rest of the discussion and may be skipped.
    Finally, section \ref{sec:interface_changing_partition_function} determines partition functions of interface 
    changing operators that act within the class of interfaces that we have proposed.
    
    \subsection{Notation and conventions} \label{sec:symmetric_product_notation}
    \newcommand{\ptn}[1]{{[#1]}} 
    \def\calC{\mathcal{C}}
    \def\STq{x}
    
        This section sets up the necessary notation to work with symmetric product orbifolds 
        $\SymN$ of a CFT $\mathcal{M}$. The Hilbert space of the 
        seed theory $\mathcal{M}$ is denoted by $\mathcal{H}_\mathcal{M}$. It comes 
        equipped with two commuting actions of the Virasoro algebra along with possibly further modes
        of some extended chiral algebra $\mathcal{W}$. Under the action of this chiral algebra, 
        the state space decomposes into sectors which we label by letters $i,j,\dots$ of the Roman alphabet. 
        In a unitary CFT, these are lowest weight representations. The partition function of the seed theory 
        $\mathcal{M}$ is given by  
        \begin{align} \label{eq:seedPFbulk} 
            Z(\tst, \bar \tst) = \text{Tr}_{\mathcal{H}_\mathcal{M}}[\STq^{L_0 -\frac{c_\mathcal{M}}{24}}
            \bar \STq^{\bar L_0 -\frac{c_\mathcal{M}}{24}}],\quad \STq = e^{2\pi i \tst}, \quad \bar 
            \STq = e^{-2\pi i \bar\tst}.
        \end{align}
        Here, $c_\mathcal{M}$ denotes the central charge of the Virasoro algebra, as usual. With 
        these basic ingredients of the seed theory set up, we move on to the symmetric product 
        orbifold.  

        States of the symmetric product orbifold $\SymN$ are obtained by the standard rules of the 
        orbifold construction. In a first step, we construct twisted sectors $\mathcal{H}^g$ for the 
        $N^{th}$ power $\mathcal{M}^{\otimes N}$ of the seed 
        theory. These are labelled by group elements $g \in S_N$ of the symmetric group. For a 
        given element $g$, we define the centraliser  subgroup 
        \begin{equation} 
            \calC_g = \calC_g^N  = \{ \, \sigma \in S_N \, | \, \sigma g \sigma^{-1} = g \, \} \, 
            \subseteq  S_N  . 
        \end{equation} 
        The twisted sector $\mathcal{H}^g$ carries a representation of $\calC_g$ and, as part of 
        the orbifold procedure, we are instructed to project to its $\calC_g$ invariant subspace. 
        The projection can be performed by averaging over the orbits of $\calC_g$. Hence, the 
        torus partition function of the symmetric product orbifold is given by\footnote{The normalisation 
        factor is $1/N!$ because we divide both by the size of the centraliser  $|\mathcal{C}_g|$ (due 
        to the averaging over $\mathcal{C}_g$ orbits) as well as by the size of the conjugacy class 
        of $g$.}
        \begin{align} \label{eq:defPFN}
            Z_N(\tst, \bar \tst) :=  \frac{1}{N!}\sum\limits_{g \in S_N}\sum\limits_{h \in 
            \calC_g}\text{Tr}_{\mathcal{H}^{g}}[ h \STq^{L_0 -\frac{Nc_\mathcal{M}}{24}}\bar \STq^{\bar L_0 
            -\frac{Nc_\mathcal{M}}{24}}].
        \end{align}
        One can straightforwardly express the partition function of the orbifold theory in terms 
        of the seed theory partition function by directly evaluating the trace. The most elegant 
        formula for the partition function \cite{Dijkgraaf:1996xw} passes through the 
        grand canonical ensemble  
        \begin{align}\label{eq:torus_grand_canonical}
            \mathcal{Z}(\kappa;\tst,\bar \tst) := \sum\limits_{N=0}^\infty \kappa^N Z_N(\tst, \bar \tst) 
            = \exp\left[\sum\limits_{k=1}^\infty \kappa^k T_k Z(\tst, \bar \tst) \right].
        \end{align}
        The terms in the exponent on the right hand side are obtained from the partition function 
        $Z(\tst, \bar \tst)$ of the seed theory through application of the following Hecke operators
        \begin{align}\label{eq:def_Hecke_operator}
            T_k Z(\tst , \bar \tst) := \frac{1}{k} \sum\limits_{w|k} \sum\limits_{j=0}^{w-1} Z
            \left(\frac{k \tst}{w^2}+\frac{j}{w},\frac{k \bar \tst}{w^2}+\frac{j}{w}\right).
        \end{align}
        By expanding the exponent on the right hand side in powers of $\kappa$, it is 
        straightforward to obtain the partition functions $Z_N$ of symmetric product orbifolds for any finite 
        integer $N$. 
        \smallskip 
    
        It is important to discuss formula \eqref{eq:defPFN} in more detail. We have written the 
        formula as a sum over all elements $g$ of the symmetric group $S_N$. But many of the 
        twisted sectors $\mathcal{H}^g$ we are summing over give rise to equivalent representations 
        of the chiral algebra. 
        By construction, two twisted sectors $\mathcal{H}^{g_1}$ and $\mathcal{H}^{g_2}$ give rise to 
        equivalent representations if $g_1$ and $g_2$ belong to the same conjugacy class. We
        denote the conjugacy class of $g \in S_N$ by 
        \begin{equation} 
            \ptn{g} := \{ h g h^{-1} \, : \, h\in S_N \, \} \subseteq S_N   
        \end{equation} 
        and more generally
        \begin{align}\label{eq:def_conjugacy_classes}
            [g_1,g_2, \dots ,g_{n}] := \{(h g_1 h^{-1}, h g_2 h^{-1}, \dots, h g_{n}h^{-1}) \, : \, h \in S_N\}.
        \end{align}
        Famously, the conjugacy classes $[g]$ of the symmetric group $S_N$ are in one to one correspondence with integer 
        partitions $(w_1, \dots,w_\ell)$ of $ N = \sum_{r=1}^\ell w_i$ or, equivalently, Young 
        diagrams $Y$ with $N$ boxes. The correspondence is easy to spell out more explicitly. Any 
        element $g \in S_N$ possesses a unique (up to reordering of the factors) decomposition 
        \begin{equation}  \label{eq:gfactorisation}
            g = \omega_1 \omega_2 \, \cdots\,  \omega_\ell 
        \end{equation} 
        into cyclic permutations $\omega_r$ of length $w_r$ such that the cycles constitute a partition  
        $\{\omega_r\}_{r=1}^\ell$ of $\{1,\dots,N\}$ into ordered sets. 
        The integer partition $\{w_r\}_{r=1}^\ell$ of $N$ is graphically represented by the Young diagram $Y_g$ 
        associated to $g \in S_N$. Conversely, any two 
        group elements that possess a factorisation of the form \eqref{eq:gfactorisation} with 
        $\ell$ cycles of order $w_r$ are conjugate to each other. Given a Young diagram $Y$ with 
        $N$ boxes we can always pick some representative $g_Y$ of the corresponding conjugacy 
        class of $S_N$.  
        
        With the help of the factorisation \eqref{eq:gfactorisation} of the group element $g$ into 
        a product of cyclic permutations, it is easy to give a more explicit description of the centraliser  group
        $\calC_g^N$. Concretely, $\calC_g^N$ is given by 
        \begin{equation} \label{eq:calCvs2} 
            \calC_g^N := \{ \, \sigma = \pi \prod\limits_{r=1}^\ell \omega_r^{\nu_r} |  
            \nu_r \in \mathbb{Z} \text{ and } \, \pi \in S_\ell \text{ such that } w_{\pi r}=w_r\} 
            \simeq \prod\limits_{w=1}^N  S_{m_w} \ltimes \mathbb{Z}_w^{m_w},
        \end{equation} 
        where the sequence $(m_w)_{w\in \mathbb{N}}$ is the weight of the diagram $Y_g$ i.e.~$m_w$ is the 
        number of rows of length $w$. By the orbit-stabiliser theorem, the order $|S_N| = N!$ of the symmetric 
        group factorises as
        \begin{equation}
            |S_N| = |\calC^N_g| \cdot |Y_g| ,
        \end{equation} 
        where $|Y_g|$ denotes the number of elements in the conjugacy class $Y_g$ 
        of $g$. After this preparation, 
        it is natural to rewrite our formula \eqref{eq:defPFN} for the partition function of the 
        symmetric product orbifold as 
        \begin{align} \label{eq:defPFNvs2}
            Z_N(\tst, \bar\tst) =  \sum\limits_{|Y|=N} 
            \text{Tr}_{\mathcal{H}^{g_Y}}[ \Pi_0^{g_Y} \STq^{L_0 -\frac{Nc_\mathcal{M}}{24}}
            \bar \STq^{\bar L_0 
            -\frac{Nc_\mathcal{M}}{24}}] \quad \textrm{where} \quad \Pi_0^{g_Y} := 
            \frac{1}{|\calC^N_{g_Y}|} \sum_{\sigma \in \calC^N_{g_Y}} \sigma  .   
        \end{align}
        In performing the sum over Young diagrams $Y$ with $N$ boxes, we can pick a single representative 
        $g = g_Y \in S_N$ and then construct the twisted sector $\mathcal{H}^g$, the centraliser  $\mathcal{C}_g$ 
        and the projection $\Pi^g_0$ for that representative. The result is 
        obviously independent of the representative we choose. Thereby, we have rewritten the 
        partition function of the symmetric product orbifold as a sum over partitions of $N$ rather 
        than group elements $g \in S_N$. 
    
        In order to illustrate the difference between the formulas \eqref{eq:defPFN} and \eqref{eq:defPFNvs2},
        let us briefly discuss the example of $N=3$. In this case, the symmetric group $S_3$ has six 
        elements and the original sum in eq.~\eqref{eq:defPFN} runs over all these six elements. But the six group elements fall 
        into only three conjugacy classes which are associated with the three possible partitions of $N=3$. The 
        conjugacy class for the partition $Y = [1,1,1]$ consists of the unit element 
        $g = e$ and its centraliser  group $\calC^3_e \simeq S_3 \ltimes \mathbb{Z}_1^3$ is the entire symmetric group $S_3$. Next, let us 
        consider the partition $Y=[1,2]$. The associated conjugacy class contains three elements, namely 
        $g_1 = (1)(23)$, $g_2 = (2)(13)$ and $g_3 = (3)(12)$. Their centraliser  subgroups $\calC_{g_i}^3 \cong 
        \mathbb{Z}_2$ contain two elements each. Finally, for the trivial partition $Y=[3]$, the conjugacy 
        class consists of the elements $\omega_1 = (123)$ and $\omega_2 = (132)$. Their centraliser  group, 
        on the other hand, is now given by $\calC_{\omega_i}^3 \cong \mathbb{Z}_3$. 
        \smallskip 
    
        In the following, we shall continue to think of the twisted sectors $\mathcal{H}^g$ as being 
        associated with elements $g \in S_N$ rather than with the Young diagrams $Y_g$. 
        This is correct as long as we always remember not to sum but to average over the elements of a conjugacy class.
        Now given a group element $g$ with 
        the factorisation \eqref{eq:gfactorisation} of length $\ell$, states of the associated twisted 
        sector $\mathcal{H}^g$ take the form  
        \begin{equation} \label{eq:twistedsectorstate}
            |(\psi_1, \dots, \psi_\ell)\rangle_g  \quad \textrm{ where } \quad 
            \psi_r \in \mathcal{H}_\mathcal{M}
        \end{equation} 
        are $\ell$ states in the seed theory. By abuse of notation, we simply write
        \begin{align}
            |\psi_r\rangle_g \quad \text{instead of} \quad |(\psi_r)_{r=1}^\ell\rangle_g.
        \end{align}
        Let us assume that these $\ell$ states $\psi_r$, $r=1, \dots, \ell$ 
        of the seed theory have been chosen to be eigenstates of $(L_0,\bar L_0)$ with eigenvalues $(h_r, 
        \bar h_r)$. Then the twisted sector state \eqref{eq:twistedsectorstate} is an eigenstate of 
        the symmetric product orbifold Virasoro generators $(L_0,\bar L_0)$ with eigenvalues 
         \begin{align}\label{eq:orbifold_spectrum}
             h^g  = \sum\limits_{r=1}^\ell \left(\frac{h_r }{w_r} + \frac{c_{\mathcal{M}}}{24}\left(w_r-\frac{1}{w_r}
            \right)\right) && \text{and} && 
            \bar h^g = \sum\limits_{r=1}^\ell \left(\frac{\bar h_r}{w_r} + \frac{c_{\mathcal{M}}}{24}\left(w_r-\frac{1}{w_r}\right)\right).
        \end{align}
        In particular, its spin is given by 
        \begin{align}
            J^g = \bar h^g-h^g = \sum\limits_{r=1}^\ell 
            \left(\frac{\bar h_r - h_r}{w_r} \right).
        \end{align}
        Only states \eqref{eq:twistedsectorstate} for which the spin is an integer correspond to local operators of 
        the symmetric product orbifold. 
        In order to find the subset of physical states within the twisted sector 
        $\mathcal{H}^g$, one needs to impose the projection to $\calC_g$ invariants, see the construction in $\Pi_0$
        in eq.~\eqref{eq:defPFNvs2}. 
        The action of $\sigma \in \calC_g$ on twisted sector states 
        \eqref{eq:twistedsectorstate} is given by 
        \begin{align}\label{eq:sigma_action}
            \sigma |\psi_r\rangle_g  =
            \prod\limits_{r=1}^\ell e^{2 \pi i\frac{\nu_r}{w_r}(\bar h_{r} - h_{r} )}
            |\psi_{\pi r}\rangle_{g}=
            |e^{2 \pi i \nu_{\pi r} (\bar L_0- L_0)} \psi_{\pi r}\rangle_{g}.
        \end{align}
        Here, we have represented $\sigma\in \calC_g$ as a product $\sigma= \pi \omega_1^{\nu_1} \cdots 
        \omega_\ell^{\nu_\ell}$ as explained in eq.~\eqref{eq:calCvs2}. Furthermore, the Virasoro elements 
        on the right hand side~are understood to be those of the symmetric product orbifold i.e.~operators with a spectrum as given in \eqref{eq:orbifold_spectrum}. The exponents $\nu_r$ of the single 
        cycles $\omega_r$ only enter the phase factor, while $\pi$ permutes the $r$ labels.
        In order to help the reader to get used to the notation for symmetric product orbifolds introduced in 
        this section, we use it in appendix \ref{app:torus_partition_function} to compute the torus partition 
        function.
        
    \subsection{Definition of the interfaces}\label{sec:def_sym_orbifold_boundary_states}
    
        This section constructs a class of interfaces between Sym$^{N_-}(\mathcal{M})$ and Sym$^{N_+}(\mathcal{M})$.
        We think of these interfaces as separating the upper and lower half of the complex plane with 
        Sym$^{N_-}(\mathcal{M})$ living in the lower half plane and Sym$^{N_+}(\mathcal{M})$ in the upper. 
        The definition of our interfaces 
        involves selecting a pair of boundary states $|a_\pm\rangle$ of the seed theory $\mathcal{M}$. These boundary 
        states of the seed theory give rise to a purely reflecting interface, in which all the $N_-[N_+]$ 
        copies of the seed theory in the lower[upper] half plane are simply reflected\footnote{See section
        \ref{sec:reflectivity_transmissitivty} below for a more precise definition of reflectivity and 
        transmissivity in this context.} back. This is obviously a bit too trivial to be interesting. 
        Instead, we propose to consider interfaces in which $p\leq \textrm{min}(N_-,N_+)$ components 
        can pass through the interface while the remaining ones are reflected. 

        \paragraph{Qualitative overview of the construction.}
        
        Before we discuss the precise construction of the interfaces, let us first give a qualitative description 
        of all steps that constitute it. To spell out a concrete formula for the interfaces $\mathcal{I} = \Ipa$, we utilize the 
        folding trick and describe the interfaces as boundary states of
        \begin{align}
            \text{Sym}^{N_-}(\mathcal{M}) \otimes \text{Sym}^{N_+}(\mathcal{M}) =
            \mathcal{M}^{N_-+N_+}/S_{N_-}\times S_{N_+}.
        \end{align}
        In our construction, we first choose a gauge in which $p$ components in each of the theories that we want 
        to be transmitting are singled out. This procedure manifestly breaks the $S_{N_-}\times S_{N_+}$ symmetry of 
        the folded theory down to $S_{N_- - p} \times S_p\times S_p \times S_{N_+-p}$.
        
        After this first gauge choice, we impose the reflecting boundary condition $|a_-\rangle$ 
        along $N_--p$ components from the first factor and likewise $|a_+\rangle$ along $N_+-p$ components from the second factor. 
        The remaining copies are glued together with transmitting boundary conditions. Such a boundary condition
        is stabilised by the subgroup 
        \begin{equation} \label{eq:groupembedding} 
             S_{N_--p} \times S_{p,\text{diag}} \times S_{N_+-p} \subseteq S_{N_- - p} \times S_p\times S_p 
             \times S_{N_+-p} \subseteq S_{N_-} \times S_{N_+} \  
         \end{equation} 
        and thus the selection of a particular way to transmit the selected $p$ copies of the seed theory in the lower 
        half plane to $p$ copies in the upper half plane constitutes a second gauge choice that breaks $S_p \times S_p$ 
        to the diagonal subgroup $S_{p,\text{diag}}:= \{(g,g)|g\in S_p\} \subseteq S_p \times S_p$. To restore the full 
        $S_{N_-}\times S_{N_+}$ symmetry, we finally need to average over the gauge orbits.
        
        In the construction 
        of boundary states of the folded theories, we can select characters for each of the three factors of 
        the stabiliser subgroup within the orbifold group $S_{N_-} \times S_{N_+}$. We denote these 
        characters by $(\chi_-,\chi_p,\chi_+)$. The boundary states we are about to construct are 
        denoted by
        \begin{equation}  \label{eq:boundarystateschi}
            | p,a_\pm;\chi_\pm,\chi_p\rangle =  |p,a_\pm;\chi_\pm,\chi_p\rangle^{N_\pm}_\mathcal{M} . 
        \end{equation} 
        We shall often drop the sub- and superscripts that refer to the bulk data. In addition, throughout 
        most of our discussion we will set the characters $\chi_\pm, \chi_p$ to be trivial, drop them 
        from the arguments and write 
        \begin{equation} \label{eq:boundarystates1}
            |p,a_\pm;N_\pm\rangle = |p,a_\pm;\chi_\pm = {\bf{1}}, \chi_p = {\bf{1}} \rangle^{N_\pm}_{\mathcal{M}} .   
        \end{equation} 
        The interfaces obtained from the trivial representation of the stabiliser subgroup turn out to be the ones 
        that are relevant for holography.
        This is not merely an a 
        posteriori observation, but rather to be expected from first principles: From the string perspective, the 
        representations associated to the characters $\chi_\pm$ and $\chi_p$ are the representations of the symmetric 
        group that govern the statistics of multi-string states. If we would like to describe multi-string 
        ensembles with Bose statistics, we should hence make use only of the trivial characters in the dual 
        symmetric orbifold, see also \cite{Knighton:2024noc}. Moreover, the restriction to fully symmetric 
        representations of the `gauge group' $S_N$ is also consistent with the situation for Maldacena-Wilson 
        lines in four-dimensional $N=4$ SYM 
        theory. In the higher dimensional context, Wilson lines in other than symmetric traceless tensor 
        representations require to consider multiple $D3$ branes in the bulk \cite{Gomis:2006sb} (or $D5$ 
        branes).

        \paragraph{Reflective part of the interfaces.}After this first qualitative description of our interfaces, 
        let us now proceed with their systematic construction. The boundary states $|a_\pm\rangle$ of the seed theory 
        constitute the most non-trivial data that enters the construction. 
        Let us recall that the sectors of the seed theory, i.e.~the irreducible
        representations of its chiral algebra $\mathcal{W}$, are labelled by some index $j$.
        Since the gluing condition for chiral fields relates the label of the right- and left-moving 
        representations, the index $j$ also labels the Ishibashi states $|j\rrangle$ that can 
        contribute to the boundary state.\footnote{To be more precise, the sectors of the seed theory 
        are labelled by pairs $(j,\bar\jmath)$ of representations for the (anti-)holomorphic chiral 
        fields. The specification of a boundary state involves picking a gluing 
        automorphism $\Omega$ for the chiral algebra. This automorphism induces a map on representation
        labels $j \mapsto j_\Omega$. Given the choice of $\Omega$, Ishibashi states only exist for 
        those sectors of the theory for which $\bar \jmath = j_\Omega$. In this sense, we only need 
        to specify a single representation label in order to specify the Ishibashi state, see 
        \cite{Recknagel_Schomerus_2013} for details. \label{footnote:Omega}} Thus, a boundary state $|a\rangle$ 
        of the seed theory is described by a set of coefficients $a_j$ in the expansion 
        \begin{equation} 
            |a\rangle = \sum_j \, a_j |j\rrangle  .  
        \end{equation} 
        We can uplift this boundary state of the seed theory to a boundary state of the symmetric 
        product orbifold $\textrm{Sym}^M(\mathcal{M})$ for any $M\in\mathbb{N}$. Specifically we are 
        interested in the cases $M = N_- -p$ or $M = N_+ -p$. 
        Once again, the boundary state of the symmetric product orbifold is a sum over Ishibashi states.
        For a fixed gauge $\rho \in S_M$, the relevant Ishibashi state is given by
        \begin{equation}
            |j_r\rrangle^M_\rho : = |\{ |j_r\rrangle \} \rangle^M_\rho .
        \end{equation}
        Here, $\rho$ is assumed to possess a factorisation of the form \eqref{eq:gfactorisation} and  
        the twisted sector states that appear on the right hand side are obtained from $\ell$ 
        Ishibashi states $|j_r\rrangle$, $r = 1, \dots \ell$ of the seed theory in the spirit of 
        eq.~\eqref{eq:twistedsectorstate}. The only difference is that Ishibashi states live in some 
        appropriate completion of $\mathcal{H}^\rho$ rather than $\mathcal{H}^\rho$ itself.\footnote{For 
        instance, they may be viewed as discontinuous linear functionals on $\mathcal{H}$.} The action 
        of elements $\sigma \in \mathcal{C}_\rho$ in the centraliser  subgroup on the Ishibashi states
        takes the form 
        \begin{equation} 
            \sigma |j_r\rrangle_\rho = |j_{\pi r}\rrangle_\rho \, ,
        \end{equation} 
        where $\pi = \pi_\sigma$ is a permutation that exchanges two cycles of the same length. Note 
        that the action of the factors $\omega_r^{\nu_r}$ in the factorisation formula for $\sigma$ 
        is trivial since contributions from holomorphic and anti-holomorphic components cancel each 
        other. The overlap of any two of these Ishibashi states is given by 
        \begin{equation} \label{eq:overlapIshi1}
            \ _{\rho'}\llangle j_r' |\sigma \hat \STq^{\frac12(L_0 + \bar L_0  - \frac{c}{12})} |j_r \rrangle_{\rho} 
            = \delta^{(\ell)}_{j_{r}', j_{\pi r}} \, \delta_{\rho',\rho} \, 
            \prod_{r=1}^\ell \, \chi_{j_r}(\tfrac{\hat t}{w_r}) .   
        \end{equation}
        Here, $\sigma \in \mathcal{C}^M_\rho$ is in the centraliser  subgroup of $\rho$ and $\pi = 
        \pi_\sigma$, as before.
        Furthermore\footnote{In accordance with footnote \ref{footnote:Omega}, we could also have 
        written $j_\Omega$ instead of $\bar \jmath$ in eq.~\eqref{eq:chi_j(t)}.},
        \begin{align}\label{eq:chi_j(t)}
            \chi_{j} (t) := \text{Tr}_{\mathcal{H}_{j,\bar \jmath}} {\STq}^{\frac{1}{2}(L_0 + \bar L_0  - \frac{c}{12})}.
        \end{align}
        Let us stress that the Ishibashi states $|j_r\rrangle_\rho$ we have 
        introduced are not yet projected to the subspace of $\mathcal{C}_\rho$ invariant states. 
        Given our Ishibashi states $|j_r\rrangle_\rho$, we construct the linear combination 
        \begin{align} \label{eq:atwisted} 
            |a\rangle_{\rho} = \sum_{\{j_r\}} \ \frac{a_{j_1} \cdots a_{j_\ell}}{\sqrt{|\mathcal{C}_\rho^M|}}\,  |j_r \rrangle^M_{\rho}  
        \end{align}
        with coefficients formed from products of the coefficients that appeared in the boundary 
        state $|a\rangle$ of the seed theory. The overlap of any two of these states can be 
        easily computed from the overlap \eqref{eq:overlapIshi1}. 
        Note that the states $|a\rangle_\rho$ are actually within the subspace of $\mathcal{C}_\rho$ invariant elements, 
        i.e.~the projector $\Pi^\rho_0$ acts trivially on these states. 
        The normalisation we chose is the one that is appropriate for Cardy consistent boundary states, 
        see e.g.~\cite{Belin:2021nck} or Section 4.3 of \cite{Gaberdiel:2021kkp}. The states
        \eqref{eq:atwisted} constitute the first ingredient we shall use in constructing our 
        interfaces. 
        
        \paragraph{Transmissive part of the interfaces.}As a second ingredient of our construction, we need to discuss conformal interfaces for 
        $\mathrm{Sym}^p(\mathcal{M})$. After the folding trick, such an interface is described as
        a boundary state in which the holomorphic fields of the theory on the upper half plane are 
        glued to the anti-holomorphic fields of the second theory that was folded up from the lower 
        half plane and vice versa. Ishibashi states for the ``permutation boundary states'' \cite{Recknagel:2002qq} in
        $\textrm{Sym}^p(\mathcal{M}) \times \textrm{Sym}^p(\mathcal{M})$ will be denoted by 
        \begin{align} 
            |j_r,i_s\rrangle^{p,p}_{\tau_-,\tau_+}
        \end{align} 
        with $\tau_\pm \in S_p$ that are conjugate to each other, i.e.~$[\tau_-] = 
        [\tau_+]$. 
        More concretely, if we formally\footnote{Of course $|j_r\rrangle_{\tau}$ is not a pure tensor product of left and right movers, but a sum of such products. Our notation captures the structure of the individual pure tensor summands.} write
        \begin{align}
            |j_r\rrangle_{\tau} =|j_r\rangle_{\tau} \overline{|j_r\rangle}_{\tau},
        \end{align}
        then
        \begin{align}\label{eq:formal_split_of_reflecting_boundary}
            |j_r,i_s\rrangle^{p,p}_{\tau_-,\tau_+} = |j_r\rangle_{\tau_-} \overline{|i_s\rangle}_{\tau_-}|i_s\rangle_{\tau_+} \overline{|j_r\rangle}_{\tau_+}.
        \end{align}  
        The action of the centraliser  of $\tau_+$ and $\tau_-$ on these 
        Ishibashi states is given by 
        \begin{equation}\label{eq:sigma_acting_on_ji}
            \sigma_+ \sigma_- \, |j_r,i_s\rrangle^{p,p}_{\tau_-,\tau_+} =  |\alpha^{\nu_{\pi^- r}^-}j_{\pi^-r}\rangle_{\tau_-} \overline{|\alpha^{\nu^-_{\pi^-s}} i_{\pi^-s}\rangle}_{\tau_-}|\alpha^{\nu^+_{\pi^+ s}}i_{\pi^+ s}\rangle_{\tau_+} \overline{|\alpha^{\nu^+_{\pi^+r}}j_{\pi^+ r}\rangle}_{\tau_+}  ,
        \end{equation} 
        where 
        \begin{align}
             \alpha = e^{-2\pi i L_0 } .
        \end{align}
        The overlaps of these Ishibashi states take the form 
        \begin{equation} \label{eq:overlapIshi2}
            \ _{\tau'_\pm}\llangle j'_r,i'_s |\sigma_-\sigma_+ \hat \STq^{L_0 + \bar L_0  - \frac{c}{12}} 
            |j_r,i_s \rrangle_{\tau_\pm} 
            \hspace{- 0.1 cm}=  
            \delta^{(2\ell)}_{i'_{k},i_{\pi_\pm k}}
            \delta^{(2\ell)}_{j'_{k},j_{\pi_\pm k}}
            \delta^{(2)}_{\tau'_\pm,\tau_\pm} 
            \chi_{i,\sigma_{+},\sigma_{-}}(t)  \chi_{j,\sigma_-,\sigma_+}(t).
        \end{equation}
        In particular, they are only non-vanishing if $j_{\pi^- r} = j_{\pi^+ r}$. 
        To spell out $\chi_{j,\sigma_{-},\sigma_+}(t)$, recall that $\pi^\pm$ is a product of permutations that shuffle twisted sectors associated to cyclic permutations of length $w$.
        Hence, if we write $\pi^\pm$ as a product
        $\pi^\pm = \pi_1^\pm \pi_2^\pm \dots$ of cyclic permutations, then each permutation $\pi_{a}^\pm$ has two integers associated to it, namely the winding $w_{\pi_a^\pm}$ of the sectors it permutes and the length $\ell_{\pi_a^\pm}=|\pi_a^\pm|$.
        
        The same is true for $\pi:=\pi_+\pi_-^{-1}= \pi_1 \dots \pi_m$, i.e.~for each $\pi_a$ we have pair $(w_a,\ell_a)$ of integers.
        The condition $j_{\pi^- r} = j_{\pi^+ r}$ is equivalent to the statement that  $j_{r} = j_{\pi r}$ i.e.~$j_{r} = j_{\pi_{a} r}$ for all $a$. 
        In particular, if we denote by $\mathbf{a}$ the set of labels appearing in the cycle $\pi_a$, then we can introduce $j_{\mathbf{a}}$ as being equal to $j_r$ for some not further specified $r\in \mathbf{a}$ and this is a well defined prescription.
        In terms of 
        \begin{align}
            \chi_{j}(t , \bar t \, ) := \text{Tr}_{\mathcal{H}_{j,\bar \jmath}}\ \STq^{L_0 - \frac{c}{24}} \bar{\STq}^{\bar L_0 - \frac{c}{24}},
        \end{align}
        $\chi_{j,\sigma_-,\sigma_+}(t)$ can then be expressed as
        \begin{align}
            \chi_{j,\sigma_-,\sigma_+}(t) = \prod\limits_{a=1}^m \chi_{j_{\mathbf{a}}}\left(\tfrac{\ell_a t + \sum\limits_{k \in \mathbf{a}} \nu_k^- }{w_a},\tfrac{\ell_a t+ \sum\limits_{k \in \mathbf{a}} \nu_k^+}{w_a}\right).
        \end{align}
        Once again, we introduce a linear combination of these twisted Ishibashi states by summing 
        over the labels $j_r,i_s$. But in this case, the sum over the labels is not sufficient in 
        order to ensure $\mathcal{C_{\tau_\pm}}$ invariance. Hence, we also have to perform the 
        relevant projection 
        \begin{align} \label{eq:1twisted}
            |\mathbb{I}\rrangle^{p,p}_{\tau_-,\tau_+} = \Pi^{\tau_+}_0 \Pi^{\tau_-}_0 \ \sum\limits_{j_r,j_s}  
            |j_r,i_s\rrangle^{p,p}_{\tau_-,\tau_+} \qquad \textrm{ where } \qquad 
            \Pi^{\tau_\pm}_0 = \frac{1}{|\mathcal{C}^p_{\tau_\pm}|} \, 
            \sum_{\sigma_\pm \in \mathcal{C}^p_{\tau_\pm}}  \ \sigma_\pm  . 
        \end{align}
        The overlap between any two such states can be computed with the help of 
        formula \eqref{eq:overlapIshi2}. This completes the discussion of the second ingredient for 
        our construction of interfaces. 
        
        \paragraph{Averaging over gauge orbits.} As the final step in the construction of the boundary states \eqref{eq:boundarystateschi}, 
        we now perform the averaging over gauge orbits. This gives the gauge invariant state
        \begin{align}\label{eq:patwist}
            |p,a_\pm \rangle^{N_\pm}_{\mathscr{O}_\pm} = \tfrac{1}{\sqrt{|\mathscr{O}_-||\mathscr{O}_+|}}\sum\limits_{(\rho_\pm,\tau_\pm) \in \mathscr{O}_\pm}
            \hspace{-0.2 cm}| a_-\rangle_{\rho_-} \cdot |\mathbb{I}\rangle^{p,p}_{\tau_-, \tau_+}
             \cdot |a_+\rangle_{\rho_+}  ,
        \end{align}
        where 
        \begin{align} \label{eq:gauge_orbit}
            \mathscr{O}_\pm := [\tau,\rho_\pm]_p^{N_\pm} = \{(h \tau h^{-1},h \rho_\pm h^{-1}) : h \in S_{N_\pm}\}
        \end{align}
        are $S_{N_\pm}$ gauge orbits associated to a choice of $\rho_\pm \in S_{N_\pm - p}$ and $\tau \in S_p$. We added the subscript $p$ to remind ourselves, that $\tau$ and $\rho_\pm$ are elements of $S_{N_\pm - p}$ and $S_p$ and that the associated subgroup is an extra piece of data that we explicitly keep track of. That is, we distinguish for instance between $id_{S_p} \in S_p$ and $h \cdot id_{S_p}  \cdot h^{-1} = id_{h S_p} \in h S_p$.        
        
        Let us briefly comment on the choice of normalisation that we made in the definition of $|p,a_\pm \rangle^{N_\pm}_{\mathscr{O}_\pm}$. 
        Ultimately, we do not have an a priori principle that tells us what the ``correct'' normalisation should be. 
        We can only justify our choice by showing that its overlaps lead to the physical partition functions that we are 
        interested in, which indeed will turn out to be the case. However, we can at least argue, already before any computation, that 
        our guess is very natural one: Since we normalised appropriately the mutually orthogonal states $|a_\pm\rangle_{\rho_\pm}$ and
        $|\mathbb{I}\rangle_{\tau_-,\tau_+}$ to represent reflective/transmissive boundaries, we should simply divide out the square 
        root of the size of the orbits that we sum over such that the sum does not alter the already correct norm. 
        
        Now that we have constructed the boundary states $|p,a_\pm \rangle^{N_\pm}_{\mathscr{O}_\pm}$ associated to some choice of $S_{N_\pm}$ orbits, we finally perform a weighted sum over all choices of $\mathscr{O}_\pm$ to obtain 
        the boundary state 
        \begin{tcolorbox}[left=-30pt,right=0pt,top=-10pt,bottom=0pt]
            \begin{align}\label{eq:def:pra}
                |p,a_\pm;\chi_-,\chi_p,\chi_+\rangle = \sum\limits_{\tau \in S_p} 
                \sum\limits_{\rho_\pm \in S_{N_\pm -p}} \frac{\chi_-(\rho_-) \chi_p(\tau) \chi_+(\rho_+)}{|[\tau]||[\rho_+]||[\rho_-]|}\, 
                \,  
                |p,a_\pm \rangle^{N_\pm}_{\mathscr{O}_\pm}  .
            \end{align}
        \end{tcolorbox}
        \noindent As advertised with eq.~\eqref{eq:boundarystateschi} at the end of our first qualitative description of the construction,  
        formulating eq.~\eqref{eq:def:pra} is the purpose of this section. 
        
        \paragraph{Summary of the construction.} Let us finally summarise all ingredients of eq.~\eqref{eq:def:pra} in a concise manner. $\chi_\pm,\chi_p$ are 
        characters of the $S_{N_\pm-p}$ and $S_p$, respectively. The 
        normalising prefactors involving the numbers $|[\rho_\pm]|$ and $|[\tau]|$ of elements in 
        the conjugacy classes of $\rho_\pm \in S_{N_\pm -p}$ and $\tau \in S_p$, respectively, appear here because 
        we are summing over group elements rather than conjugacy classes. The states we sum over are
        defined in eq.~\eqref{eq:patwist} through a weighted average over elements $g_\pm$ of $
        S_{N_\pm}$. Their definition involves a product of three states. Two of them are the states $|a_\pm\rrangle_{\rho'_\pm}$, 
        which are purely reflective with reflection coefficients that are specified by the coefficients 
        of boundary states $|a_\pm\rangle$ in the seed theory, see eq.~\eqref{eq:atwisted}.
        The third factor in the states we average over is purely transmitting, see eq.~\eqref{eq:1twisted}. 

    \subsection{Reflectivity and transmissivity of the interfaces}\label{sec:reflectivity_transmissitivty}

        Now that we have defined the interfaces in eq.~\eqref{eq:def:pra} let us pause for a moment and 
        compute a first physical quantity that may help to better understand them. Concretely,  we determine the (stress-energy) reflectivity and transmissivity as 
        proposed in \cite{Quella:2006de}. It measures e.g.~how much energy gets transmitted through the 
        interface or, after passing to the folded setup, how much energy gets passes from e.g.~from  
        $+$-components of the folded CFT to the $-$-component and vice versa. The interpretation of the 
        reflectivity is similar. More concretely, after folding the theory to the upper half plane we need 
        to compute correlations of the product $T^\eta(z) \bar T^\eta(\bar z)$ with $\eta \in \{+,-\}$ in the presence 
        of the interface boundary state \eqref{eq:def:pra}. Replacing the Virasoro fields by the corresponding 
        states, the quantity in question is 
        \begin{align}\label{def:R}
            R_{\eta\eta'} = \frac{\langle 0|L_2^\eta \bar L_2^{\eta'}|p,a_\pm;\chi_-,\chi_p,\chi_+\rangle}
            {\langle 0|p,a_\pm;\chi_-,\chi_p,\chi_+\rangle}. 
        \end{align}
        The overlaps that appear in this expressions are not that difficult to compute. The simplest is clearly 
        the overlap in the denominator which is obviously given by 
        \begin{align} 
            \langle 0|p,a_\pm;\chi_-,\chi_p,\chi_+\rangle =  \chi_{r_-}(id) \chi_{p}(id) \chi_{r_+}(id) 
            \sqrt{\binom{N_-}{p}\binom{N_+}{p}}\frac{ (a_-)^{N_--p}_0(a_+)^{N_+-p}_0}{\sqrt{(N_--p)!(N_+-p)!}}.
        \end{align}
        The overlap in the numerator is not that much harder to compute. Note that the Virasoro fields $T^\pm$
        and $\bar T^\pm$ of the two symmetric product orbifold CFTs are in the untwisted sector and hence the 
        overlap in the numerator sees essentially the same coefficients as that in the denominator. In formulas 
        this means only the term with $\tau = \rho_\pm = id$ in the expansion of the state \eqref{eq:def:pra} 
        can contribute to give
        \begin{align}
            \langle 0|L_2^\eta \bar L_2^{\eta'}|p,a_\pm;\chi_-,\chi_p,\chi_+\rangle = \chi_{r_-}(id) \chi_{p}(id) 
            \chi_{r_+}(id)  \langle 0|L_2^i \bar L_2^{j}|p,a_\pm\rangle^{N_\pm}_{[id,id]_p^{N_\pm}}.
        \end{align}
        The state on the right hand side is a special case of eq.~\eqref{eq:patwist} which consists of a 
        single term only and has trivial coefficient, 
        \begin{align}
            |p,a_\pm\rangle^{N_\pm}_{[id,id]_p^{N_\pm}} = 
            |a_-\rangle \cdot |\mathbb{I}\rangle^{p,p} \cdot |a_+\rangle .            
        \end{align}
        Let us now first address the off-diagonal elements of the reflection matrix $R$, i.e.~the 
        matrix elements $R_{+-} = R_{-+}$. Each of the $N_- -p$ components within the state $|a_\pm\rangle$ contributes 
        a factor of $(a_\pm)_0$. 
        Moreover, 
        every transmitting component gives rise to a term $c_\mathcal{M} 2(2^2-1)/12 = c_\mathcal{M}/2$ from the 
        commutations relations of the Virasoro generators in the seed theory. Since we need to sum over all 
        the transmitting components of which there are $p$, we find the simple relation 
        \begin{align} 
            \langle 0|L_2^\pm \bar L_2^{\mp}|p,a_\pm;\chi\rangle = p \frac{c_\mathcal{M}}{2} \, 
            \langle 0|p,a_\pm;\chi\rangle.
        \end{align} 
        between the 
        overlaps in the numerator and denominator.
        Here, we used the shorthand $\chi = (\chi_-,\chi_p,\chi_+)$ to denote the triple of characters. For the 
        diagonal elements of $R$, the transmissive boundaries do not contribute at all. The contribution of all 
        the reflective factors, on the other hand,  sum to
        \begin{align}
            \langle 0|L_2^\pm \bar L_2^{\pm}|p,a_\pm;\chi\rangle = (N_\pm - p)\frac{c_{\mathcal{M}}}{2} 
            \langle 0|p,a_\pm;\chi \rangle  .
        \end{align}
        Plugging the previous two relations back into the definition \eqref{def:R} of the reflection matrix $R$
        we conclude that 
        \begin{align}\label{eq:R_matrix}
            R = \frac{c_\mathcal{M}}{2} \left(
            \begin{array}{cc}
            N_--p & p \\
             p & N_+-p \\
            \end{array}
            \right). 
        \end{align}
        From the matrix elements, we read off the reflectivity of our interface, which is given by  
        \begin{align} \label{eq:reflectivity} 
            \mathcal{R} := \frac{2}{c_{\mathcal{M}}(N_-+N_+)}(R_{--} + R_{++}) = 1-\frac{2p}{N_-+N_+}, 
        \end{align}
        and the transmissivity  
        \begin{align} \label{eq:transmittivity}
            \mathcal{T} = \frac{2p}{N_-+N_+}.
        \end{align}
        Note that the ``unitarity condition''  $\mathcal{R}+\mathcal{T}=1$ is indeed satisfied. We also observe that 
        $p=0$ corresponds to a purely reflective interface and $N_- = N_+ = p$ is purely transmissive, as expected.
        Let us end this short interlude with three minor comments on the result. 

        First, note that the formula \eqref{eq:R_matrix} for $R$ may be brought into its standard form
        \begin{align}
            R = \frac{c_{\mathcal{M}}}{2} \frac{N_-N_+}{N_-+N_+}
            \left[
                \left(
                    \begin{array}{cc}
                        \frac{N_-}{N_+} & 1 \\
                        1 & \frac{N_+}{N_-} \\
                    \end{array}
                \right)
                +
                \omega_b
                \left(
                    \begin{array}{cc}
                        1 & -1 \\
                       -1 & 1 \\
                    \end{array}
                \right)
            \right]
        \end{align}
        from which we read off 
        \begin{align}
            \omega_b = 1 - \frac{p}{N_-} - \frac{p}{N_+}.
        \end{align}
        Second, note that the reflectivity and transmissivity coefficients are insensitive to most of the data that we used to construct 
        the interfaces in the previous subsection. In fact, only the number $p$ of transmissive components and 
        the numbers $N_\pm-p$ of reflective components enter. The choice of the boundary conditions $a_\pm$ 
        or the characters $\chi_\pm$ and $\chi_p$ on the other hand has no effect. The reason for this is rather clear from the 
        derivation: The stress-energy reflectivity and transmissivity only receive contributions from the 
        untwisted sector of the symmetric product orbifold. 
        In this sense, $\mathcal{R}$ and $\mathcal{T}$ are very coarse grained characterisations of the interfaces $\Ipa$.
        
        Let us finally make a somewhat related observation. We decided to determine the matrix $R$ above 
        from the concrete formula \eqref{eq:def:pra}. However, the authors of \cite{Quella:2006de} noted already that for 
        a certain class of interfaces, the matrix $R$ is rather easy to compute by more general arguments. They considered a situation in 
        which the chiral algebras $\mathcal{A}_-$ and $\mathcal{A}_+$ of the two CFTs the interface interpolates
        between possess a common subalgebra $\mathcal{C}$. If the interface $\mathcal{I}$ in question is completely 
        transmissive for the common subalgebra $\mathcal{C}$ and reflective for all other degrees of freedom, then 
        the chiral algebra of the folded theory is broken down to a product 
        \begin{align} \label{eq:AAfactorisation}
            \mathcal{A}_- \otimes \mathcal{A}_+ \rightarrow     \mathcal{A}_-/\mathcal{C} \otimes \mathcal{C} \otimes  \mathcal{C} \otimes \mathcal{A}_+/\mathcal{C} .
        \end{align}
        by the presence of the interface $\mathcal{I}$. Denoting the corresponding boundary state by 
        $|\mathcal{I}\rangle$, we can easily compute 
        \begin{align}
            \langle 0 | L_2^{\pm} \bar L_2^{\mp} |\mathcal{I}\rangle&= \langle 0 | (L_2^{(\mathcal{A}/\mathcal{C})_\pm}+
            L_2^{\mathcal{C}_\pm})(\bar L_{2}^{(\mathcal{A}/\mathcal{C})_\mp}+\bar L_{2}^{\mathcal{C}_\mp})  
            |\mathcal{I}\rangle \nonumber \\[2mm]
            &= \langle 0 | (L_2^{(\mathcal{A}/\mathcal{C})_\pm}+L_2^{\mathcal{C}_\pm})
            ( L_{-2}^{(\mathcal{A}/\mathcal{C})_\mp}+ L_{-2}^{\mathcal{C}_\pm})  |\mathcal{I}\rangle = \frac{c}{2}
        \end{align}
        and
        \begin{align}
            \langle 0 | L_2^{\pm} \bar L_2^{\pm} |\mathcal{I}\rangle&= \langle 0 | 
            (L_2^{(\mathcal{A}/\mathcal{C})_\pm}+L_2^{\mathcal{C}_\pm})(\bar L_{2}^{(\mathcal{A}/\mathcal{C})_\pm}+\bar L_{2}^{\mathcal{C}_\pm})  |\mathcal{I}\rangle \nonumber \\[2mm]
            &= \langle 0 | (L_2^{(\mathcal{A}/\mathcal{C})_\pm}+L_2^{\mathcal{C}_\pm})
            ( L_{-2}^{(\mathcal{A}/\mathcal{C})_\pm}+ L_{-2}^{\mathcal{C}_\mp})  |\mathcal{I}\rangle = \frac{c_\pm-c}{2},
        \end{align}
        where $c_\pm$ are the central charges of $\mathcal{A}_\pm$ and $c$ the central charge of the chiral subalgebra 
        $\mathcal{C}$. In the two short calculations we have split the Virasoro elements $L^\pm$ and $\bar L^\pm$ of the chiral 
        algebras $\mathcal{A}_\pm$ according the factorisation \eqref{eq:AAfactorisation}. Then, we used the Ishibashi
        conditions of the interface boundary state $|\mathcal{I}\rangle$ to replace anti-holomorphic components by 
        holomorphic ones in passing from the first to the second line. In the last step we used commutation relations 
        of the Virasoro elements to evaluate the overlap in terms of the central charges. In our case, the chiral 
        algebras $\mathcal{A}_\pm$ are those of the symmetric product orbifolds $\textrm{Sym}^{N_\pm}(\mathcal{M})$
        while the common subalgebra $\mathcal{C}$ that is preserved by the interface is the chiral algebra of 
        $\textrm{Sym}^p(\mathcal{M})$. Plugging in the associated values of the central charges $c_\pm = c_{\mathcal{M}}
        N_\pm$ and $c= c_{\mathcal{M}}p$ reproduces eq.~\eqref{eq:R_matrix}.
   
    \subsection{Partition functions of interface changing operators}\label{sec:interface_changing_partition_function}
        
        A very interesting object related to the interfaces that we defined in section \ref{sec:def_sym_orbifold_boundary_states} is the partition 
        function that enumerates interface changing operators. To be concrete, let us choose one interface $|p^R,a^R_\pm;
        \chi^R\rangle$  to the right of the origin and another $|p^L,a^L_\pm;\chi^L\rangle$ to the left. The associated 
        partition function for interface changing operators is obtained from the associated overlap as follows 
        \begin{align}\label{eq:left_right_overlap}
            \mathcal{Z}^{N_\pm}_{(p^L,a^L_\pm;\chi^L),(p^R,a^R_\pm,\chi^R)}(\tst) =  \text{Tr}[\STq^{L_0-\frac{c}{24}}] =  
            \langle p^L, a^L_\pm;\chi^L| \hat{\STq}^{\frac{1}{2}(L_0 + \bar L_0 -\frac{c}{12})} |p^R,a^R_\pm;\chi^R\rangle. 
        \end{align}
        Here, $c = (N_+ + N_-)c_{\mathcal{M}}$ is the total central charge of the folded theory and the parameter 
        $\hat \STq$ on the right is related to $\STq = \exp(2 \pi i \tst)$ by modular transformation $\hat \STq = \exp(2 \pi 
        i\hat\tst)$ with $\hat \tst = -1/\tst$. We continue to use the shorthand 
        $\chi$ for the entire triple $(\chi_-,\chi_p,\chi_+)$ of characters. These overlaps 
        could certainly be computed in full generality, but since we have no need for such a general result, we 
        restrict to the case in which all the characters $\chi^{L/R}$ are trivial. This turns out to be the 
        only one that is relevant for the holographic relation with $AdS_2$ branes. 
        
        As usual, the partition functions of the type defined in eq.~\eqref{eq:left_right_overlap} are somewhat inconvenient to describe individually, but as whole organise into a rather simple grand canonical partition function.
        Concretely, we propose that the grand canonical partition function 
        \begin{align}\label{eq:GCZ}
            &\mathcal{Z}_{a^{L/R}_\pm}[\mu_\pm,\rho_{L/R};\tst]  := 
            \hspace{-0.2 cm}
            \sum\limits_{N_\pm=0}^\infty \sum\limits_{p^{L/R}=0}^{\textrm{min}(N_\pm)} \hspace{-0.2 cm}
            \mu_+^{N_+} \mu_-^{N_-} \rho_L^{N_++N_--2p^L} \rho_R^{N_++N_--2p^R} 
              \mathcal{Z}^{N_\pm}_{(p^L,a^L_\pm;{\bf 1}),(p^R,a^R_\pm,{\bf 1})}(\tst) 
        \end{align}
        is given by the exponential
        \begin{align}\label{eq:grand_canonical_exponential}
            \mathcal{Z}_{a^{L/R}_\pm}[\mu_\pm,\rho_{L/R};\tst] = \exp(\mathcal{Z}_{C}[\mu_\pm,\rho_{L/R};\tst]+
            \mathcal{Z}_{O}[\mu_\pm,\rho_{L/R};\tst])
        \end{align}
        of the sum of a ``connected\footnote{The term \emph{connected} is motivated in this context by the covering map approach to symmetric product orbifolds, in which the exponent of the grand canonical partition function is generated by covering maps with a connected covering space.} closed string'' part
        \begin{align}\label{eq:Z_orbifold_closed} 
            \mathcal{Z}_{C}[\mu_\pm,\rho_{L/R};\tst] = \sum\limits_{k=1}^\infty \mu_-^k\mu_+^kT_k Z_c(\tst)
        \end{align}
        and a ``connected open string'' part
        \begin{tcolorbox}[left=0pt,right=0pt,top=-10pt,bottom=0pt] 
            \begin{align}\label{eq:Z_orbifold_open}  
                 \mathcal{Z}_{O}[\mu_\pm,\rho_{L/R};\tst]
                 = \hspace{-0.6 cm}\sum\limits_{A,B\in\{L,R\}}
                 \sum\limits_{k=1}^\infty\sum\limits_{w|k} \rho_A^w\rho_B^w\mu_-^{k-w\delta_A^R\delta_B^L}
                 \mu_+^{k-w\delta_A^L\delta_B^R}                
                 \frac{1}{w}\hat{Z}^{AB}_o\left(\tfrac{(2k - w + w\delta_A^B)\hat\tst}{w^2}\right).
            \end{align}
        \end{tcolorbox}
        \noindent
        $\mathcal{Z}_{C}$ and $\mathcal{Z}_{O}$ involve two types of partition functions for the bulk theory. On 
        the one hand, there is the ``closed string'' partition function $Z_c$ defined as
        \begin{equation} \label{eq:seedZc} 
            Z_c(\tst) = Z_\mathcal{M} (\tst,\bar \tst)_{\tst = \bar \tst}  
        \end{equation} 
        which is the 
        modular invariant bulk partition function of the seed theory $\mathcal{M}$ restricted to the diagonal 
        $\tst = \bar \tst$. 
        On the other hand, there are the open string partition functions $Z^{AB}_o$ defined as
        \begin{align} \label{eq:seedZo}
            Z^{LL}_o = Z_{a^L_-,(a^L_+)^*}, \quad 
            Z^{LR}_o = Z_{a^L_-,a^R_-}, \quad 
            Z^{RL}_o = Z_{a^L_+,a^R_+} \quad 
            \text{and}
            \quad
            Z^{RR}_o = Z_{(a^R_+)^*,a^R_-} 
        \end{align}
        where $Z_{a,b}(\tst) = Z_{b*,a*}(\tst)$ is the partition function of the annulus with boundary 
        conditions $a$ and $b$ imposed along the two components of the boundary. It is famously related to 
        the overlap of the boundary states in the seed theory through 
        \begin{equation}
            Z_{a,b}(t)  = \langle a |\, \hat \STq^{\frac12 (L_0 + \bar L_0 - \frac{c_\mathcal{M}}{12})}
            \, |b\rangle =  \hat Z_{a,b}(\hat \tst)  .  
        \end{equation} 
        with $\hat t = - 1/t$ as usual. Strictly speaking, formula \eqref{eq:grand_canonical_exponential} is a conjecture.
        We decided against a rigorous derivation of the full formula in this work. However, we do prove it carefully for 
        the special case $\rho_L = 0$ in appendix \ref{app:Overlap_full_computation}. Considering this special case is 
        necessary in order to verify that we made the correct normalisation choices in section
        \ref{sec:def_sym_orbifold_boundary_states}. 
        It is also sufficient in the sense that the case of $\rho_L = 0$ is fully sensitive to all normalisations that 
        we chose in the construction 
        of the interface \eqref{eq:def:pra}.
        Assuming that, with the correct choice of normalisation, the grand canonical partition function exponentiates,
        eq.~\eqref{eq:grand_canonical_exponential} then follows by computing the ``connected world sheet'' sums $\mathcal{Z}_C$ 
        and $\mathcal{Z}_O$ in the exponent. 
        The calculation of these ``connected world sheet'' sums is the most illuminating part of the computation. 
        This is why we dedicate the remainder of the current section to this task
        \smallskip  
        
\begin{figure}
            \centering
            \includegraphics[width= \linewidth]{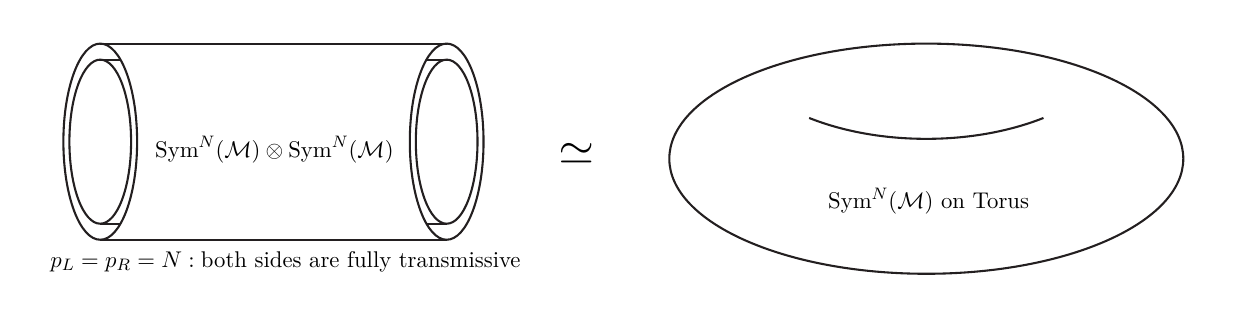}
            \caption{Coincidence of partition function of fully transmissive boundaries and torus partition function. }
            \label{fig:Trans2Torus}
        \end{figure}

        Let us start by discussing the limit $\rho_{L}=\rho_R = 0 $ in which only transmissive boundaries are present.  
        When both $\rho$ parameters vanish, so does $\mathcal{Z}_O$.
        This means that only the first term in the exponent contributes in the limit. 
        Hence, our result \eqref{eq:grand_canonical_exponential} simplifies to
        \begin{align}\label{eq:grand_canonical_torus_Hecke}
             \sum\limits_{N=0}^\infty \kappa^N \langle N|\STq^{\frac{1}{2}(L_0^{(1)} + \bar{L}_0^{(2)} - 
             \frac{c}{12})} \bar{\STq}^{\frac{1}{2}(\bar{L}_0^{(1)} + L_0^{(2)} - \frac{c}{12})}|N\rangle = 
             \exp\left[\sum\limits_{k=1}^\infty 
             \mu_-^k\mu_+^k T_k Z(\tst, \bar \tst) \right],
        \end{align}
        where $Z= Z_c$ is the torus partition function of the seed theory, as before, $T_k$ is the usual Hecke 
        operator (see eq.~\eqref{eq:def_Hecke_operator}), and $|N\rangle$ is the fully transmissive boundary state 
        \begin{align}
            |N\rangle = |p=N,a_\pm;N_+ = N = N_-\rangle .
        \end{align}
        Obviously, this state does not depend on the choice of $|a_\pm\rangle$ since we do not allow 
        for any reflection. The state $|N\rangle$ is associated with the trivial defect line of the symmetric 
        product orbifold. Hence, we expect the partition function to coincide with the torus partition function
        \eqref{eq:torus_grand_canonical}, as shown in figure \ref{fig:Trans2Torus}. This is indeed what we observe. 
        \smallskip
        
        To verify eq.~\eqref{eq:grand_canonical_torus_Hecke}, we start by plugging in the definition
        \eqref{eq:def:pra} of $|N\rangle$ on the left hand side of the equation and simplifying, which gives
        \begin{align}
            \langle N|\STq^{\frac{1}{2}(L_0^{(1)} + \bar{L}_0^{(2)})} \bar{\STq}^{\frac{1}{2}(\bar{L}_0^{(1)} + L_0^{(2)})}| N \rangle 
            = \frac{1}{N!}\sum\limits_{\tau \in S_N}  
            \sum_{\sigma \in \mathcal{C}^N_{\tau}}  \sum\limits_{j_r,i_s} 
             \,^{N,N}_{\tau,\tau}\llangle j_r,i_s|  \STq^{L_0}\bar \STq^{\bar L_0} \sigma |j_r,i_s\rrangle^{N,N}_{\tau,\tau}. \label{eq:overlap_simplified}
        \end{align}
        This formula for the overlap manifestly is a sum of contributions each associated to a permutation 
        $\tau \in S_N$ and an element $\sigma$ of the centraliser of $\tau$. As explained in section
        \ref{sec:symmetric_product_notation}, $\sigma$ includes a factor $\pi_{\sigma}$ that permutes $j$ and $i$ labels associated to
        cycles of the same length within the permutation $\tau$. 
        Only those terms in the sum for which $j_r = j_{\pi_\sigma r}$ and $j_s = j_{\pi_\sigma s}$ are non-vanishing.
        The full answer factorises into individual contributions associated to the cycles of $\tau$ and  
        $\pi_{\sigma}$. This suggests\footnote{As mentioned before, the computation in 
        appendix~\ref{app:Overlap_full_computation} proves the claim rigorously.} that the full grand 
        canonical partition function is an exponential of sums of connected contributions $Z_{(w,\ell)}$ 
        associated to a $\pi_{\sigma}$ that is a single cycle of length $\ell$ and a $\tau$ that is a product 
        of $\ell$ cycles of length $w$ such that $N = w \ell$. To compute the connected contributions 
        $Z_{(w,\ell)}$, let us define 
        \begin{align}
            \tau_{(w,\ell)} = \prod\limits_{k=0}^{\ell-1}(kw+1,kw+2,\dots, kw+w).
        \end{align}
        From eq.~\eqref{eq:overlap_simplified}, we can then directly conclude that
        \begin{align}
            Z_{(w,\ell)} =&  \frac{|[\tau_{(w,\ell)}]|}{N!}
            \sum_{\substack{\sigma \in \mathcal{C}_{\tau_{(w,\ell)} }\\ |\pi_{\sigma}|=\ell}}  \sum\limits_{j_r,i_s} 
             \,^{N,N}_{\tau_{(w,\ell)},\tau_{(w,\ell)}}\llangle j_r,i_s|  \STq^{L_0-\tfrac{c}{24}}\bar \STq^{\bar L_0-\tfrac{c}{24}} \sigma |j_r,i_s\rrangle^{N,N}_{\tau_{(w,\ell)},\tau_{(w,\ell)}}. 
        \end{align}
        The action of $\sigma$ on the states was spelled out in eq.~\eqref{eq:sigma_acting_on_ji} which leads to
        \begin{align}
            Z_{(w,\ell)} =& \tfrac{1}{w^\ell \ell!} (\ell-1)! \sum\limits_{i_1=0}^{w-1}\dots\sum\limits_{i_\ell=0}^{w-1} Z_\mathcal{M}\left(\tfrac{\ell \tst + \sum\limits_{k=1}^\ell i_k}{w},\tfrac{\ell \bar \tst + \sum\limits_{k=1}^\ell i_k}{w}\right).
        \end{align}
        Using that the partition function $Z_\mathcal{M}$ of the bulk theory is invariant under modular $T$ transformations, 
        i.e.~under shifts of the argument by arbitrary integers, the previous formula simplifies to
        \begin{align}
            Z_{(w,\ell)} =& \tfrac{1}{w \ell} \sum\limits_{i=0}^{w-1} Z_\mathcal{M}\left(\tfrac{\ell \tst +i}{w},\tfrac{\ell \bar \tst + i}{w}\right) = \frac{1}{N} \sum\limits_{i=0}^{w-1} Z_\mathcal{M}\left(\tfrac{N \tst}{w^2} + \tfrac{i}{w},\tfrac{N \tst}{w^2} + \tfrac{i}{w}\right).
        \end{align}
        This is what we wanted to show. 
        \medskip 

        Let us now look at the more generic case, where only $\rho_L$ is sent to $0$ in our general formula \eqref{eq:grand_canonical_exponential}. In this case, we obtain
        \begin{align}\label{eq:ZthermSC2}
            \mathcal{Z}_{a^{L/R}}[\mu_\pm,0,\rho_R;\hat \tst] = \exp\left(\sum\limits_{k=1}^\infty \mu^k_-\mu^k_+T_k Z_c(\hat\tst) + \sum\limits_{k=1}^\infty\sum\limits_{w|k} \mu^k_-\mu^k_+\rho_R^{2w} \frac{1}{w} \hat{Z}^{RR}_o\left(2\tfrac{k\hat\tst}{w^2}\right)\right).
        \end{align}
        We now sketch how this result is derived. Again, just by plugging in the definition \eqref{eq:def:pra} and performing a 
        few elementary simplification steps, one can deduce that
        \begin{align}
            \langle N |{\hat \STq}^{L_0}|p,a\rangle 
            = \frac{1}{p!(N-p)!}\sum\limits_{\tau \in S_{p}}\sum\limits_{\rho \in S_{N-p}}  
            \sum_{\sigma \in \mathcal{C}^N_{\tau\rho}} \sum\limits_{j_r,i_s} 
            \,^{N,N}_{\rho\tau,\tau\hspace{- 1 pt}\rho}\llangle j_r,i_s|  {\hat \STq}^{L_0} \sigma | a\rangle_{\rho} |\mathbb{I}\rangle_{\tau, \tau}
            |a\rangle_{\rho}  \label{eq:N_p_Simplified},
        \end{align} 
        where $\sigma$ acts only on the states of the symmetric orbifold living in the lower half plane.
        At this point, we can also unfold using the notation introduced in eq.~\eqref{eq:formal_split_of_reflecting_boundary} and obtain
        \begin{align}
            \langle N |{\hat \STq}^{L_0}|p,a\rangle 
            = \frac{1}{p!(N-p)!}\sum\limits_{\tau \in S_{p}}\sum\limits_{\rho \in S_{N-p}}  
            \sum_{\sigma \in \mathcal{C}^N_{\tau\rho}} \sum\limits_{j_r,i_s} 
            \,_\rho\langle a| \langle i_s|^p_{\tau} \overline{\langle j_r|}^p_{\tau}  {\hat \STq}^{L_0} \sigma 
            |i_s \rangle^p_{\tau} \overline{|j_r \rangle}^p_{\tau} 
             |a\rangle_{\rho}  \label{eq:N_p_unfolded}.
        \end{align}
        The overlap can be interpreted as a sum of products of ``open string'' contributions (which involve reflecting 
        boundaries) and ``closed string'' contributions (which do not involve reflecting boundaries) that arise from 
        the interplay of different $\tau$, $\rho$ and $\sigma$. In the special case $p=0$, there are no closed string
        contributions.

        Furthermore, for $p=0$, the only connected contributions come from the conjugacy class $[\rho] = [(1 \dots N)]$
        of the maximal cyclic permutation. The choice of $\sigma$ does not matter: every element of the commutator acts trivially. Accordingly, these contributions to the partition are simply
        \begin{align}
            \frac{|[\rho]||\mathcal{C}_\rho|}{N!}\,_\rho\langle a| {\hat \STq}^{2 L_0 - \tfrac{c}{12}} |a\rangle_\rho =\,_\rho\langle a| {\hat \STq}^{2 L_0 - \tfrac{c}{12}} |a\rangle_\rho= \frac{1}{N}\hat Z_o\left(\tfrac{2\tst}{N}\right),
        \end{align}
        where $Z_o$ is defined in eq.~\eqref{eq:seedZo}.
        The factor $1/N$ on the right hand side of the equation originates from our choice to normalise $|a\rangle_\rho$ 
        by $1/\sqrt{\mathcal{C}_\rho}$, see eq.~\eqref{eq:atwisted}.
    
        More generally, we can find connected contributions 
        at $p=w(\ell-1)$ for $N = w \ell$. They arise for
        \begin{align}
            |\pi_\sigma| = \ell , \quad [\tau] = [\tau_{(w,\ell-1)}] , \quad [\rho] = [((\ell-1)w+1 \dots \ell w)].
        \end{align}
        Their contribution to the partition function is
        \begin{align}
            \frac{|[\tau]||[\rho]||\{\sigma \in \mathcal{C}_{\tau\rho}^N:|\pi_\sigma|=\ell\}|}{p!(N-p)!|\mathcal{C}_\rho^{N-p}|} \hat Z_o\left(\tfrac{2\ell\tst}{w}\right) = \frac{|\{\sigma \in \mathcal{C}_{\tau\rho}^N:|\pi_\sigma|=
            \ell\}|}{|\mathcal{C}_\tau^{p}||\mathcal{C}_\rho^{N-p}|^2} \hat Z_o\left(\tfrac{2\ell\tst}{w}\right) 
            = \frac{1}{w}\hat Z_o\left(\tfrac{2N\tst}{w^2}\right).
        \end{align}
        Assuming that the grand canonical partition function is obtained by the exponentiation of these connected contributions, we obtain eq.~\eqref{eq:ZthermSC2}. Appendix \ref{app:Overlap_full_computation}
        provides a detailed computation that establishes this result more rigorously.
        \smallskip

        Before we close this section, let us add a few comments concerning the formula \eqref{eq:Z_orbifold_open}.
        First of all, we have introduced $\mu_\pm$ and $\rho_{L/R}$ in order to unify the discussion of the various special cases obtained by sending some of the chemical potentials to zero in this section.
        However, we will ultimately be especially interested in the case
        \begin{align}  
             \mathcal{Z}_{O}[\mu,\hat\tst]
             := \mathcal{Z}_{O}[\mu,\mu,1,1,\hat\tst]
             = \hspace{-0.6 cm}\sum\limits_{A,B\in\{L,R\}}\sum\limits_{k=1}^\infty\sum\limits_{w|k}\frac{1}{w}\mu^{2k+w\delta_A^B-w}\hat{Z}^{AB}_o\left(\tfrac{(2k - w + w\delta_A^B)\hat\tst}{w^2}\right),
        \end{align}
        which is more conveniently described by summing over
        \begin{align}
            \ell = 2\tfrac{k}{w}+\delta_A^B-1       
        \end{align}
        instead of $k$. Indeed,
        \begin{tcolorbox}[left=0pt,right=0pt,top=-10pt,bottom=0pt]  
            \begin{align}\label{eq:ZO[t]}  
                 \mathcal{Z}_{O}[\mu,\hat\tst]
                 = \hspace{-0.6 cm}\sum\limits_{A,B\in\{L,R\}}\sum\limits_{\ell,w=1}^\infty \tfrac{1}{2w}(1-e^{i\pi(\ell + \delta_A^B)})\mu^{w \ell}\hat{Z}^{AB}_o\left(\tfrac{\ell\hat\tst}{w}\right).
            \end{align}
        \end{tcolorbox}
        \noindent
        In this equation, as in all other partition functions above, we have used the ``closed string'' modular parameter $\hat \tst$ as appropriate for the overlap of boundary states. Given the interpretation as a counting function for boundary/interface 
        changing operators, it may be more natural to perform a modular $S$ transformation and rewrite the grand canonical 
        partition function in terms of the dual modular parameter $\tst = - 1/{\hat\tst}$. Using $Z_o(\tst) = \hat Z_o(\hat \tst)$,
        we obtain 
        \begin{align} \label{eq:ZO[t]MT}  
             \mathcal{Z}_{O}[\tst]
             = \hspace{-0.6 cm}\sum\limits_{A,B\in\{L,R\}}\sum\limits_{\ell,w=1}^\infty \tfrac{1}{2\ell}(1-e^{i\pi(w + \delta_A^B)}){Z}^{AB}_o\left(\tfrac{\ell\tst}{w}\right).
        \end{align}
        In rewriting the expression we also exchanged $w$ and $\ell$. 
        
        In the remainder of this work, our focus is on the holographic relation with $AdS_2$ branes for tensionless
        superstring theory in $AdS_3 \times S^3 \times \mathbb{T}^4$. In this theory, the spacetime symmetry is 
        enhanced from the usual Virasoro algebra to the small $\mathcal{N} = 4$ superconformal algebra and hence 
        states of the CFT can be distinguished by their R-symmetry charge that is given by the eigenvalue of the 
        Cartan generators $K_0$ and $\bar K_0$ of the R-symmetry group. It is then natural to include these charges 
        in the counting functions for states. At the level of the seed theory, this means that the bulk and boundary 
        partition functions \eqref{eq:seedZc} and \eqref{eq:seedZo} become functions of another variable $\zeta$ in 
        addition to $t$, i.e. 
        \begin{equation} 
            Z_c(t) \ \rightarrow \  Z_c(t,\zeta)  , \quad 
            Z^{AB}_o(t) \ \rightarrow \  Z^{AB}_o(t,\zeta) . 
        \end{equation} 
        The corresponding change to formula \eqref{eq:Z_orbifold_open} is given by the substitution 
        \begin{eqnarray}
            \hat  Z ^{AB}_o \left(\frac{\ell \hat t}{w} \right) \ \longrightarrow\ 
            \hat Z ^{AB}_o \left(\frac{\ell \hat t}{w}, \ell \hat \zeta\right).  
        \end{eqnarray}
        The function $\hat Z$ is obtained from the partition function $Z$ through the prescription 
        \begin{equation}
            \hat Z (\hat t,\hat \zeta) = Z (-1/\hat t,\hat \zeta/\hat t) = Z(t,\zeta)  .   
        \end{equation} 
        We complement these qualitative remarks with more explicit formulas in section \ref{sec:partition_function_match} after establishing some more background on the supersymmetric four-torus in section \ref{sec:global_ads3_partition_function}. 
        
\section{\texorpdfstring{$AdS_2$}{AdS2} branes in tensionless \texorpdfstring{$AdS_3$}{AdS3} backgrounds} 
\label{sec:tensionless_strings}    
    
    In this section, we turn to the dual side and discuss the string theory of $AdS_2$ branes in $AdS_3$, 
    or more concretely, in type IIB superstring theory on $AdS_3 \times S^3 \times \mathbb{T}^4$ with one 
    unit of NS-NS flux. In the hybrid formulation or Berkovits-Vafa-Witten \cite{Berkovits:1999im}, the latter 
    involves a $\mathfrak{psu}(1,1|2)_k$ WZNW model at level $k=1$ along with a topologically twisted four-torus and some ghosts, 
    see below. 
    
    In the first subsection, section \ref{sec:boundary_states_for_ads2}, we review the relevant boundary states 
    for $AdS_2$ branes in the $\psu$ WZNW model following \cite{Gaberdiel:2021kkp}.  After adding the remaining 
    factors of the worldsheet model, i.e.~the topologically twisted $\mathbb{T}^4$ and the ghost factors, we, 
    in section \ref{sec:global_ads3_partition_function}, compute the overlaps of the full boundary states and 
    interpret the resulting quantity as a counting function for boundary operators. The worldsheet partition 
    function we end up with counts physical open string vertex operators in the tensionless 
    $AdS_3 \times S^3 \times \mathbb{T}^4$ background. In order to compute the partition function of the 
    associated spacetime string theory, we must integrate over the modulus $\tws$ of the worldsheet torus. 
    We do this first for 
    global $AdS_3$ and then, anticipating our comparison with the dual CFT we studied in section \ref{sec:symmetric_product}, for thermal 
    $AdS_3$ in section \ref{sec:thermal_ads3_partition_function}. Note that the boundary of global $AdS_3$ is a sphere whereas the boundary of thermal $AdS_3$ is a 
    torus. The latter is the geometry that is relevant for comparison with the CFT partition functions we 
    computed above. 

    \subsection{Boundary states for \texorpdfstring{$AdS_2$}{AdS2} branes in tensionless \texorpdfstring{$AdS_3$}{AdS3}}\label{sec:boundary_states_for_ads2}
    
        According to the work of Berkovits, Vafa and Witten \cite{Berkovits:1999im}, the description of superstrings 
        in $AdS_3 \times S^3$ in the hybrid formulation involves a $\psu$ WZNW model. The latter has been studied 
        in the past, see in particular \cite{Gotz:2006qp,Quella:2007hr} and references therein to earlier work in 
        string theory and quantum Hall plateaux transitions, as well as \cite{Gaberdiel:2011vf} for some later 
        extensions. Here, we restrict to the case of $k=1$. Our exposition is purposefully kept minimalistic and 
        we refer the reader to sections 3 and 4 of \cite{Gaberdiel:2021kkp} for more detail.

        The even part of the $\psu$ current algebra is generated by the affine currents $J^a, a=1,2,3,$ of an
        $\sl$ WZNW model along with the currents $K^a, a=1,2,3,$ of an $\su$ WZNW model. These six bosonic currents
        are associated with the six directions in the bosonic base of the supergroup PSU$(1,1|2)$. In 
        addition, the supergroup also has eight fermionic directions which give rise to the fermionic currents 
        $S^{\alpha\beta\gamma}$ in the WZNW model. Here $\alpha,\beta,\gamma$ are spinor indices that take the 
        values $\alpha = \pm 1$, respectively. These currents satisfy the relations of an affine $\psu$ Kac-Moody 
        algebra at level $k=1$. Since we do not need these relations, we do not spell them out here, see e.g.~section 3 of 
        \cite{Eberhardt:2018ouy} for a complete list. 
        
\begin{figure}
            \centering
            \begin{minipage}{0.47\textwidth}
                \centering
                \includegraphics[width=0.8\textwidth]{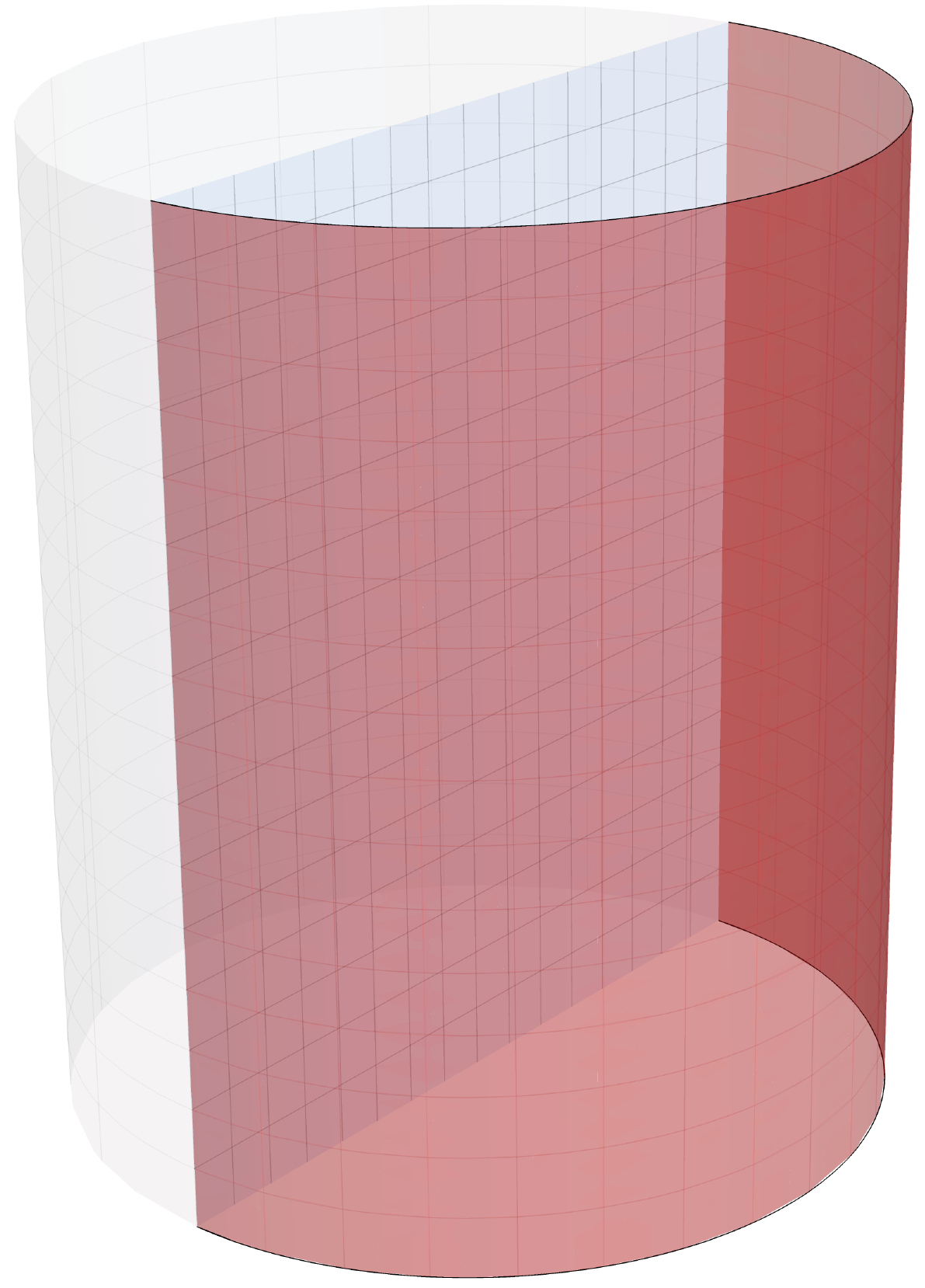}
                \vspace{-0.1 cm}
                \caption{Open string winding half way around $AdS_3$ at $k=1$. Since the worldsheet (red) is pinned to $\partial AdS_3$, the end points of the string remain on either of the two asymptotics boundaries of the $AdS_2$ brane (blue).}
                \label{fig:long_open_strings_tensionless}
            \end{minipage}\hfill
            \begin{minipage}{0.47\textwidth}
                \centering
                \vspace{-0.2 cm}
                \includegraphics[width=0.8\textwidth]{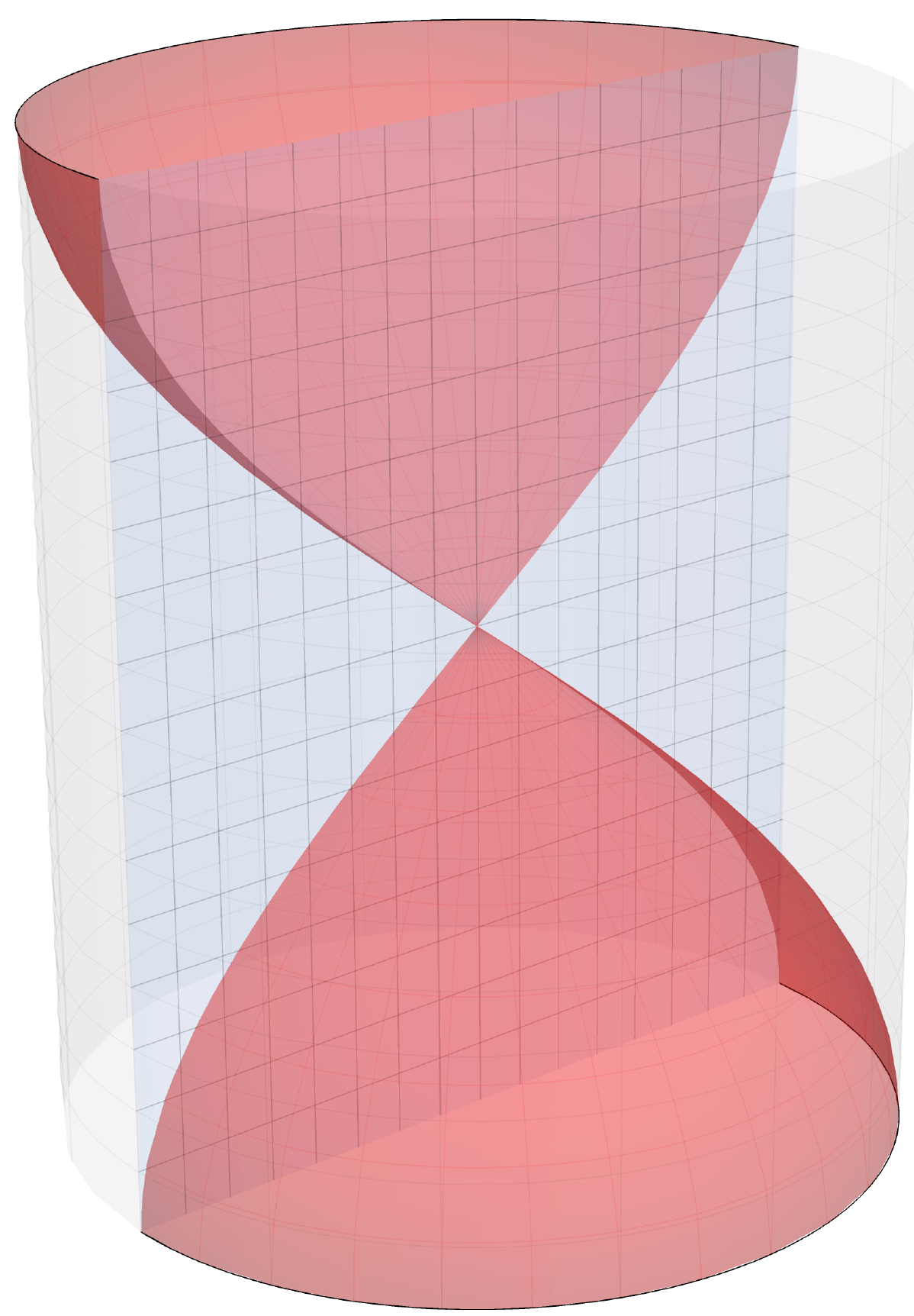}
                \vspace{-0.2 cm}
                \caption{Open string winding half way around $AdS_3$ for large $k$. The string can fall into the interior of $AdS_3$ and re-emerge with its end-point on opposite asymptotic boundaries of the $AdS_2$ brane.}
                \label{fig:long_open_strings_classical}
            \end{minipage}
        \end{figure}

        The state space of the bulk theory is a direct sum of representations of this worldsheet current algebra. 
        Following standard conventions, we shall denote by $\Fla$ the representations of the $\psu$ algebra whose 
        spectrum of conformal weights is bounded from below. Here, the index $\lambda \in [0,1[$ determines the 
        quantization of the zero mode $J^3_0$ of the current $J^3$ of the non-compact current algebra $\sl$. For 
        generic level $k$, the representations $\Fla$ would carry other labels that keep track of angular momenta 
        but these are all removed by the null-vectors at $k=1$. In addition to the representations $\Fla$, 
        the worldsheet model also includes representations $\Flaw$ that are obtained by application of the 
        spectral flow automorphism $\sigma^w$, that is $\Flaw = \sigma^w(\Fla)$. Recall that the space of ground 
        states from which $\mathcal{F}_\lambda$ is obtained by the action of the negative $\mathfrak{psu}(1,1|2)_1$ 
        modes is the short multiplet
        \begin{align}
            \label{eq:Short}
            \begin{array}{ccc}
                & (\mathscr{C}^\frac{1}{2}_{\lambda},\mathbf{2}) &\\
                (\mathscr{C}^1_{\lambda+\frac{1}{2}},\mathbf{1}) &  & (\mathscr{C}^0_{\lambda+\frac{1}{2}},\mathbf{1}) .
            \end{array}
        \end{align}
        Here, the symbol $(\mathscr{C}^j_{\lambda},\mathbf{m})$ denotes an irreducible representation of the subalgebra $\sl \oplus \su$
        which is the maximal bosonic subalgebra of $\mathfrak{psu}(1,1|2)_1$. These representations are constructed 
        from the $\mathbf{m}$-dimensional highest weight representation $\mathbf{m}$ of the compact algebra $\su$, 
        along with the continuous series representations $\mathscr{C}^j_{\lambda}$ of the non-compact $\sl$. The 
        labels $j$ and $\lambda$ of $\mathscr{C}^j_{\lambda}$ capture the eigenvalue $-j(j-1)$ of the quadratic 
        Casimir element and the fractional part of the $J^3_0$ eigenvalues, respectively. As we stressed before, 
        in the case $k=1$, the label $j$ is entirely fixed and hence we dropped it from the symbol $\Fla$. In terms 
        of representations of the $\psu$ current algebra, the Hilbert space of the $\mathfrak{psu}(1,1|2)_1$ WZNW 
        is\footnote{Strictly speaking the Hilbert space of the WZNW model does not directly include $\mathcal{F}_\lambda$ for $\lambda = \tfrac{1}{2}$. The representation \eqref{eq:bulkdecomposition} is however equivalent to the true Hilbert space in the Grothendieck ring. Since we only compute partition functions in this work, the special nature of the $\lambda = \tfrac{1}{2}$ representation is not relevant for our analysis. See \cite{Eberhardt:2018ouy} for more details on the atypical $\lambda = \tfrac{1}{2}$ representation.} the direct sum
        \begin{align} \label{eq:bulkdecomposition} 
            \mathcal{H}^{\textrm{WZNW}} = \bigoplus\limits_{w \in \mathbb{Z}}\  \int\limits_{0}^1 d\lambda 
            \   \sigma^w(\mathcal{F}_\lambda) \otimes \overline{\sigma^w(\mathcal{F}_\lambda)} . 
        \end{align}
        Let us now turn to the boundary theory. The gluing conditions for the $\mathfrak{sl}(2,\mathbb{R})$ 
        currents $J^a(z)$, the $\mathfrak{su}(2)$ currents $K^a(z)$ and the Fermionic currents $S^{\alpha\beta
        \gamma}(z)$ associated to the $AdS_2$ branes take the form 
       \begin{align}\label{eq:glueing_condition}
             J^a(z) = \bar{J}^a(\bar z) , \quad 
             K^a(z) = \bar{K}^a(\bar z) , \quad 
             S^{\alpha\beta\gamma}(z) = \varepsilon \bar{S}^{\alpha\beta\gamma}(\bar z)
       \end{align}
        at $z = \bar z$. The parameter $\varepsilon$ that enters the gluing conditions we impose 
        on the fermionic currents is a sign i.e.~$\varepsilon \in \{+,-\}$. The corresponding Ishibashi states $|w,\lambda,\varepsilon\rrangle$ in the 
        sectors of the bulk decomposition \eqref{eq:bulkdecomposition} 
        are characterised by
        \begin{align}
            (J_n^a+\bar{J}_{-n}^a)|w,\lambda,\varepsilon\rrangle=0 , \quad 
            (K_n^a+\bar{K}_{-n}^a)|w,\lambda,\varepsilon\rrangle=0 , \quad
            (S_n^{\alpha\beta\gamma}+ \varepsilon\, \bar{S}_{-n}^{\alpha\beta\gamma})
            |w,\lambda,\varepsilon\rrangle&=0.
        \end{align}
        These conditions imply $w = 0$ and $\lambda \in \left\{0, 1/2\right\}$. In \cite{Gaberdiel:2021kkp}, 
        Gaberdiel et al.~suggested to consider the following two linear combinations of Ishibashi states
        \begin{align}\label{eq:def_adsbrane_WS}
            \|\varepsilon\rrangle_A = \frac{1}{\sqrt{2}} \sum\limits_{\lambda = 0,1/2} 
            e^{2 \pi i(\lambda -1/2) \delta_A^L} |0,\lambda,\varepsilon\rrangle,
         \end{align}
        where the label\footnote{In \cite{Gaberdiel:2021kkp}, this label is called $\Theta$ and chosen from 
        $\{0,1\}$.} $A$ is chosen from $A \in \{L,R\}$. The two boundary states impose Dirichlet conditions for 
        the angular coordinate of $AdS_3$ forcing open strings to end ``left'' or ``right'' of the centre of 
        the $AdS_2$ brane respectively. In this sense, they only describe half branes and the worldsheet 
        boundary state corresponding to the full $AdS_2$ brane is
        \begin{align}\label{eq:splitting_of_AdS_brane}
            \|\varepsilon\rrangle = \|\varepsilon\rrangle_L + \|\varepsilon\rrangle_R.
        \end{align}
        For generic $k>1$, the splitting of $\|\varepsilon\rrangle$ into $\|\varepsilon\rrangle_L$ and 
        $\|\varepsilon\rrangle_R$ is unnatural since the ends of open strings can cross the centre of $AdS_3$.
        It is a special feature of the tensionless limit, that it is useful to keep track of $\|\varepsilon
        \rrangle_L$ and $\|\varepsilon\rrangle_R$ separately. Indeed, since long open strings in the 
        tensionless limit cannot probe the centre of $AdS_3$, an open string end on the left/right asymptotic 
        end of the $AdS_2$ brane will always stay on the left/right. We illustrate the situation in figures 
        \ref{fig:long_open_strings_tensionless} and \ref{fig:long_open_strings_classical}.
        
        As we stressed  above, the full worldsheet description of superstrings in the hybrid formulation also 
        involves a topologically twisted four-torus and ghosts. Hence, the boundary states of the $\psu$ WZNW model 
        we have described here must still be multiplied with further contributions from these additional 
        sectors to obtain the following boundary states for $AdS_2$ branes in $AdS_3 \times S^3 \times 
        \mathbb{T}^4$, 
        \begin{align}
            \|u,\varepsilon\rrangle
            = \|\varepsilon\rrangle\, \|u,\mathrm{R},\varepsilon\rrangle_{\mathbb{T}^4}\,
            \|\text{ghost},\varepsilon\rrangle .\label{eq:BS}
        \end{align} 
        Corresponding to the splitting \eqref{eq:splitting_of_AdS_brane}, we also define
        \begin{align}
            \|u,\varepsilon\rrangle_A
            = \|\varepsilon\rrangle_A\, \|u,\mathrm{R},\varepsilon\rrangle_{\mathbb{T}^4}\,
            \|\text{ghost},\varepsilon\rrangle .\label{eq:BS_split}
        \end{align} 
        The additional factors are discussed in detail in appendix B.1 of \cite{Gaberdiel:2021kkp}. Here, it 
        suffices to say that the letter $u$ refers to the freedom we have in selecting a boundary condition 
        on the four-torus $\mathbb{T}^4$, such as the brane's dimension, orientation and position. It is not 
        necessary for us to specify this choice any further. 
        
        Regarding the R label, recall that in the hybrid formalism, the fermions of the original sigma model on $\mathbb{T}^4$ with small $\mathcal{N}=(4,4)$ supersymmetry mix with the currents of the $\mathfrak{sl}(2,\mathbb{R})_1$ and $\mathfrak{su}(2)_1$ WZNW models to generate the supercurrents of the $\mathfrak{psu}(1,1|2)_1$ model. 
        The subsequent decoupling of torus degrees of freedom from the $\mathfrak{psu}(1,1|2)_1$ WZNW model leaves us with a topologically twisted $\mathbb{T}^4$.
        While the bosonic degrees of freedom of the topologically twisted sigma model coincide with that of the untwisted theory, part of the fermionic degrees of freedom are removed and we are essentially only left with the RR sector. 
        Consequentially, the boundary condition that we pick for the topologically twisted sector is a boundary condition for RR fermions.
        The purpose of the label R is to remind us of this fact. 
        
    \subsection{Partition function for \texorpdfstring{$AdS_2$}{AdS2} branes in global \texorpdfstring{$AdS_3$}{AdS3}}
    \label{sec:global_ads3_partition_function}
    
        Our goal in this subsection is to compute the overlap of any two of the boundary states \eqref{eq:BS}
        for $AdS_2$ branes and to apply a modular transformation in order to interpret this overlap in terms 
        of the boundary spectrum. More precisely, the quantity in question is 
        \begin{align} 
            \hat{Z}_{u|v}^\text{WS}(\hat{\tst},\hat{\zeta};\hat{\tws}):=
            \llangle u,\mp\|\hat{q}^{\frac{1}{2}(L_0+\bar{L}_0-\frac{c}{12})}
            \hat{x}^{\frac{1}{2}(J_0^3-\bar{J}_0^3)}\hat{y}^{\frac{1}{2}(K_0^3-\bar{K}_0^3)}
            \|v,\pm\rrangle =\hspace{-0.5 cm} \sum\limits_{A,B\in\{L,R\} }\hspace{-0.3 cm} \hat{Z}_{A,u|B,v}^\text{WS}(\hat{\tst},\hat{\zeta};\hat{\tws}),\label{eq:P_all}
        \end{align}
        where on the right hand side, we split the brane into a left and right half according to \eqref{eq:splitting_of_AdS_brane}, that is
        \begin{align} 
            \hat{Z}_{A,u|B,v}^\text{WS}(\hat{\tst},\hat{\zeta};\hat{\tws}):=
            \,_A\llangle u,\mp\|\hat{q}^{\frac{1}{2}(L_0+\bar{L}_0-\frac{c}{12})}
            \hat{x}^{\frac{1}{2}(J_0^3-\bar{J}_0^3)}\hat{y}^{\frac{1}{2}(K_0^3-\bar{K}_0^3)}
            \|v,\pm\rrangle_B\label{eq:P}
        \end{align}
        counts the open strings that stretch between the $A$-half of the $u$-brane and the $B$-half of the $v$-brane.
        The overlap factorises into three contributions from the $\psu$ WZNW model, the four-torus 
        $\mathbb{T}^4$ and the ghosts respectively. In order to spell these out explicitly, we need a bit 
        of notation. It is convenient to use the conventions of \cite{Blumenhagen:2013fgp} for Jacobi 
        theta-functions. That is, we define
        \begin{align} \label{eq:Jacobi}
            \vartheta\begin{bmatrix}
            \alpha \\
            \beta
            \end{bmatrix}(\zeta|\tws) := \sum\limits_{n\in\mathbb{Z}} 
            \exp\left(i \pi (n + \alpha)^2 \tws + 2\pi i (n+ \alpha)(\zeta + \beta)\right) .
        \end{align}
        Together with the standard Dedekind $\eta$ function, one can use these theta-functions to compute 
        the supercharacter \cite{Gaberdiel:2021kkp}
        \begin{align}
        \begin{aligned}
            \widetilde{\mathrm{ch}}[\mathcal{F}_\lambda](t,z;\tws)
            :=\mathrm{tr}_{\mathcal{F}_\lambda} \big[(-1)^F q^{L_0 - \frac{c}{24}}x^{J_0^3}y^{K_0^3}
            \big] =\hspace{-2 mm}\sum_{r\in \mathbb{Z}+\lambda} \hspace{-1 mm} \frac{x^r}{\eta(\tws)^4}
            \vartheta\begin{bmatrix}
            \tfrac{1}{2} \\
            \tfrac{1}{2}
            \end{bmatrix}\left(\tfrac{t+\zeta}{2}{\Big|}\tws\right) 
            \vartheta\begin{bmatrix}
            \tfrac{1}{2} \\
            \tfrac{1}{2}
            \end{bmatrix}\left(\tfrac{t-\zeta}{2}{\Big|}\tws\right) ,
        \end{aligned}
        \end{align}
        which by performing the sum over $r$ can also be written as
        \begin{align}
            \widetilde{\mathrm{ch}}[\mathcal{F}_\lambda](t,z;\tws) =  x^\lambda \sum\limits_{m\in\mathbb{Z}} \frac{\delta(t- m)}{\eta(\tws)^4}
            \vartheta\begin{bmatrix}
            \tfrac{1}{2} \\
            \tfrac{1}{2}
            \end{bmatrix}\left(\tfrac{t+\zeta}{2}{\Big|}\tws\right) 
            \vartheta\begin{bmatrix}
            \tfrac{1}{2} \\
            \tfrac{1}{2}
            \end{bmatrix}\left(\tfrac{t-\zeta}{2}{\Big|}\tws\right) .
        \end{align}
        By the identity
        \begin{align}
            \vartheta\begin{bmatrix}
            \tfrac{1}{2} \\
            \tfrac{1}{2}
            \end{bmatrix}\left(\tfrac{m\pm\zeta}{2}{\Big|}\tws\right) =  
            e^{i \pi \left\lceil \tfrac{m}{2} \right\rceil} \vartheta\begin{bmatrix}
            \tfrac{1}{2} \\
            \tfrac{1+e^{i\pi m}}{4} 
            \end{bmatrix}\left(\pm\tfrac{\zeta}{2}{\Big|}\tws\right),
        \end{align}
        this simplifies further to
        \begin{align}\label{eq:char}
            \widetilde{\mathrm{ch}}[\mathcal{F}_\lambda](t,z;\tws) =  \sum\limits_{m\in\mathbb{Z}} e^{2\pi i \lambda m}\frac{\delta(t-m)}{\eta(\tws)^4}
            \vartheta\begin{bmatrix}
            \tfrac{1}{2} \\
            \tfrac{1+e^{i\pi m}}{4} 
            \end{bmatrix}\left(\tfrac{\zeta}{2}{\Big|}\tws\right)
            \vartheta\begin{bmatrix}
            \tfrac{1}{2} \\
            \tfrac{1+e^{i\pi m}}{4} 
            \end{bmatrix}\left(-\tfrac{\zeta}{2}{\Big|}\tws\right).
        \end{align}
        The supercharacter can be used to determine the overlap of the boundary states for $AdS_2$ branes in the $\psu$ WZNW model,
       \begin{align}
            \llangle 0,\lambda',\mp |\, \hat{q}^{\frac{1}{2}(L_0+\bar{L}_0-\frac{c}{12})}
            \hat{x}^{\frac{1}{2}(J_0^3-\bar{J}_0^3)}\hat{y}^{\frac{1}{2}(K_0^3-\bar{K}_0^3)}\, |0,
            \lambda,\pm\rrangle&=
            \delta_{\lambda,\lambda'}\,
            \widetilde{\mathrm{ch}}[\mathcal{F}_\lambda](\hat t, \hat \zeta;\hat{\tws}) .
        \label{eq:overlapAdS}
        \end{align}
        Combining the explicit expression \eqref{eq:char} for the super character with the 
        definition \eqref{eq:def_adsbrane_WS} of the boundary states representing the $AdS_2$ brane, we 
        obtain
        \begin{align}
        \begin{aligned}
            \hat{Z}_{A|B}^{\mathfrak{psu}}(\hat{\tst},\hat{\zeta};\hat{\tws}) =& 
            \frac{1}{2} \sum\limits_{\lambda = 0,1/2} e^{2\pi i(1/2-\lambda)(\delta_A^L-\delta_B^L)} 
            \, \widetilde{\mathrm{ch}}[\mathcal{F}_\lambda](\hat t, \hat \zeta;\hat{\tws}) \\[2mm]
            =& \sum_{m=1}^\infty \frac{e^{i\pi m}-e^{i\pi \delta_{A}^B }}{2 \eta(\hat \tws)^4}  
            \delta(\hat t-m) \, \vartheta\begin{bmatrix}
            \tfrac{1}{2} \\
            \tfrac{\delta_{A,B}}{2}
            \end{bmatrix}\left(-\tfrac{\hat \zeta}{2}{\Big|} \hat \tws\right)\, \vartheta
            \begin{bmatrix}
            \tfrac{1}{2} \\
            \tfrac{\delta_{A,B}}{2}
            \end{bmatrix}\left(+\tfrac{\hat \zeta}{2}{\Big|} \hat \tws\right).
        \end{aligned}
            \label{eq:psu_factor}
        \end{align}
        The two remaining contributions from the overlaps of the boundary states in the four-torus and 
        the ghost factor are the same as in the case of spherical branes that were fully analysed 
        in \cite{Gaberdiel:2021kkp}. We use the same conventions and only slightly different notation. 
        For instance, we write the four-torus factor in the overlap using the notation for 
        Jacobi theta-functions introduced in eq.~\eqref{eq:Jacobi} as
        \begin{align} \label{eq:ZTuvhat}
            \hat{Z}^{\mathbb{T}^4}_{u|v}\begin{bmatrix}
            \alpha \\
            \beta
            \end{bmatrix}(\hat \zeta| \hat t) := \frac{\hat\Theta^{\mathbb{T}^4}_{u|v}(\hat t)}{\eta(\hat t)^6} \vartheta\begin{bmatrix}
            \alpha \\
            \beta
            \end{bmatrix}\left(+\tfrac{\hat \zeta}{2}{\Big|}\hat t \right) 
            \vartheta\begin{bmatrix}
            \alpha \\
            \beta
            \end{bmatrix}\left(-\tfrac{\hat \zeta}{2}{\Big|}\hat t\right).
        \end{align}
        Here, the quantity $\hat\Theta^{\mathbb{T}^4}_{u|v}$ in the numerator of the first factor stems from the bosonic 
        directions on the four-torus and it depends on the choice of boundary states on $\mathbb{T}^4$, 
        see appendix~B.1 of \cite{Gaberdiel:2021kkp} for details. Instead of the bracket notation it is also 
        common to write, as Gaberdiel et al.~do in \cite{Gaberdiel:2021kkp}, 
        \begin{align}
            \hat Z^{\mathbb{T}^4}_{u|v}\begin{bmatrix}
            \tfrac{1}{2} \\
            \tfrac{1}{2}
            \end{bmatrix} = \hat Z^{\mathbb{T}^4}_{u|v, \tilde{\text{R}}}, && \hat Z^{\mathbb{T}^4}_{u|v}\begin{bmatrix}
            \tfrac{1}{2} \\
            0
            \end{bmatrix} = \hat Z^{\mathbb{T}^4}_{u|v, \text{R}}, &&
            \hat Z^{\mathbb{T}^4}_{u|v}\begin{bmatrix}
            0 \\
            0
            \end{bmatrix} = \hat Z^{\mathbb{T}^4}_{u|v, \text{NS}}, &&
            \hat Z^{\mathbb{T}^4}_{u|v}\begin{bmatrix}
            0 \\
            \tfrac{1}{2}
            \end{bmatrix} = \hat Z^{\mathbb{T}^4}_{u|v, \tilde{\text{NS}}},
        \end{align}
        where R and NS refer to the Ramond and Neveu-Schwarz sector of the fermions, respectively, and the 
        tilde on top indicates whether we count fermions with additional signs or not, i.e.~whether we 
        insert a factor $(-1)^F$ in the trace. 
        Since
        \begin{align}
            \|
            u,\mathrm{R},\mp
            \rrangle = (-1)^F \|
            u,\mathrm{R},\pm
            \rrangle  ,
        \end{align}
        the insertion of $(-1)^F$ is effectively achieved by choosing opposite $\varepsilon$ for the two states whose overlap we compute.
        With the notations in place, we can now state that the
        relevant overlap of the boundary states on the four-torus is 
        \begin{align}
            \llangle u,\mathrm{R},\mp \|  \hat q^{\frac12(L_0 + \bar L_0 - \frac{c}{12})} 
            \|
            v,\mathrm{R},\pm
            \rrangle =\frac{\hat\Theta^{\mathbb{T}^4}_{u|v}(\hat \tau)}{\eta(\hat \tau)^6} \vartheta\begin{bmatrix}
            \tfrac{1}{2} \\
            \tfrac{1}{2}
            \end{bmatrix}\left(0{|} \hat \tau \right) 
            \vartheta\begin{bmatrix}
            \tfrac{1}{2} \\
            \tfrac{1}{2}
            \end{bmatrix}\left(0{|}\hat \tau \right)  .\label{eq:T4_factor}
        \end{align}
        The last contribution we need is that from the overlap of boundary states in the ghost sector
        which we also take from \cite{Gaberdiel:2021kkp}
        \begin{align}
            \hat{Z}_{\text{ghost}}(\hat{\tws})= \frac{\eta(\hat{\tws})^4}
            {\vartheta\begin{bmatrix}
            \tfrac{1}{2} \\
            \tfrac{1}{2}
            \end{bmatrix}\left(0{|} \hat \tau \right) 
            \vartheta\begin{bmatrix}
            \tfrac{1}{2} \\
            \tfrac{1}{2}
            \end{bmatrix}\left(0{|}\hat \tau \right) }\ \label{eq:ghost_factor}.
        \end{align}
        Multiplying eqs.~\eqref{eq:ghost_factor}, \eqref{eq:T4_factor} and \eqref{eq:psu_factor}, we  
        conclude that the overlap \eqref{eq:P} is given by 
        \begin{align}\label{eq:closed_string_ZWS}
            \hat{Z}_{A,u|B,v}^\text{WS}(\hat{\tst},\hat{\zeta};\hat{\tws})   
            = \sum_{m=1}^\infty \tfrac{1}{2}(e^{i\pi m}-e^{i\pi \delta_{A}^B })
            \delta(\hat t-m)\hat{Z}^{\mathbb{T}^4}_{u|v}\begin{bmatrix}
            \tfrac{1}{2} \\
            \tfrac{\delta_{A,B}}{2}
            \end{bmatrix}(\hat \zeta| \hat \tws)   .
        \end{align}
        We placed a superscript WS on this quantity to remind us that this is a partition of the 
        worldsheet theory that describes strings in $AdS_3 \times S^3 \times \mathbb{T}^4$. In particular, 
        the modular parameter $\hat \tws = - 1/\tws$ is that of the annulus worldsheet. 
        \smallskip 
        
        The overlap we have just computed is still written in terms of closed string parameters $\hat t, \hat \zeta$ and $\hat \tws$. In order to reinterpret the 
        overlap in terms of open string vertex operators, we need to perform a modular $S$ transformation. This 
        transformation acts as 
        \begin{align}
            \hat{\tws} = -\frac{1}{\tws} , \qquad \hat{\tst} = \frac{\tst}{\tws} , \qquad \hat{\zeta} = 
            \frac{\zeta}{  \tws}
        \end{align}
        on the chemical potentials. We can rewrite our result \eqref{eq:closed_string_ZWS} as a function of 
        the dual variables with the help of the following transformation formula for the overlap of boundaries 
        on the four-torus 
        \begin{align}\label{eq:modular_transform_torus}
            \hat{Z}^{\mathbb{T}^4}_{u|v}\begin{bmatrix}
            \alpha \\
            \beta
            \end{bmatrix}(\hat \zeta|\hat \tws) = 
            e^{\frac{\pi i\zeta^2}{2\tws}}
            Z^{\mathbb{T}^4}_{u|v}\begin{bmatrix}
            \beta \\
            \alpha
            \end{bmatrix}(\zeta| \tws)
        \end{align}
        with $\alpha,\beta \in \{0,\tfrac{1}{2}\}$.
        Here, following \cite{Gaberdiel:2021kkp}, we absorb the sign from the modular transformation of the theta functions into the definition
        \begin{align} \label{eq:ZTuv}
            {Z}^{\mathbb{T}^4}_{u|v}\begin{bmatrix}
            \alpha \\
            \beta
            \end{bmatrix}( \zeta| t) := e^{4\pi i \alpha \beta }\frac{\Theta^{\mathbb{T}^4}_{u|v}( t)}{\eta( t)^6} \vartheta\begin{bmatrix}
            \alpha \\
            \beta
            \end{bmatrix}\left(+\tfrac{\zeta}{2}{\Big|} t \right) 
            \vartheta\begin{bmatrix}
            \alpha \\
            \beta
            \end{bmatrix}\left(-\tfrac{ \zeta}{2}{\Big|} t\right).
        \end{align}
        This allows us to rewrite eq.~\eqref{eq:closed_string_ZWS} as 
        \begin{align} \label{eq:Zopen} 
            {\hat Z}^\mathrm{WS}_{A,u|B,v}(\hat t,\hat \zeta;\hat \tws) =& 
            \sum_{m=1}^\infty \tfrac{1}{2}(e^{i\pi m}-e^{i\pi\delta_{A}^B})
            \delta(\tfrac{\tst}{\tws}-m)
            e^{\frac{\pi i\zeta^2}{2\tws}}
            {Z}^{\mathbb{T}^4}_{u|v}\begin{bmatrix}
            \tfrac{\delta_{A,B}}{2} \\ 
            \tfrac{1}{2}
            \end{bmatrix}( \zeta| \tws).
        \end{align}
        As usual, we interpret the right hand side~as the partition function that counts operators that can be 
        inserted along the boundary of the worldsheet or, equivalently, the state space of the theory in an 
        infinite strip, i.e. 
        \begin{align}\label{eq:ZwhatZw}
        \begin{aligned}
            {Z}^\mathrm{WS}_{A,u|B,v}(t,\zeta; \tws) :=&\mathrm{Tr}_{\mathcal{H}^{\textrm{WZNW}}_{A,u|
            B,v}} \Bigl( (-1)^F \, q^{L_0-\frac{c}{24}}\, e^{2\pi i t J_0^3} \, e^{2\pi i  \zeta K_0^3} 
            \Bigr) \\[2mm]
            =&  \tfrac{\tws^2}{\tst}\sum_{w=1}^\infty \tfrac{1}{2}(e^{i\pi w}-e^{i\pi\delta_{A}^B})\delta(\tfrac{\tst}{w}-\tws)
            e^{\frac{\pi i\zeta^2}{2\tws}}
            {Z}^{\mathbb{T}^4}_{u|v}\begin{bmatrix}
            \tfrac{\delta_{A,B}}{2} \\
            \tfrac{1}{2}
            \end{bmatrix}( \zeta| \tws) . 
        \end{aligned}
        \end{align}
        Relative to the previous result in eq.~\eqref{eq:Zopen}, there is a prefactor $\tfrac{\tau^2}{t}$ that results 
        from rewriting the argument of the $\delta$ function. 
        
        Integrating over the worldsheet modulus $\tws$  using the integral  
        measure $d\hat \tws = d\tws/\tws^2$ yields  
        \begin{align} \label{eq:intZwhatZw}
            \int\limits_{0}^{i\infty} \frac{d\tws}{\tws^2} {Z}^\mathrm{WS}_{A,u|B,v}(t,\zeta;\tws)=   \frac{1}{\tst} \sum_{w=1}^\infty \tfrac{1}{2}(e^{i\pi w}-e^{i\pi\delta_{A}^B}) 
            e^{\frac{\pi iw\zeta^2}{2t}}
            {Z}^{\mathbb{T}^4}_{u|v}\begin{bmatrix}
            \tfrac{\delta_{A, B}}{2} \\
            \tfrac{1}{2}
            \end{bmatrix}( \zeta| \tfrac{\tst}{w}).
        \end{align}       
        Formulas \eqref{eq:ZwhatZw} and its integrated cousin \eqref{eq:intZwhatZw} are the main result of this 
        subsection. We discuss these further in section \ref{sec:matching_thermal_ads_with_sym_orbifold} when we compare with the expressions for 
        the partition function of the spacetime CFT. 
        
    \subsection{Partition function for \texorpdfstring{$AdS_2$}{AdS2} brane in thermal \texorpdfstring{$AdS_3$}{AdS3}}
    \label{sec:thermal_ads3_partition_function}
    
        We finally compute the boundary spectrum of $AdS_2$ branes in thermal $AdS_3$ along the lines of 
        \cite{Eberhardt:2020bgq}. As is well known, thermal AdS is obtained from the global one through an
        orbifold construction with the orbifold group $\mathbb{Z}$. The $\mathbb{Z}$ action enforces a 
        periodic identification along the Euclidean time direction. Our starting point for the 
        construction is the boundary partition function we computed in eq.~\eqref{eq:ZwhatZw}. 
        The associated orbifold partition function is obtained as a sum over twisted sector 
        contributions
        \begin{align}
            \raisebox{-0.5\height}{%
             \begin{tikzpicture}[scale=0.4]
        		\node [style=none] (0) at (1, 0) {};
        		\node [style=none] (1) at (3, 0) {};
        		\node [style=none] (2) at (-1, 0) {};
        		\node [style=none] (3) at (-3, 0) {};
        		\node [style=none] (4) at (0, 1) {};
        		\node [style=none] (5) at (0, -1) {};
        		\node [style=none] (6) at (0, -3) {};
        		\node [style=none] (8) at (0, 3) {};
        		\node [style=none] (9) at (2, 0.45) {$\ell$};
                    \draw [style=Dashed] (0.center) to (1.center);
        		\draw [style=Orange, bend left=45] (8.center) to (1.center);
        		\draw [style=Orange, bend left=45] (1.center) to (6.center);
        		\draw [style=Orange, bend right=315] (6.center) to (3.center);
        		\draw [style=Orange, bend left=45] (3.center) to (8.center);
        		\draw [style=Cyan, bend left=45] (4.center) to (0.center);
        		\draw [style=Cyan, bend left=45] (0.center) to (5.center);
        		\draw [style=Cyan, bend right=315] (5.center) to (2.center);
        		\draw [style=Cyan, bend left=45] (2.center) to (4.center);
            \end{tikzpicture}
            }
            = \mathcal{N} {Z}_{A,u|B,v}^\text{WS}({\ell t},{\ell \zeta};{\tws}).
        \end{align}
        Here, the orange and cyan solid lines are the two boundaries of the world sheet annulus on which the states labelled by $(A,u)$ and $(B,v)$ are inserted.
        Contributions with different $\ell$ correspond (after integration over $\tau$) to a trace of a segment of global Euclidean $AdS_3$ of length $\ell t$ over the physical open string Hilbert space
        \footnote{  
        Note that this can be rephrased as the statement that we compute the annulus partition function describing the boundary spectrum of a brane in the $\mathbb{Z}$-orbifolded world sheet theory that has additional couplings to $\mathbb{Z}$-twisted states i.e.~thermal winding modes.
        }. 
        In the equation, we introduced a normalisation constant $\mathcal{N}$ that we shall determine later. Summing over 
        $\ell$, we obtain the worldsheet partition function
        \begin{align} \label{eq:thermalAdSWSPF}
              {Z}_{A,u|B,v}^\text{WS \therm}(t,\zeta;\tws)
               &= \mathcal{N}\tfrac{\tws^2}{\ell \tst}\sum\limits_{w,\ell = 1}^\infty 
               \tfrac{1}{2}(e^{i\pi w}-e^{i\pi\delta_{A}^B}) \delta(\tfrac{\ell}{w}t-\tws)   
               e^{\frac{\pi i\ell^2\zeta^2}{2\tws}}
               {Z}^{\mathbb{T}^4}_{u|v}\begin{bmatrix}
               \tfrac{\delta_{A, B}}{2} \\
               \tfrac{1}{2}
        \end{bmatrix}( \ell\zeta| \tws). 
        \end{align}
        Upon integration over the modulus $\tws$ of the worldsheet, we obtain
        \begin{align}
        {Z}_{A,u|B,v}^\text{ST \therm}(t,\zeta)
            &=   \tfrac{\mathcal{N}}{\tst}\sum\limits_{w,\ell = 1}^\infty   \tfrac{1}{2\ell}(e^{i\pi w}-e^{i\pi\delta_{A}^B}) 
            e^{\frac{\pi i w\ell\zeta^2}{2t}}
            {Z}^{\mathbb{T}^4}_{u|v}\begin{bmatrix}
            \tfrac{\delta_{A,B}}{2} \\
            \tfrac{1}{2}
            \end{bmatrix}( \ell\zeta| \tfrac{\ell}{w} t). 
        \end{align}
        In terms of $N = w \ell$, we can also rewrite this as
        \begin{align}
            {Z}_{A,u|B,v}^\text{ST \therm}(t,\zeta)
            &=  \tfrac{\mathcal{N}}{\tst} \sum\limits_{N=1}^\infty \sum\limits_{w|N}  \tfrac{w}{N} 
            \tfrac{1}{2}(e^{i\pi w}-e^{i\pi\delta_{A}^B})  e^{\frac{\pi i N \zeta^2}{2t}}  
            {Z}^{\mathbb{T}^4}_{u|v}\begin{bmatrix}      \tfrac{\delta_{A,B}}{2} \\
            \tfrac{1}{2}
            \end{bmatrix}( \tfrac{N}{w}\zeta| \tfrac{N}{w^2} t).
        \end{align}
        Finally, we conclude that the full open string partition function, counting strings that can start and end on both halves of the $AdS_2$ brane, is
        \begin{tcolorbox}[left=0pt,right=0pt,top=-10pt,bottom=0pt] 
            \begin{align}
                \label{eq:STspectrum}{Z}_{u|v}^\text{ST \therm}(t,\zeta)
                &=   \tfrac{\mathcal{N}}{\tst}\sum\limits_{A,B\in\{L,R\}}\sum\limits_{w,\ell = 1}^\infty   \tfrac{1}{2\ell}(e^{i\pi w}-e^{i\pi\delta_{A}^B}) 
                e^{\frac{\pi i w\ell\zeta^2}{2t}}
                {Z}^{\mathbb{T}^4}_{u|v}\begin{bmatrix}
                \tfrac{\delta_{A,B}}{2} \\
                \tfrac{1}{2}
                \end{bmatrix}( \ell\zeta| \tfrac{\ell}{w} t). 
            \end{align}
        \end{tcolorbox}
        \noindent
        Here, $t$ is the modular parameter of the spacetime torus and $\zeta$ is a chemical potential associated with 
        the R-charge. The only non-trivial ingredient in the summands are the functions $Z^{\mathbb{T}^4}_{u|v}$ that 
        we defined in eq.~\eqref{eq:ZTuv}. 
        \smallskip 
        
        Before carrying on and matching the result of this string computation to the symmetric orbifold 
        grand canonical partition function in section \ref{sec:matching_thermal_ads_with_sym_orbifold}, let us 
        pause and give a geometric interpretation of the formulas we obtained. Looking back at our formula 
        \eqref{eq:thermalAdSWSPF}, we infer from the argument of the $\delta$ function that the worldsheet 
        modulus $\tws$ localises to $\frac{\ell}{w} t$ with two integers $\ell$ and $w$. We can relate this 
        observation to the geometric interpretation that was given to a similar closed string computation in 
        \cite{Eberhardt:2020bgq}. Eberhardt explained that the one-to-one correspondence between solutions 
        to the equation 
        \begin{align}
            t = \frac{a \tws+b}{b\tws +d} , \quad  \tws = \frac{dt -b}{-ct +a}
        \end{align}
        with integer coefficients $a,b,c,d \in  \mathbb{Z}$ and holomorphic coverings of the $t$ torus 
        by the $\tws$ torus can be understood by realising that such a covering must lift to an affine 
        map $\tilde \Gamma: \mathbb{C}^2 \rightarrow \mathbb{C}^2$, $z \mapsto \alpha z + \beta$ such 
        that $\tilde \Gamma(\Lambda_\tws) \subseteq \Lambda_t$ holds. The latter condition in particular 
        implies $\alpha,\beta \in \Lambda$ and $\alpha \tws +\beta \in \Lambda_t$ and hence enforces 
        that $\tws$ is a fraction of elements of $\Lambda_t$. Consequentially, $\tws$ can be written 
        as
        \begin{align}
            \tws = \frac{dt -b}{-ct +a}.
        \end{align}
        For the situation at hand, we describe coverings of a cylinder by a cylinder obtained by 
        cutting the tori along the imaginary axis 
         (because the connected components of the boundary of the $AdS_2$ brane are parallel thermal cycles in the boundary torus of thermal $AdS_3$). 
         Thus, we want that $\tilde \Gamma(i\mathbb{R}) 
        \subseteq i \mathbb{R}$, which implies $\alpha \in \mathbb{R}$ and $\beta \in i \mathbb{R}$. 
        However, in combination with the fact that $t$ is purely imaginary, $\alpha \in  \mathbb{R}$ 
        implies that $c$ is zero. But if $c$ is zero and $t,\tws$ are purely imaginary then $b$ must 
        also be zero. Hence, we are restricted to
        \begin{align}
            \tws = \frac{d}{a} t.
        \end{align}
        Furthermore, the prefactor of $\tfrac{1}{\ell}$ in the partition function is to be expected.
        $w$ counts the wrapping of the world sheet around the spatial cycle, while $\ell$ counts the wrapping
        around the thermal cycle. The translation symmetry along the spatial cycle is broken by the presence 
        of the $AdS_2$ brane. But we still have full translation symmetry along the thermal cycle. Hence, 
        there is a $\mathbb{Z}_{\ell}$ symmetry that the partition function needs to reflect by a factor of $\tfrac{1}{\ell}$.
                
\section{Holographic matching of interfaces with \texorpdfstring{$AdS_2$}{AdS2} branes}\label{sec:matching_thermal_ads_with_sym_orbifold}

    The goal of this section is to collect evidence supporting the claim that the interfaces we have 
    constructed in section \ref{sec:symmetric_product} indeed provide a holographic description of 
    $AdS_2$ branes in a theory of tensionless type IIB superstrings in $AdS_3 \times S^3 \times \mathbb{T}^4$. 
    In order to do so, we first, in section \ref{sec:partition_function_match}, compare our results on open 
    string spectra, most notably formula \eqref{eq:STspectrum} for a thermal partition function of open strings ending on an $AdS_2$ brane, with the counting function \eqref{eq:Z_orbifold_open} 
    for interface changing operators. 
    To fully establish the correspondence, it is necessary to furthermore match the interface symmetric orbifold correlators with string scattering amplitudes in the presence of the brane.
    In section \ref{sec:rev_bulk_N_expansion}, we review the holographic match in the absence of branes. 
    We then, in section \ref{sec:boundary_N_expansion} take first steps towards establishing the holographic duality between amplitudes involving open and closed strings in the presence of the $AdS_2$ brane and corresponding correlation 
    functions of the symmetric orbifold that involve certain interface changing operators. 
    Concretely, we show that 
    the large $N$ expansion of correlation functions takes
    the form of a string theoretic genus expansion, for a certain class of our interface changing operators. 
    
    \subsection{Partition functions and string amplitudes}\label{sec:partition_function_match}

        As promised at the end of section \ref{sec:interface_changing_partition_function}, we now use the 
        notation for partition functions of the supersymmetric four-torus established in 
        section \ref{sec:global_ads3_partition_function}, to spell out eq.~\eqref{eq:ZO[t]} more concretely 
        for the special case in which the seed theory $\mathcal{M}$ is given by the supersymmetric four-torus
        $\mathbb{T}^4$. This entails making a particular choice for the boundary conditions $a_\pm^{L/R}$.
        Our proposal is
        \begin{align}
            |a_-^L\rangle = \|u,\mathrm{R},\varepsilon\rrangle_{\mathbb{T}^4} 
            \quad \text{and} \quad
            |a_+^L\rangle = \|v^*,\mathrm{R},\varepsilon\rrangle_{\mathbb{T}^4} 
        \end{align}
        as well as 
        \begin{align}    
            \hspace{3 mm}|a_-^R\rangle = -\|v,\mathrm{R},-\varepsilon\rrangle_{\mathbb{T}^4} 
            \quad \text{and} \quad
            |a_+^R\rangle = -\|u^*,\mathrm{R},-\varepsilon\rrangle_{\mathbb{T}^4},
        \end{align}
        which leads to
        \begin{align}
            \hat Z_o^{LL}(\hat t,\hat \zeta) =
            \hat Z_o^{RR}(\hat t,\hat \zeta) = \hat{Z}^{\mathbb{T}^4}_{u|v}\begin{bmatrix}      \tfrac{1}{2} \\
              \tfrac{1}{2}    \end{bmatrix} (\hat \zeta|\hat t)  && \text{and} && \hat Z_o^{LR}(\hat t,\hat \zeta) = \hat Z_o^{RL}(\hat t,\hat \zeta) = -\hat{Z}^{\mathbb{T}^4}_{u|v}\begin{bmatrix}      \tfrac{1}{2}  \\
              0    \end{bmatrix} (\hat \zeta|\hat t).
        \end{align}
        We thus conclude that        
        \begin{align}
            \mathcal{Z}_{O}[\hat\tst,\hat\zeta]
            = \hspace{-0.6 cm}\sum\limits_{A,B\in\{L,R\}}\sum\limits_{\ell,w=1}^\infty \tfrac{1}{2w}(e^{i\pi \ell}-e^{i\pi\delta_{A}^B})\hat{Z}^{\mathbb{T}^4}_{u|v}\begin{bmatrix}      \tfrac{1}{2}\vspace{1mm}   \\
            \tfrac{\delta_{A,B}}{2}
            \end{bmatrix} (\ell \hat \zeta|\tfrac{\ell }{w} \hat t)\ . 
        \end{align}
        Applying eq.~\eqref{eq:modular_transform_torus}, as well as swapping $w$ and $\ell$, gives
        \begin{align} \label{eq:ZOspectrum}
            \mathcal{Z}_{O}[\tst,\zeta]
            = \hspace{-0.6 cm}\sum\limits_{A,B\in\{L,R\}}\sum\limits_{\ell,w=1}^\infty  \tfrac{1}{2\ell}(e^{i\pi \ell}-e^{i\pi\delta_{A}^B})
            e^{\frac{\pi i w \ell \zeta^2}{2t}}{Z}^{\mathbb{T}^4}_{u|v}\begin{bmatrix}      \tfrac{\delta_{A,B}}{2}  \\
            \tfrac{1}{2}
            \end{bmatrix} (\ell \zeta|\tfrac{\ell}{w}t).
        \end{align}
        This partition function is manifestly identical to eq.~\eqref{eq:STspectrum}
        if we choose the normalisation $\mathcal{N}=t$. Furthermore, eq.~\eqref{eq:Z_orbifold_closed} directly 
        tells us that $\mathcal{Z}_{C}$ is identical to the $t=\bar t$ restriction of the space time CFT torus 
        partition function.
        Making the appropriate choice of spin structure, the latter has been matched with the thermal $AdS_3$ 
        closed string partition function in \cite{Eberhardt:2020bgq}. 
        The rest of this section is devoted to a qualitative 
        description of the holographic duality underlying this equality.
        \smallskip
        
        The equality of partition functions is equivalent to the statement that the spectrum (i.e.~the conformal weights 
        and the eigenvalues of $K_0$ as well as the dimensions of the eigenspaces) of open strings stretching between two 
        $AdS_2$ branes coincides with the spectrum of interface changing operators of our symmetric orbifold interfaces. 
        This observation leads to a natural proposal for the holographic description of the tensionless string dynamics in the presence of an 
        $AdS_2$ brane. 
        Let us first sketch this proposal in a rough qualitative manner and then explore its precise 
        quantitative implications.
        
        \begin{tcolorbox}[left=0pt,right=0pt,top=0pt,bottom=0pt]
            \label{central_box}
            1. We recall from \cite{Gaberdiel:2018rqv} that in the tensionless limit, long \textit{closed superstrings} of type 
            IIB theory on $AdS_3 \times S^3 \times \mathbb{T}^4$ possess a holographic description in terms of twisted sector 
            bulk operators in the symmetric product orbifolds $\mathrm{Sym}^N(\mathbb{T}^4)$. 
            \smallskip
            
            2. Based on the equality between the partition functions \eqref{eq:STspectrum} and \eqref{eq:ZOspectrum},
            we conjecture that the $AdS_2$ \emph{brane} is dual to a weighted sum of the interfaces of $\textrm{Sym}^{N_{\pm}}(\mathbb{T}^4)$ constructed in section \ref{sec:def_sym_orbifold_boundary_states}.
            \smallskip
            
            3. We furthermore propose that the \textit{open strings} that end on the $AdS_2$ brane are dual to interface changing operators
            between those interfaces.
        \end{tcolorbox}
        In order to simplify the navigation through the rest of this section, we divided it into short paragraphs that 
        provide the interested reader with further details on various aspects of our proposal. The paragraphs can be 
        read somewhat independently.  
        
        \paragraph{Thermal versus global  \texorpdfstring{$AdS_3$}{AdS3}.}
        In the three claims above, we do not specify whether we consider string theory on thermal or global $AdS_3$. 
        Let us start our discussion of the claims by filling in some more detail on this. 
        Recall that we constructed the open string partition function on thermal $AdS_3$ from the partition function of global $AdS_3$, which we computed in eq.~\eqref{eq:intZwhatZw}, by including additional contributions originating from world sheets that wrap the thermal cycle $\ell = N/w$ times.
        Vice versa, we can pass from the thermal $AdS_3$ open string partition function \eqref{eq:STspectrum} to its 
        analogue for global $AdS_3$ simply by dropping all the $\ell > 1$ terms\footnote{As also explained at the end of section \ref{sec:thermal_ads3_partition_function}, the situation for the open strings is in this sense simpler than the 
        closed string analogue discussed in \cite{Eberhardt:2020bgq}. Indeed, for the closed strings it is not true that the global $AdS_3$ partition function is obtained from the thermal $AdS_3$ one by simply dropping the $\ell>1$ terms.}. Because of this, the above 
        holographic matching of the thermal $AdS_3$ partition function directly implies that we can also obtain the 
        global $AdS_3$ partition function from our interface construction. What this requires is to adjust the 
        boundary states \eqref{eq:def:pra} that describe our interfaces or, more precisely, their building blocks 
        defined in eq.~\eqref{eq:patwist}. These states involve the boundary states \eqref{eq:atwisted} which contain 
        a sum in the $\rho$-twisted sectors. In order to recover the partition function of global $AdS_2$, we need to 
        restrict the sum over $\rho_\pm$ to the trivial term $\rho_\pm = id$. This manifestly removes the $\ell>1$ 
        terms in the partition function of defect changing operators.

        \paragraph{Absence of disk contributions.}  
        Note that, in sharp contrast to the bulk discussion \cite{Eberhardt:2019qcl} as well as that of spherical branes \cite{Gaberdiel:2021kkp}, our holographic matching of the open string partition function did not rely on the extrapolation of large $k$ results for (disconnected) genus 0 contributions to $k=1$.
        Instead, the one-loop (annulus) string computation directly reproduces the exact symmetric orbifold partition function of interface changing operators (up to an overall normalisation factor)\footnote{We thank Andrea Dei for encouraging us to emphasise this point more explicitly.}.
        In fact, this remarkable absence of the somewhat problematic disk contributions is to be expected on simple geometric grounds. 
        As opposed to the spherical branes, the $AdS_2$ branes are not contractible inside thermal $AdS_3$. 
        Thus, the existence of contractible world-sheets whose boundary winds one of the boundaries of the $AdS_2$ brane is topologically obstructed.

        \paragraph{The set of string scattering states.}
        Having commented on the absence of disk contributions and clarified the roles of thermal and global $AdS_3$, let us now expand more generally on part 
        three of our proposal. In order to formulate an actual prescription for holographic computations of open 
        string scattering amplitudes in the tensionless limit of superstring theory on $AdS_3 \times S^3 \times 
        \mathbb{T}^4$, the next paragraphs describe how to label and construct the interface changing operators 
        dual to string scattering states. Recall from eqs.~(\ref{eq:grand_canonical_exponential} - 
        \ref{eq:seedZo}) that these operators fall into two classes, corresponding to the two classes of 
        string states. 
        On the one hand, the string theory contains long closed strings.
        Their quantum numbers include the winding number $w$, which counts how many times the string winds the spatial cycle of the boundary torus.
        In addition, the string theory also contains long 
        open strings. These possess two new labels $A,B\in\{L,R\}$ as well as the familiar winding number $w$. 
        The labels $A, B \in \{L,R\}$ refer to the two components of the boundary of 
        the $AdS_2$ brane to which two endpoints of the open strings are attached. While the interface 
        changing operators that are dual to long closed strings are simply obtained by restricting  
        familiar bulk operators to the one-dimensional locus of the interface, those that are associated 
        with long open strings are new. In the following few paragraphs we will therefore mostly focus on the latter. 
            
        \paragraph{Interface changing operators dual to open strings.}
        When we discussed bulk fields of symmetric product orbifolds in section \ref{sec:symmetric_product}, 
        we pointed out that it is useful to work with a larger set of twist fields that are labelled by group 
        elements of the orbifold group rather than the physical fields which are labelled by conjugacy classes. 
        The physical operators are obtained from the former by averaging over the orbits of the $S_{N_\pm}$ 
        action. We follow a similar strategy for the interface changing operators and describe the physical 
        operators 
        \begin{equation} \label{eq:sigmaABw}
            \sigma^{AB}_{w} = \mathcal{N} \, \sum\limits_{\gamma}\,  \sigma^{AB}_{\gamma}
        \end{equation} 
        as a sum over integer sequences $\gamma = (\gamma_i)_{i=1}^{w}$ of length $w$. More precisely, 
        $\gamma$ is a sequence of entries that we choose alternating between the two sets $\underline 
        N_\pm = \{ 1, \dots, N_\pm\}$ with the `boundary' conditions
        \begin{align}
            \gamma_1 \in \underline N_- \text{ if } A=L  \quad &\text{and} \quad \gamma_1 \in \underline 
            N_+ \text{ if } A=R \\[2mm]
            \gamma_w \in \underline N_- \text{ if } B=R  \quad &\text{and} \quad \gamma_w \in \underline 
            N_+ \text{ if } B=L.
        \end{align}
        Repeated entries are not allowed, i.e.\ $\gamma_i \neq \gamma_j$ for $|i-j|$ even. Let us also 
        note that the boundary condition on $\gamma_w$ can only be satisfied for sequences of even length 
        $w$ if $A = B$ and for sequences of odd length $w$ if $A \neq B$. The labelling of open string states 
        through sequences of integers was described in \cite{Martinec:2022ofs} already. Martinec used red and 
        blue colours to distinguish between elements of $\underline{N}_\pm$.  
        \smallskip 

        Having described the label $\gamma$, it is rather straightforward to state how the actual operator 
        $\sigma^{AB}_{\gamma}$ acts. To do so it makes sense to first briefly reconsider the construction of 
        the interfaces on which $\sigma_w^{AB}$ is supposed to be inserted. Recall that in section 
        \ref{sec:symmetric_product} the interfaces are described as gauge invariant sums \eqref{eq:patwist} 
        of states that couple to one specific pair $g_\pm = \rho_\pm\tau_\pm \in S_{N_\pm}$ of twisted 
        sectors in the bulk of the product theory. To each cycle in $g_-$, these gauge fixed states either 
        associate a reflective boundary condition or a cycle in $g_+$ to which it is transmitted. 
        The individual gauge fixed contributions to the interface hence contain two pieces of data: The 
        distribution of elements of $\underline{N}_\pm$ into cycles of permutations in $S_{N_\pm}$ and 
        the attribution of the property ``reflective'' or ``transmissive''. As we shall argue in the 
        next paragraph, the detailed distribution into cycles is actually not important to decide 
        whether a given summand $\sigma^{AB}_\gamma$ from the sum \eqref{eq:sigmaABw} can contribute
        to a correlation function. What is important are the subsets of elements $T_+ \subseteq \underline{N}_+$ and 
        $T_- \subseteq \underline{N}_-$ of colours that are transmitted along the interface as well as the precise 
        pairing of colours. We encode the latter in bijections $\pi_\pm: T_{\pm}\rightarrow T_\mp$ with 
        $\pi_+$ and $\pi_-$ inverse to each other, i.e.\ $\pi_- \circ \pi_+ = \textit{id}$. In the 
        following we shall often drop the lower index on $\pi$ that tells us on which of the two 
        subsets $T_\pm$ it acts. We construct one triple $(T_+,T_-,\pi)$ for each summand in the 
        gauge invariant sums \eqref{eq:patwist}. Even though these triples forget the information 
        about the distribution of colours in cycles, i.e.~about the elements $g_\pm$, we now argue 
        that $T_\pm \subseteq \underline{N}_\pm$ and $\pi$ contain enough information to decide whether 
        $\sigma^{AB}_\gamma$ can be inserted at the transition point between two defects. 

        From a path integral perspective, correlation functions of the interface symmetric product 
        orbifold are computed by (locally) integrating over $N_\pm$ copies of the fields of the seed 
        theory on the upper and lower half plane labelled by the elements of $\underline{N}_\pm$.
        On fields whose $\underline{N}_\pm$ labels are elements of reflected cycles, i.e.~contained in $\underline{N}_\pm \setminus T_\pm$, reflecting boundary conditions are imposed locally at 
        the interface. On fields whose $\underline{N}_\pm$ labels are elements of transmissive cycles, 
        i.e.~contained in $T_\pm\subset \underline{N}_\pm$,  we locally impose transmissive boundary 
        conditions. How the transmitted and reflected elements of $\underline{N}_\pm$ are distributed 
        among different cycles has no direct impact on the local boundary conditions imposed at the 
        interface, but affects the global structure of how the different copies of the seed theory 
        fields are glued together\footnote{From a global perspective the path integral is performed 
        over sections of $S_{N_\pm}$ fibre bundles that only in local trivialisations can be identified 
        with  $N_\pm$ copies of the fields of the seed theory. The distribution of the 
        $\underline{N}_\pm$ labels among different $S_{N_\pm}$ cycles described the monodromy of these 
        fibre bundles along the interface.}.
        Local interface changing operators describe how the boundary conditions imposed at the interface 
        change locally at their insertion point. Hence, we do not need to reference the global structure 
        encoded in the distribution of $\underline{N}_\pm$ labels among different $S_{N_\pm}$ cycles to 
        define them.

        Let us now choose two interfaces $\mathcal{I}^{p^L}$ and $\mathcal{I}^{p^R}$ to the left 
        (i.e.~on the $L$ interface) and the right (i.e.~on the $R$ interface) of some interface 
        changing operator $\sigma_w^{AB}$. As we argued above, we can associate a triple $(T_+,T_-,
        \pi)$ to each term in the sum \eqref{eq:patwist}. Since we have an interfaces on both sides of 
        the insertion, we have two triples $(T^L_+,T_-^L,\pi^L)$ and $(T^R_+,T^R_-,\pi^R)$. A summand 
        $\sigma^{AB}_\gamma$ can contribute to the corresponding correlation functions provided 
        the following conditions are satisfied 
        \begin{itemize}
            \item \textbf{Reflective end points.} The first label $\gamma_1$ must be in the set $\underline{N}_\pm
            \setminus T^A_\pm$ of colours that are reflected on the $A$ interface. Similarly, the last entry 
            $\gamma_w$ must be taken from the set $\underline{N}_\pm \setminus T_\pm^B$ that are reflected on the $B$ 
            interface.
            \item \textbf{Transmissive interior.} If $A = L$ then the entries $\gamma_i$ and $\gamma_{i+1}$ 
            for odd integer $i$ are glued together at the $R$ interface, i.e.~$\gamma_i \in T^R_-$ and 
            $\gamma_{i+1} = \pi^R(\gamma_i) \in T^R_+$. For even $i$ they are glued at the $L$ interface, i.e.~$\gamma_i \in T^L_+$ and $\gamma_{i+1} = \pi^L(\gamma_i) \in T^L_-$. For $A = R$ the prescription is reversed.
            \item \textbf{Unaffected complement.} The boundary condition imposed on colours that do 
            not appear in $\gamma$ are unchanged at the insertion point of $\sigma_w^{AB}$. If we 
            denote the set of colours from $\underline{N}_\pm$ that appear in $\gamma$ by 
            $S^\pm_\gamma$, the formal condition reads 
            \begin{equation}
            T_\pm^L \setminus S^\pm_\gamma  =  
            T_\pm^R \setminus S^\pm_\gamma \ . 
            \end{equation}
            In addition the restriction of $\pi^L$ and $\pi^R$ must agree on the two intersections. 
        \end{itemize}
        
        \begin{figure}
            \centering
            \includegraphics[width=1\linewidth]{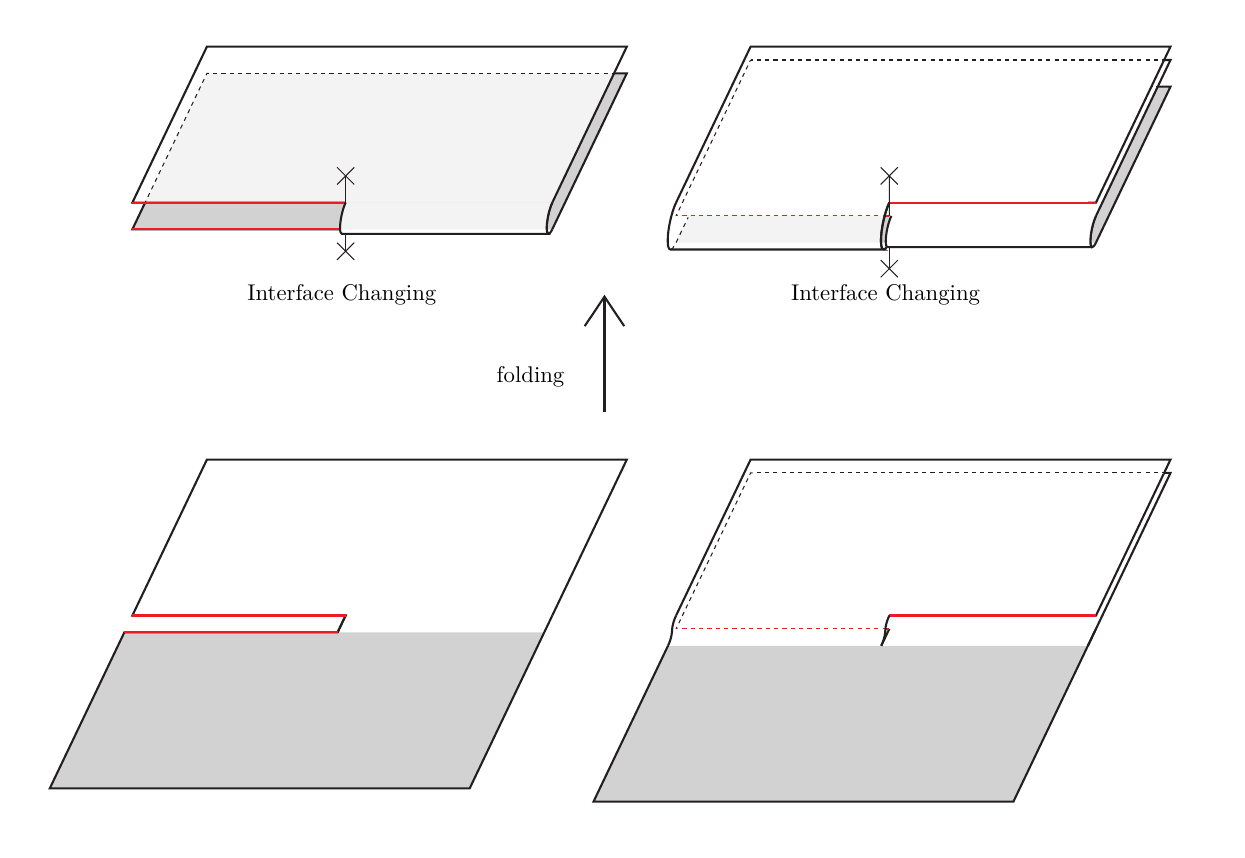}
            \caption{Visualisation of interface changing operators. Red lines indicate reflective boundaries. Gray 
            sheets are representing the lower half-plane on which $\SymNm$ lives. 
            White sheets are upper half-planes that carry $\SymNp$. 
            On the left hand side, we portray the operators $\sigma^{LL}_{2}$. The right hand side 
            represents $\sigma^{RL}_{3}$.
            } 
            \label{fig:IntrfaceChanging}
        \end{figure} 
            
  \paragraph{Interface changing operators dual to closed strings.}

       Now that we have completed our description of the interface changing operators we propose to be 
       dual to open strings, let us briefly comment on the operators $\sigma_{w}$ dual to closed strings. These are more familiar. 
      
       Again, they possess a representation akin to
       eq.~\eqref{eq:sigmaABw} in terms of a sum of gauge fixed operators $\sigma_\gamma$. 
       The only difference between $\sigma_\gamma$ and $\sigma_\gamma^{AB}$ is that $\sigma_\gamma$ describes no reflective i.e.~only transmissive boundary conditions. 
       More concretely, it is obtained by dropping the ``reflective end points'' condition formulated in the previous paragraph and instead use the ``transmissive interior'' condition with the periodic identification
       $\gamma_{w+1}=\gamma_1$. 
       Upon performing such a periodic identification any choice of the labels $A$ and $B$ describes the same operator and hence these labels become obsolete. 
       Notice that since we had chosen neighbouring entries from the upper 
       and lower half plane in an alternating fashion, $w$ must be even in order to admit 
       the periodic identification. 
       The fields described in this manner are essentially bulk twist fields with twist $w/2$ inserted on the interface.
            
        \paragraph{Taking stock.}
        Let us pause a moment to take stock of what we have formulated in the previous paragraphs.
        We have constructed two types of interface changing operators $\sigma^{AB}_{w}$ and 
        $\sigma_{w}$ that carry the same labels as open strings ending on $AdS_2$ branes 
        and closed strings winding the boundary of $AdS_3$ respectively.
        All interface changing operators between any two of the interfaces of
        section \ref{sec:symmetric_product} can be obtained by taking products of the elementary 
        operators we described. 
        
        We have seen that the string partition function computed by summing over connected world sheets describing open strings ending the $AdS_2$ brane and propagating around the non-contractible cycle of thermal $AdS_3$ matches the connected partition function $\mathcal{Z}_O$ of interface changing operators in the symmetric product orbifold.
        By exponentiating both sides, we can deduce that the second quantised string partition function coincides with the trace over the Hilbert space of interface changing operators in a grand canonical ensemble of symmetric product orbifolds.
        This Hilbert space manifestly has the form of a Fock space generated from a single particle Hilbert space that is spanned by the single cycle twist fields discussed above. 
        Thereby we have provided strong evidence for the 
        conjecture that the two families of operators we have introduced in the previous paragraphs 
        are dual to the single particle open and closed string states of the bulk theory.
        
        \paragraph{General string scattering amplitudes.} Of course, the proposed duality should 
        eventually be supported through a comparison of arbitrary scattering amplitudes of open 
        and closed strings with correlation functions of the interface changing operators 
        $\sigma^{AB}_{w}$ and $\sigma_{w}$ we introduced above. At this point, we
        can at least state somewhat more explicitly what this relation is expected to look like. 
        \smallskip 
            
        On the string side, we encounter scattering amplitudes that can be written as integrals 
        over the moduli space $\mathcal{M}_{b,g,n_c,n_o}$ of Riemann surfaces $\Sigma_{b,g,n_c,n_o}$
        of genus $g$ with $b$ boundary components as well as $n_c$ punctures in the interior and 
        $n_o$ punctures on the boundary. The associated integrand is a correlation function of 
        $n_c$ closed string vertex operators and $n_o$ open string vertex operators of the 
        worldsheet theory. Thus, such a string scattering amplitude takes the form
        \begin{align}
            \mathcal{A} =  \int\limits_{\mathcal{M}_{b,g,n_c,n_o}} \hspace{-0.4 cm}\langle 
            \prod\limits_{i=1}^{n_c} V_{w_i,h_i,\bar h_i}(z_i,\bar z_i,x_i,\bar x_i) \prod\limits_{j=1}^{n_o} 
            V^{AB}_{w_j,\Delta_j}(y_j,t_j) 
            \rangle_{\Sigma_{b,g,n_c,n_o}}.
        \end{align}
        Here, $w_i$ label spectral flow of the closed string vertex operators, 
        $(h_i,\bar h_i)$ are the holomorphic and anti-holomorphic space time conformal weights, $(z_i,
        \bar z_i)$ labels a point in the interior of the Riemann surface, $(x_i, \bar x_i)$ labels a point on the holographic boundary of the target space and $t_i$ labels a point on 
        the boundary of the $AdS_2$ brane. The role of $t_i$ is completely 
        analogous to that of $(x_i, \bar x_i)$ in the $x$-basis vertex operators of 
        \cite{Eberhardt:2019ywk}, i.e.~they are introduced by an appropriate deformation of the 
        ordinary vertex operators. In the closed string scenario discussed in \cite{Eberhardt:2019ywk}, 
        the deformation is generated by the zero modes $J^+_0$ and $\bar J^+_0$ of one holomorphic 
        and one anti-holomorphic current of the $SL(2,\mathbb{R})$ WZNW model. The $AdS_2$ boundary 
        condition glues these currents together along the boundary so that only one zero modes 
        survives. 
        This zero mode generates the deformation by the parameter $t$. Moreover, the open 
        string vertex operators are inserted at the points $y_j$ on the boundary of the worldsheet 
        and they carry only a single conformal weight $\Delta_j$. 
        \smallskip 
            
        On the symmetric orbifold side, these string scattering amplitudes should be reproduced by 
        correlation functions 
        \begin{align}
            \langle \prod\limits_{i=1}^{n_c} \sigma_{w_i,h_i,\bar h_i}(x_i,\bar x_i) \prod\limits_{j=1}^{n_o} 
            \sigma^{AB}_{w_j,\Delta_j}(t_j) \rangle\Big|_{g,b}
        \end{align}
        of interface changing operators. Here, we have not only included the labels which specify 
        the ground state of the twisted sectors in which the operators live, but also added labels 
        $h_i,\bar h_i$ and $\Delta_j$ that keep track of the excitations. The label $g,b$ instructs 
        us to restrict to the contribution to the correlator computed through covering maps whose 
        covering space is an element of $\mathcal{M}_{b,g,n_c,n_o}$.

        We will formulate the conjecture in more detail and collect evidence for it in 
        forthcoming work. Here, we shall content ourselves with a few basic statements. 
        Concretely, we show in section 
        \ref{sec:boundary_N_expansion}, for a special subset of boundary operators, 
        that the $N$ scaling of the large $N$-expansion of the boundary symmetric orbifold 
        matches the scaling of the dual string scattering amplitudes in the string coupling.
            
\subsection{Bulk correlation functions and amplitudes - a review}\label{sec:rev_bulk_N_expansion}
        
         As mentioned towards the end of the last section, now that we have successfully matched 
         the string theory and space time CFT partition functions, the natural 
         next step is to show that the duality also extends to correlators. Here, we shall content ourselves with the 
         very first step of such an extension. Namely, relating the large $N$ expansion of correlation 
         functions involving bulk and interface changing operators in the presence of our interfaces to a string theoretic genus expansion. 
         This is reminiscent of `t Hooft's analysis of the large $N$ limit for gauge theories. A detailed comparison of the CFT correlators with the scattering amplitudes of string theory 
         in $AdS_3$ will be performed in a forthcoming publication\footnote{This detailed comparison will include a discussion of the interface grand-canonical ensemble. 
         In the current paper, we restrict ourselves to a fixed $N$ analysis of the interface correlators, which suffers from the same issues of the analogous fixed $N$ bulk correlators, see e.g.~\cite{Aharony:2024fid}.
         }. 
         For bulk correlations of 
         symmetric product orbifolds, the large $N$ behaviour was studied by Pakman, Rastelli and Razamat 
         \cite{Pakman:2009zz}. To set up notations and illustrate the main ideas, we shall review their 
         arguments first before including interface changing operators in the next subsection. 
     
         The covering space method of\footnote{See also \cite{HAMIDI1987465} for an earlier application of covering space methods to orbifold CFTs.} Lunin and Mathur \cite{Lunin:2000yv} computes correlation functions 
        \begin{align}\label{eq:correlator_of_twist_fields}
            \langle \sigma_{g_1}(z_1,\bar z_1) \dots  \sigma_{g_{n_c}}(z_n,\bar z_n) \rangle
        \end{align}
        of gauge fixed twist fields in (bulk) symmetric product orbifolds on some two-dimensional base space $M$ 
        from the data of branched coverings 
        \begin{align}
            \gamma: \Sigma \rightarrow M
        \end{align}
        of the base space $M$ by a covering space $\Sigma$. Here, we shall assume that the elements $g_\nu
        \in S_N$ are cyclic permutations of length $w_\nu$. The general case can be treated similarly and in 
        fact follows directly from that of cyclic $g_\nu$ by taking suitable OPEs. We place a subscript $c$ 
        on the number $n_c$ of bulk field insertions to remind us that these should be in one-to-one 
        correspondence with insertions of vertex operators of closed strings in the dual theory. 

        Let us explain how to determine the covering space topology from the choice of cyclic permutations 
        $g_i$. To this end, it is useful to introduce the concept of ``active colours'' \cite{Pakman:2009zz}. Given any 
        $g \in S_N$, the subset $A_g$ of active colours consists of all elements in $\{1,\dots,N\}$ on which $g$ acts 
        non-trivially i.e.~$A_g := \{i | g i \neq i\}$. Furthermore, we define the active colours $A_S$ of any subset 
        $S \subseteq S_N$ of the permutation group to be the union of the active colours of all its elements. This 
        definition is now applied to the set $S= \{g_1, \dots, g_{n_c}\}$ of cyclic permutations that appear in 
        the correlation function~\eqref{eq:correlator_of_twist_fields}. We call the associated set $A_S$ the \textit{set 
        of active colours of the correlation functions}. 
        
        The permutations $g_\nu$ that appear in the correlation function we want to evaluate generate a subgroup $H_S 
        \subseteq S_N$ of the permutation group. By construction, $H_S$ acts faithfully on the corresponding set 
        $A_S$ of active colours. Under this action, the set $A_S$ decomposes into orbits which we denote by 
        $\mathscr{O}_1,\dots,\mathscr{O}_m$. Once we have determined the orbits, it is rather easy to infer key features of the covering surface $\Sigma$. More specifically, it 
        turns out that 
        \begin{enumerate}
            \item The orbits $\mathscr{O}$ are in one-to-one correspondence with the connected 
            components $\Sigma_\mathscr{O}$ of the covering surface $\Sigma$. In particular, the number 
            of connected components of $\Sigma$ is equal to the number $m = m(S)$ of orbits.
            \item The genus $g_\mathscr{O}$ of the connected component $\Sigma_\mathscr{O}$ that 
            is associated to the orbit $\mathscr{O}$ can be computed in terms of the number of 
            elements $|\mathscr{O}|$ of the set $\mathscr{O}$ as   
            \begin{align}
                g_\mathscr{O} = 1 - |\mathscr{O}|
                + \tfrac{1}{2} \sum\limits_{A_{g_\nu} \subseteq \mathscr{O}} (w_\nu-1). 
            \end{align}
            The sum runs over all those indices $\nu \in \{ 1, \dots, n_c\}$ for which the set of active 
            colours $A_{g_\nu}$ of the group element $g_\nu$ is a subset of the orbit $\mathscr{O}$.
        \end{enumerate}
        Hence, the covering surface $\Sigma$ is connected if and only if there is exactly one orbit. Put 
        differently, in order for $\Sigma$ to be simply connected the subgroup $H_S \subseteq S_N$ that 
        is generated by the elements of $S = \{g_1,\dots,g_{n_c}\}$ must act transitively on the active 
        colours $A_S$ of the correlator. Denoting the number of active colours of the correlator by $|A_S| 
        = n$, the formula for the genus then becomes 
        \begin{align}\label{eq:genus_formula_bulk}
            g = 1 - n + \tfrac{1}{2} \sum\limits_{\nu=1}^{n_c} (w_\nu-1).
        \end{align}
        The physical correlation functions to consider in a symmetric product orbifold are of course not 
        the gauge fixed correlators that we have discussed so far, but rather the correlation functions 
        of gauge invariant twist fields
        \begin{align}
            \sigma_{w} = \tfrac{1}{\sqrt{|[(1\dots w)]|}}\sum\limits_{g \in [(1\dots w)]}  \sigma_g = 
            \sqrt{\tfrac{(N-w)!w}{N!}}\sum\limits_{g \in [(1\dots w)]}  \sigma_g.
        \end{align}
        Clearly, the expansion of the gauge invariant twist fields into gauge fixed twist fields allows 
        us to immediately express correlators of the former in terms of correlators of the latter,  
        \begin{align}
            \langle \prod\limits_{\nu=1}^{n_c} \sigma_{w_\nu}(z_\nu,\bar z_\nu) \rangle = 
            \left(\prod\limits_{\nu=1}^{n_c}\sqrt{\tfrac{(N-w_\nu)!w_\nu}{N!}} \right)
                \sum\limits_{g_\nu \in [(1\dots w_\nu)]} 
                \langle \sigma_{g_1}(z_1,\bar z_1) \dots \sigma_{g_{n_c}}(z_{n_c},\bar z_{n_c}) \rangle.
        \end{align}
        The connected part of such a correlator is defined as the restriction of the above sum to those terms 
        that are associated to connected covering spaces $\Sigma$. According to the comments that we made earlier 
        in this section, we can state this definition as 
        \begin{align}\label{eq:connected_part_bulk_case}
            \langle \prod\limits_{\nu=1}^{n_c} \sigma_{w_\nu}(z_\nu,\bar z_\nu) \rangle_c :=
            \left(\prod\limits_{\nu=1}^{n_c}
            \sqrt{\tfrac{(N-w_\nu)!w_\nu}{N!}} \right) 
                \sum\limits_{\substack{g_\nu \in [(1\dots w_\nu)] \\ m(\{g_1, \dots, g_{n_c}\}) =1 }}  
                \langle \sigma_{g_1}(z_1,\bar z_1) \dots \sigma_{g_{n_c}}(z_{n_c},\bar z_{n_c}) \rangle.
        \end{align}
        The restricted sum now runs over all sets $S = \{g_1, \dots, g_{n_c}\}$ of cyclic permutations $g_\nu$ for 
        which the active colours $A_S$ of the correlations function form a single orbit under the action of 
        $H_S$. Remarkably, the large $N$ expansion of the connected part of the correlator takes the form of a string 
        genus expansion. In order to understand the precise scaling with $N$, we point out that the restricted sum 
        on the right hand side~of eq.~\eqref{eq:connected_part_bulk_case} con\-tains combinatorial enhancements from 
        the number of ways in which one can pick $n$ active colours from the $N$ colours that are available. Thus, the 
        connected correlator scales as 
        \begin{equation} \label{eq:largeNc}
        \prod\limits_{\nu=1}^{n_c}
            \sqrt{\tfrac{(N-w_\nu)!w_\nu}{N!}}  \, \frac{N!}{n! (N-n)!}  \ \sim \ N^{n- \tfrac{1}{2} 
            \sum^{n_c}_\nu w_\nu} \sim {g_s}^{-2n+ \sum^{n_c}_\nu w_\nu} = g_s^{-2 + 2g + n_c} \ . 
        \end{equation} 
        In the first step, we used Stirling's formula to determine the leading asymptotics of the prefactor as $N$ 
        becomes large. Then, we used the standard relation $g^2_s \sim 1/N$ between the string coupling $g_s$ and 
        the total number $N$ of colours. Finally, we inserted eq.~\eqref{eq:genus_formula_bulk} in order to rewrite 
        the exponent in terms of the genus $g$ of the covering surface $\Sigma$. The right hand side of
        eq.~\eqref{eq:largeNc} can be recognised as the usual dependence of closed string scattering amplitudes 
        on the string coupling. 
         
    \subsection{Boundary genus expansion for \texorpdfstring{$AdS_2$}{AdS2} branes}\label{sec:boundary_N_expansion}
        
        In this subsection, we extend the previous analysis to correlation functions involving both bulk and 
        interface changing fields and show that the large $N$ expansion of correlation functions in the presence 
        of our interfaces organises itself in the fashion of a string theory genus expansion. Just as in section 
        \ref{sec:rev_bulk_N_expansion}, (the connected part of) a gauge invariant correlation function is a sum 
        of gauge dependent correlators each of which can be associated to a covering space with a certain genus 
        $g$ and with $n_c$ punctures associated to the (bulk) operators inserted inside the correlator. In contrast 
        to the covering spaces described in section \ref{sec:rev_bulk_N_expansion}, the covering spaces that we
        encounter in this subsection additionally have a possibly non-vanishing number $b$ of boundaries, as well 
        as a number $n_o$ of punctures on the boundary.

        In this paper, we do not provide a full analysis of all interface changing correlation functions, but 
        rather restrict to one specific class of contributions to one specific type of boundary operators. 
        The generalisation to arbitrary covering maps contributing to interface correlation functions of interface changing operators will be the subject of an upcoming publication.
        Concretely, 
        we restrict to fields $\sigma^\pm$ that interpolate between $\Ipa$ and $\mathcal{I}^{(p\pm 1)}_{|a_\pm
        \rangle}$ at $N_- = N_+ = N$ and possess an expansion of the form 
        \begin{align}\label{eq:def_sigma_pm}
            \sigma^\pm = N^{-\tfrac{1}{2}} \sum\limits_{i^{-}=1}^{N}
           \sum\limits_{i^{+}=1}^{N} \sigma^\pm_{i^{\mp}i^{\pm}}.
        \end{align}
        Here, the operator $\sigma_{i^- i^+}^-$ cuts the $i^-$th copy of the seed theory in the lower hemisphere from 
        the $i^+$th copy in the upper hemisphere and turns on an $|a_\pm\rangle$ boundary for both. Conversely, the 
        operator $\sigma_{i^+ i^-}^+$ removes two copies of the boundary $|a_\pm\rangle$ and glues the $i^-$th copy of 
        the seed theory in the lower hemisphere to the $i^+$th copy in the upper hemisphere.
        Consequently
        \begin{align}
            \sigma^{+} = \sigma^{RR}_{2} \quad \text{and} \quad  \sigma^{-} = \sigma^{LL}_{2}
        \end{align}
        in the notation used in section \ref{sec:partition_function_match}.
        Note that, in eq.~\eqref{eq:def_sigma_pm}, the order of the 
        indices $i^+$ and $i^-$ is different in $\sigma^+$ and $\sigma^-$. This is a choice of notation that is convenient 
        in describing the computation of correlation functions
        \begin{align}\label{eq:AdS2_dual_orbifold_correlator}
            \langle \prod\limits_{\nu=1}^{n_o}\sigma^{s_\nu}(t_\nu)\prod\limits_{\mu=1}^{n_c}
            \sigma_{w_\mu}(x_\mu,\bar x_\mu) \rangle \, ,
        \end{align}
        where $s_\nu \in \{-,+\}$. We assume the ordering $\nu < \mu \Rightarrow x_\nu < x_\mu$. 
        In general \eqref{eq:AdS2_dual_orbifold_correlator} is a sum over many gauge fixed contributions or said differently, of contributions arising from many different covering maps. 
        Here, we restrict to those contributions for which the boundary condition imposed at infinity is fully transmissive. 
        For gauge fixed contribution to the correlator which fall into this class,
        any boundaries that should be 
        glued together by a $\sigma^+$ first had to be cut open by the insertion of a $\sigma^-$ and thus, such a 
        correlator with $n_o$ gauge fixed interface changing operators $\sigma^\pm$ has to satisfy
        \begin{align}
            \sum\limits_{i=1}^{n_o} s_i = 0 \quad \textrm{and} \quad  \sum\limits_{i=1}^k s_i \leq 0 \quad 
       \end{align}
        for every $k < n_o$.
        As for the correlation functions of bulk twist fields, we can express the correlator 
        \eqref{eq:AdS2_dual_orbifold_correlator} as a sum over gauge fixed terms. Concretely, inserting the definition 
        \eqref{eq:def_sigma_pm} of $\sigma^\pm$, we obtain
        \begin{align} \label{eq:bulkinterfaceexpansion}
            \langle \prod\limits_{\nu=1}^{n_o}\sigma^{s_\nu}(t_\nu)\prod\limits_{\mu=1}^{n_c}
            \sigma_{w_\mu}(x_\mu,\bar x_\mu) \rangle 
            = N^{-\tfrac{n_o}{2}} \langle \prod\limits_{\nu=1}^{n_o}\sum\limits_{i_\nu^\pm}^N
            \sigma^{s_\nu}_{i_\nu^{-s_\nu},i_\nu^{s_\nu}}(t_\nu)\prod\limits_{\mu=1}^{n_c}
            \sigma_{w_\mu}(x_\mu,\bar x_\mu) \rangle
        \end{align}
        For the individual terms on the right hand side, all indices $i_\nu^\pm$ must be contracted 
        according to the following rule: 
        
        \medskip
        
        \noindent \textit{Wick contractions for $\sigma^\pm$ colour indices:} The left index $i_\nu^+$ of every boundary creation operator 
        $\sigma^-_{i_\nu^+,i_\nu^-}$ needs to be contracted with the right index $i_\mu^+$ of a boundary annihilation operator 
        $\sigma^+_{i_\mu^-,i_\mu^+}$ such that $x_\nu < x_\mu$. In complete analogy, the left index $i_\mu^-$ of a boundary 
        annihilation operator inserted at $x_\mu$ needs to be contracted with the right index $i_\nu^-$ of a boundary creation 
        operator inserted at $x_\nu < x_\mu$. 
        
        \medskip

        Following this contraction procedure, starting from an arbitrary boundary creation operator $\sigma^-_{c_1^+ c_1^-}$, 
        we zigzag back and forth along the boundary line, collecting a string of boundary creation and annihilation operators 
        $\sigma^-_{c_1^+ c_1^-} \sigma^+_{c_1^- c_2^+}\sigma^-_{c_2^+ c_3^-} \dots \sigma^+_{c_{w_{n_c}}^- c_1^+}$ that ultimately 
        has to close into some cycle bringing us back to $c_1^+$. Keeping only the $c_i^-$, we obtain a cyclic permutation 
        $g_{n_c+1} = (c_1^- c_2^-\dots c_{w_{n_c+1}}^-)\in S_{N_-}$ whose length we denote by $w_{n_c+1}$. This notation 
        reflects the fact that upon fusing all of the operators that formed the cycle, we obtain twist fields in the $g_{n_c+1}$ 
        twisted sector. The latter are analogous to the bulk operators that are indexed by $n_c$ cycles $g_\nu$ of length $w_\nu$. 
        
        Effectively, the Wick contractions among interface changing operators we just performed thus gave rise to an additional 
        cycle. If the latter did not include all the interface changing operators, we pick one of the remaining ones and form a 
        second cycle $g_{n_c+2}$ of length $w_{n_c+2}$. We continue this process until we reach the last cycle $g_{n_c+b}$ of 
        length $w_{n_c+b}$, i.e.~until no uncontracted interface changing operators remain. 
         
        Once we have completed this contraction process, we are now left with $n_c+b$ cycles $g_\nu$ of length $w_\nu$. 
        These data are identical to those that would appear if we had to calculate a $(n_c+b)$-point correlation function 
        of bulk operators. Consequently, we can now follow essentially the same steps we outlined in the previous subsection. 
        As we did there, we first form the set $S = \{g_1, g_2, \dots g_{n_c+b}\}$ and the group $H_S$ that is generated by the 
        cycles $g_\mu$. We then let $H_S$ act on the active colours. We shall say that a term in the sum on the right hand 
        side of eq.~\eqref{eq:bulkinterfaceexpansion} contributes to the connected part of the correlation functions if 
        the group $H_N$ acts transitively on the active colours. Furthermore, the terms that are contributing to the 
        connected part of our correlation functions are associated with a connected covering space whose genus $g$ is given 
        by\footnote{This follows directly from the fact that the interface changing operators that are associated to a cycle 
        $g_{n_c+i}$ fuse into $g_{n_c+i}$ twist fields upon performing the OPE, which we have already alluded to above.}
        \begin{align} \label{eq:gco}
             g = 1 - n + \tfrac{1}{2} \sum\limits_{\mu=1}^{n_c+b} (w_\mu-1)
        \end{align}
        where $n$ denotes the number of active colours as before. By construction, the cycles that are 
        associated with the interface changing operators are in one-to-one correspondence with the connected components of 
        the boundary of the covering surface. In particular, the number of connected components of the boundary is $b$.
         
        We are finally able to compute the large $N$ scaling behaviour of the connected part of the correlation function. There are 
        three factors that contribute. As in the case of bulk correlation functions, we need to select the $n$ active colours from the 
        $N$ colours that are available. In addition, we need to collect the normalisations of bulk and interface changing fields. The 
        large $N$ behaviour of the first two factors we evaluated in eq.~\eqref{eq:largeNc} already. Since the interface changing fields 
        we consider each contribute a factor of $N^{-1/2}$, the large $N$ asymptotics of the prefactor reads 
        \begin{equation}  \label{eq:largeNo1}
            \ N^{n- \tfrac{1}{2} \sum^{n_c}_\nu w_\nu - \tfrac{n_o}{2}} =    
            N^{n- \tfrac{1}{2} \sum^{n_c}_\nu w_\nu - \tfrac{1}{2} \sum_{\nu=1}^b w_{n_c+\nu} - \tfrac{n_o}{4}}\ . 
        \end{equation} 
        In going to the right hand side of this equations, we have used the fact that  
        \begin{align}
            \sum\limits_{\nu=1}^{b} w_{n_c+\nu} = \tfrac{n_o}{2},
        \end{align}
        which follows from the construction of the $w_{\mu}$ with $\mu > n_c$ since they count the number of interface changing 
        operators that are combined into a cycle  and non-vanishing correlators require that all the fields are part of some cycle. 
        Now we can proceed by inserting equation \eqref{eq:gco} to express the sums over the cycle lengths $w_\mu$
        through the genus $g$ of the branching surface. The result is 
        \begin{align}
            N^{n- \tfrac{1}{2} \sum^{n_c}_\nu w_\nu - \tfrac{1}{2} \sum_{\nu=1}^b w_{n_c+\nu} - \tfrac{n_o}{4}} = 
            N^{1-g-\tfrac{b}{2}- \tfrac{n_c}{2}-\tfrac{n_o}{4}} \sim g_s^{-2+2g+b + n_c + \tfrac{n_o}{2}}\ .
        \end{align}
        Once again, we have used that $g_s^2 \sim 1/N$ to express the final result for the large $N$ asymptotics 
        of the correlation 
        function in terms of the string coupling $g_s$ rather than $N \sim 1/g_s^2$. The final result indeed has 
        the $g_s$ dependence of a string amplitude for a surface of genus $g$ with $b$ boundary components, $n_c$ 
        bulk punctures and $n_o$ boundary punctures. This is what we wanted to show. 
        
\section{Conclusion and outlook}\label{sec:outlook}

    In this work, we have constructed a new family of interfaces $\Ipa$ between two symmetric product orbifolds
    $\SymN$ with $N = N\pm$. These interfaces are associated with a pair of boundary states $|a_\pm\rangle$ of the 
    seed theory $\mathcal{M}$. The integer $p  \leq \textrm{min}(N_-,N_+)$ controls the transmissivity of the interface. 
    More precisely, the transmissivity \eqref{eq:transmittivity} is proportional to $p$ and takes its largest value for 
    $p = \mathrm{min}(N_-,N_+)$. For $p=0$, on the other hand, the interface is purely reflecting. A precise formula 
    for the interface was given in section \ref{sec:symmetric_product}, see eq.~\eqref{eq:def:pra} and the paragraph 
    below that equation for notations. The overlaps between the associated boundary states and hence the spectrum of 
    interface changing operators was computed in section \ref{sec:interface_changing_partition_function}. As in 
    the case of the bulk spectrum of symmetric product orbifolds which is elegantly encoded in the grand canonical 
    partition function found in \cite{Dijkgraaf:1996xw}, we stated the results in terms of a generalised grand 
    canonical partition function which also involves chemical potentials for the indices $p^L$ and $p^R$ of 
    transmissivity on both sides of the interface changing operator, see eqs.~(\ref{eq:GCZ}-\ref{eq:seedZo}). The 
    modular transform of the central formula \eqref{eq:Z_orbifold_open}, or rather of its restriction \eqref{eq:ZO[t]} 
    to special values of the chemical potentials, was given in eq.~\eqref{eq:ZO[t]MT}. The latter formula contains 
    contributions from interface changing operators of arbitrary twist. 
    
    The fact that our interfaces support operators in twisted sectors with arbitrary twist is a key feature that 
    hints toward a possible holographic description in terms D-branes in $AdS_3$. Indeed, $AdS_2$ branes have 
    been argued to support long open strings of arbitrary winding numbers \cite{Lee:2001xe}. In the case of the 
    supersymmetric four-torus $\mathcal{M} = \mathbb{T}^4$, we have in fact provided significant evidence for 
    such a holographic duality with $AdS_2$ branes in type IIB theory of tensionless strings in $AdS_3 \times 
    S^3 \times \mathbb{T}^4$. In section \ref{sec:tensionless_strings}, we computed the partition function 
    \eqref{eq:STspectrum} of $AdS_2$ branes in thermal $AdS_3$. The resulting function was shown in section 
    \ref{sec:partition_function_match} to match precisely with the corresponding terms in the exponent of the 
    grand canonical partition function \eqref{eq:ZO[t]} of our interfaces in symmetric product orbifolds.
    
    We also suggested how 
    to match more general string amplitudes involving both closed and open strings, leaving a fully conclusive 
    analysis for a forthcoming publication.
    Their computation and comparison is expected to closely follow the related analysis 
    that was carried out for bulk operators in symmetric product orbifolds and closed strings, see 
    \cite{Eberhardt:2019ywk}. 
    
    While the full proof of the holographic duality for tensionless closed and open strings that we described
    in the previous paragraph relies on the explicit calculation of quantities on both sides of the correspondence,
    it would also be very interesting to uncover the mechanism of the holographic relationship in the spirit of
    \cite{Hikida:2023jyc}. In that work, two of the authors considered an extension of string theory in $AdS_3$  
    to arbitrary values $k$ of NSNS background flux. According to a proposal of Eberhardt in \cite{Eberhardt:2021vsx} 
    such a string theory is dual to some non-rational symmetric product orbifold with a certain Liouville like 
    interaction turned on, see also \cite{Balthazar:2021xeh,Chakraborty:2025nlb}. In 
    \cite{Hikida:2023jyc} this interacting CFT$_2$ was used as a starting point and its correlators were rewritten 
    (for any number of bulk insertions) in terms of scattering amplitudes of closed strings in some $AdS_3$ string 
    theory. The main idea was to uplift the Liouville direction of the interacting non-rational symmetric product 
    orbifold to the radial direction of $AdS_3$ by reversing an intriguing relation between the $H_3^+$ WZNW model 
    and Liouville field theory \cite{Ribault:2005wp,Ribault:2005ms,Hikida:2007tq,Hikida:2008pe}, see also 
    \cite{Hikida:2020kil,Knighton:2023mhq,Demulder:2023bux,Knighton:2024qxd} for related developments. Thereby 
    the holographic correspondence was established for arbitrary correlation functions without ever computing 
    a single correlation function or scattering amplitude. It should be possible to extend this type of 
    derivation to include branes and open strings along the lines of \cite{Creutzig:2010bt}.

    A particularly interesting aspect that deserves attention in going away from the case with $k=1$ arises 
    because $AdS_2$ branes are generically not unique but rather carry an additional 
    parameter $r$. In the spacetime description, the parameter $r$ is related to the angle at which the 
    brane ends on the boundary of $AdS_3$. The case of $r=0$ corresponds to an $AdS_2$ brane that runs 
    straight through the centre of $AdS_3$ so that it approaches the boundary at 90 degrees. For other 
    non-vanishing values of $r$, the angle is non-trivial. The analogy with Karch-Randall interfaces in 
    higher dimensions suggests that the parameter $r$ should be related to the difference $N_+-N_-$. In 
    \cite{Martinec:2022ofs} Martinec also proposed that $r \sim  (N_--N_+)/(N_++N_-)$. 
    
    A very interesting challenge for the type of holographic relation we described here is to make contact 
    with the geometric supergravity regime by turning on RR background flux. In the symmetric product 
    orbifold this corresponds to a marginal deformation by some particular operator, see e.g.~\cite{David:2002wn} 
    for an early review. This deformation is somewhat similar to switching on the interaction in four-dimensional 
    ${\mathcal N}=4$ SYM theory. In the latter case, the leading perturbative corrections to the spectrum of the 
    free field theory (and eventually the entire deformation into the geometric regime of infinite 't Hooft 
    coupling $\lambda$) can be computed using integrability, see e.g.~\cite{Minahan:2002ve} for first order 
    calculations (and \cite{Beisert:2010jr} for further results from integrability). For the two-dimensional 
    cousins such powerful tools to reach the geometric regime are not available (yet), even though there 
    exists a few attempts to  start an integrability based approach to the problem, see e.g.~\cite{Gaberdiel:2023lco,Gaberdiel:2024dfw,Fabri:2025rok,Frolov:2023pjw}. 
    One of the issues that complicates the analysis of the perturbative spectrum near the symmetric product 
    orbifold is the mixing of left- and right-moving modes in the bulk. It might therefore be advantageous 
    to study the deformation for open strings on $AdS_2$ probe branes instead. Note that their open string 
    spectrum is as rich as that of closed strings with long strings of arbitrary winding number $w$ and 
    one might hope that these spectra can be deformed away from the tensionless limit all the way to the 
    geometric 
    regime using the ideas developed in \cite{Mitev:2008yt,Candu:2009ep}. It would also be interesting to 
    extend other integrability based studied of string theory in $AdS_3 \times S^3 \times \mathbb{T}^4$, 
    see e.g. \cite{Cavaglia:2022xld,Frolov:2023lwd} and references therein, to line defects as was done 
    for Wilson lines in ${\mathcal N}=4$ SYM theory \cite{Gromov:2015dfa,Grabner:2020nis}. When combined with the 
    toolbox of integrability, the interfaces we introduced here could turn out to be useful probes of 
    emerging geometries. 
    
\acknowledgments

We thank Alejandra Castro, Stefan Fredenhagen, Kazuo Hosomichi, Bob Knighton, Emil Martinec, Ingo Runkel, 
Bogdan Stefanski, Yu-ki Suzuki, Gerard Watts, and especially Andrea Dei and  Torben Skrzypek for useful 
comments and discussions. This project received funding from the German Research Foundation DFG under 
Germany's Excellence Strategy - EXC 2121 Quantum Universe - 390833306 and the Collaborative Research 
Center - SFB 1624 “Higher structures, moduli spaces and integrability” - 506632645. SH is further 
supported by the Studienstiftung des Deutschen Volkes. YH is supported by JSPS KAKENHI Grant Numbers 
JP21H05187 and JP23K25867.  TT is supported by JSPS KAKENHI Grant Number JP24KJ1374. Sketches in this 
work have been created using the Adobe Illustrator plug-in LaTeX2AI (https://github.com/isteinbrecher/latex2ai).
    
\appendix

\section{Torus partition function of the symmetric product orbifold}\label{app:torus_partition_function}
    
    In this appendix, we derive eq.~\eqref{eq:torus_grand_canonical}.
    To do so, we need to acknowledge that fixed $(g,h)$ contributions to the partition function \eqref{eq:defPFN} can be identified with ways of wrapping collections of disconnected tori around the spacetime torus.
    In this picture, we should view $g \in [w_1 , \dots , w_\ell]$ as providing the data that, if we were to cut all tori open along a spatial slice, then the corresponding covering of cylinders would be one where the covering space consists of $\ell$ disconnected cylinders that are wrapped around the spacetime cylinder $w_i$ times respectively.
    $h$ on the other hand tells us how these cylinders are glued together at the spatial cycles where we performed the cut.
    The exponential in the grand canonical partition function comes from the fact that the full partition function is obtained from the connected part -- i.e.~the part corresponding to coverings with a connected covering space -- by exponentiation.
    To be precise, the contribution
    \begin{align}
        \frac{1}{N}\sum\limits_{j=0}^{w-1} Z(\tfrac{N t}{w^2} + \tfrac{j}{w})
    \end{align}
    comes from covering the space time torus with a torus that winds $w$ times around the spatial cycle and $\ell = N/w$ times the thermal cycle (where $w$ necessarily has to divide $N$). That is, it comes from terms in the trace where $g \in [g_{(w,\ell)}] \subseteq S_N$ with
    \begin{align}
        g_{(w,\ell)} = \prod\limits_{k=0}^{\ell-1} (wk+1 \, \, wk+2 \, \,  \dots \, \,  wk+w)
    \end{align}
    and where furthermore $h \in S_\ell \ltimes (\mathbb{Z}_w)^{\ell}$ has a $S_\ell$ part $\pi_h$ that is just a single cycle of length $\ell$. There are $(\ell-1)!$ such cycles $\pi_h$ and hence we get that the $(w,\ell)$ winding covers contribute
    \begin{align}
        Z_{(w,\ell)}=\frac{1}{N!}\sum\limits_{g \in [g_{(w,\ell)}]}\sum\limits_{\substack{h \in \mathcal{C}_g \\ |\pi_h| = \ell}}\text{Tr}_{\mathcal{H}^{g}}[ h q^{L_0 -\frac{Nc_\mathcal{M}}{24}}] =
        \frac{|[g_{(w,\ell)}]|}{N!} (\ell-1)!\sum\limits_{j_1,\dots,j_\ell=0}^{w-1} Z\left(\tfrac{\ell t + \sum\limits_{k=1}^\ell j_k}{w}\right).
    \end{align}
    But $|[g_{(w,\ell)}]| = \frac{N!}{\ell! w^\ell}$ and, since the spectrum only contains integer spin states, the seed partition function is $\mathbb{Z}$ periodic. Hence,
    \begin{align}
        Z_{(w,\ell)} =
        \frac{1}{\ell w} \sum\limits_{j=0}^{w-1} Z\left(\frac{\ell t + j}{w}\right)
    \end{align}
    and summing over $\ell$ and $w$ as well as exponentiating, we get the grand canonical partition function
    \begin{align}
        \sum\limits_{N=0}^\infty \kappa^N Z_N = \exp\left(\sum\limits_{\ell,w = 1}^\infty Z_{(w,\ell)} \kappa^{w \ell}\right),
    \end{align}
    which upon replacing $\ell$ by $N/w$ is the same as eq.~\eqref{eq:torus_grand_canonical}.

\section{Interface changing operator partition function: detailed computation}\label{app:Overlap_full_computation}
    In this appendix, we provide a more detailed derivation of eq.~\eqref{eq:ZthermSC2}. The starting point is eq.~\eqref{eq:N_p_Simplified}, which written as a sum over conjugacy classes instead of individual group elements, takes the form
    \begin{align}
        \langle N |{\hat \STq}^{L_0}|p,a\rangle 
        = \sum\limits_{[\tau] \in [S_{p}]}\sum\limits_{[\rho] \in [S_{N-p}]}
        \sum_{\sigma \in \mathcal{C}^N_{\tau\rho}} \sum\limits_{j_b,i_d} \frac{1}{|\mathcal{C}_\rho^{N-p} ||\mathcal{C}_\tau^p|}
        \,^{N,N}_{\rho\tau,\tau\hspace{- 1 pt}\rho}\llangle j_b,i_d|  {\hat \STq}^{L_0} \sigma | a\rangle_{\rho} |\mathbb{I}\rangle_{\tau, \tau}
         |a\rangle_{\rho}.
    \end{align} 
    We can describe equivalence classes $[\tau]$ and $[\rho]$ with sequences\footnote{Since the modular parameter relevant for this section will always be $\hat \tst$ and not the $S$ transformed modular parameter $t$, we hope that the reader will not be offended by our choice to instead use $t$ in this appendix to refer to the sequence associated to $\tau$.} $({t}_w)_{w\in \mathbb{N}}$ and $(r_w)_{w \in \mathbb{N}}$ such that $t_w$ gives the number of cycles of length $w$ in $[\tau]$. In this context, let
    \begin{align}
        |t| := \sum\limits_{w=1}^\infty t_w w.
    \end{align}
    Using this notation,
    \begin{align}
        \mathcal{Z}_{a}[\mu_\pm,0,\rho_R;\hat \tst] =  \sum\limits_{t,r} \mu^{|t+r|} \rho_R^{2|r|}  \sum\limits_{\sigma \in \mathcal{C}_{r+t}} \frac{1}{|\mathcal{C}_{t}||\mathcal{C}_r|}  \sum\limits_{j_b,i_d}   \,^{N,N}_{\rho\tau,\tau\hspace{- 1 pt}\rho}\llangle j_b,i_d|  {\hat \STq}^{L_0} \sigma | a\rangle_{\rho} |\mathbb{I}\rangle_{\tau, \tau}
         |a\rangle_{\rho},
    \end{align}
    where $\mu:= \mu_- \mu_+$ and
    \begin{align}
        \mathcal{C}_u := \prod\limits_{w=1}^\infty \left( S_{u_w}\ltimes \mathbb{Z}_w^{u_w} \right),
    \end{align}
    while $\rho$ and $\tau$ are arbitrary representatives of the conjugacy class determined by $r$ and $t$. Now the contribution of fixed $t$ and $r$ factorises into a product of components that correspond to different cycle lengths according to
    \begin{align}
        \mathcal{Z}_{a} =  \sum\limits_{t,r} \prod\limits_{w=1}^\infty \left( \mu^{w(t_w+r_w)} \rho_R^{2wr_w}  \hspace{-1 cm} \sum\limits_{\sigma_w \in S_{r_w+t_w}\ltimes \mathbb{Z}_w^{r_w+t_w}}  \hspace{-0.8 cm} \frac{\prod\limits_{s \in S}Z_c\left(\tfrac{\ell_s^c\hat\tst+\sum\limits_{i=1}^{\ell_s^c} z_{i}}{w}\right)\prod\limits_{j=1}^{r_w} \hat Z_o\left(\tfrac{2\ell_j^o\hat\tst}{w}\right) }{t_w! w^{t_w}r_w!^2 w^{2r_w}}  \right).
    \end{align}
    Here, for $\sigma_w = (\pi,(z_1,\dots,z_{r_w +t_w}))$, define $\ell_j^o$ with $j=1,\dots,r_w$ as the smallest integer $k>0$ such that $\pi^k j \leq r_w$. Furthermore, let $\pi = \pi_1 \dots \pi_L$ be the decomposition of $\pi$ into cycles. 
    We denote the elements of the cycle $\pi_s$ by $(\pi_s^1 \dots \pi_s^{\ell_s^c})$ 
    and let $S\subseteq\{1,\dots,L\}$ be the set of those indices $s$ for which $\{\pi_s^1 , \dots, \pi_s^{\ell_i^c}\}$ has an empty overlap with $\{1,\dots, r_w\}$. Note that
    \begin{align}\label{eq:sum_r_t_identity}
        \sum\limits_{s \in S} \ell_s^c + \sum\limits_{j=1}^{r_w} \ell_j^o = r_w+t_w.
    \end{align}
    To simplify further, we can now perform the summation over $\mathbb{Z}_w^{r_w+t_w}$. This gives
    \begin{align}
        \mathcal{Z}_{a} &=    \sum\limits_{t,r} \prod\limits_{w=1}^\infty \left( \mu^{w(t_w+r_w)} \rho_R^{2wr_w}  \hspace{-0.5 cm}\sum\limits_{\sigma_w \in S_{r_w+t_w}} \hspace{-0.5 cm} \frac{w^{t_w-|S|}\prod\limits_{s \in S} Z_c\left(\tfrac{\ell_s^c\hat\tst+\sum\limits_{i=0}^{w-1} i}{w}\right)w^{r_w}\prod\limits_{j=1}^{r_w}\left( \hat Z_o\left(\tfrac{2\ell_j^o\hat\tst}{w}\right)\right)}{t_w! w^{t_w}r_w!^2 w^{2r_w}}  \right). 
    \end{align}
    By the use of eq.~\eqref{eq:sum_r_t_identity}, this can be simplified further to
    \begin{align}
        \mathcal{Z}_{a} &= \sum\limits_{t,r} \prod\limits_{w=1}^\infty \left(   \sum\limits_{\sigma_w \in S_{r_w+t_w}}  \hspace{-0.5 cm}\frac{\prod\limits_{s \in S} \frac{\mu^{w\ell_s^c}}{w} Z_c\left(\tfrac{\ell_s^c\hat\tst+\sum\limits_{i=0}^{w-1} i}{w}\right)\prod\limits_{j=1}^{r_w}  \frac{\mu^{w\ell_j^o}\rho_R^{2w}}{w}\hat Z_o\left(\tfrac{2\ell_j^o\hat\tst}{w}\right)}{t_w!r_w!^2 }  \right).
    \end{align}
    Now we perform the sum over $S_{r_w+t_w}$. As a first step towards this goal, we should replace the labels $\ell^o_j$ and $\ell^c_s$ of individual cycle lengths by labels $L^o$ and $L^c$ that just count how many cycles of certain length are present. That is
    \begin{align}
        L^o_\ell := |\{j \in \{1,\dots ,r_w\}:\ell^o_j = \ell\}| , \quad L^c_\ell := |\{s \in S:\ell^c_s = \ell\}|.
    \end{align}
    Just as we notationally suppressed the dependence of $\ell^o,\ell^c$ on $\sigma$, we will suppress the dependence of $L^o,L^c$ on $\sigma$. In terms of the new labels, the sum becomes
    \begin{align}
        \mathcal{Z}_{a} = \sum\limits_{t,r} \prod\limits_{w=1}^\infty \left(   \sum\limits_{\sigma_w \in S_{r_w+t_w}}  \hspace{-0.5 cm}\frac{\prod\limits_{\ell=1}^\infty\left( \frac{\mu^{w\ell}}{w} Z_c\left(\tfrac{\ell\hat\tst+\sum\limits_{i=0}^{w-1} i}{w}\right)\right)^{L_\ell^c}\prod\limits_{\ell=1}^{\infty} \left( \frac{\mu^{w\ell}\rho_R^{2w}}{w}\hat Z_o\left(\tfrac{2\ell\hat\tst}{w}\right)\right)^{L_\ell^o}}{t_w!r_w!^2 }  \right).
    \end{align}
    Now, since the terms which we sum only depend on the $L^o$ and $L^c$ labels, we can introduce an equivalence relation $\sim$ on $S_{r_w +t_w}$ that identifies permutations with equal $L^o,L^c$ labels, trading the sum over $\sigma_w$ with a sum over equivalence classes $[L^oL^c]$. To correctly compute the sum, we have to determine the size of $[L^oL^c]$. Let us first take a step back and try to count permutations with fixed $\ell^o$. They need to satisfy $\sigma^k_w(1)> r_w$ for all $0<k<\ell^o_1$ and $\sigma^{\ell^o_1}_w(1) \leq r_w$. This leads to $\tfrac{t_w!}{(t_w-\ell^o_1+1)!} r_w$ choices for $\{\sigma^k_w(1)\}_{k=1}^{\ell^o_1}$. Continuing a counting like this for all $\{\sigma^k_w(i)\}_{k=1}^{\ell_i^o}$ with $1 \leq i \leq r$ gives a factor of
    \begin{align}
        r_w!\tfrac{t_w!}{\left(t_w-\sum\limits_{i=1}^{r_w} (\ell_i^o-1)\right)!}  = r_w!\tfrac{t_w!}{\left(t_w-\sum\limits_{\ell=0}^{\infty} L^o_\ell(\ell-1)\right)!}
    \end{align}
    contributing to the size of $[L^oL^c]$. Additionally, we need to consider how many ways there are to realise the label $L^o$ with labels $\ell^o$. This leads to an extra factor of $\tfrac{r_w!}{\prod\limits_\ell L^o_\ell!}$. Hence, the choice of $L^o$ contributes a factor of
    \begin{align}
        \frac{r_w!^2}{\prod\limits_\ell L^o_\ell!}\frac{t_w!}{\left(t_w-\sum\limits_{\ell=0}^{\infty} L^o_\ell(\ell-1)\right)!}
    \end{align}
    to the size of $|[L^oL^c]|$. Finally, we need to multiply by the size of the conjugacy class labelled by $L^c$ to obtain
    \begin{align}
        |[L^oL^c]| = \frac{r_w!^2}{\prod\limits_{\ell=1}^\infty L^o_\ell!}\frac{t_w!}{\left(t_w-\sum\limits_{\ell=1}^{\infty} L^o_\ell(\ell-1)\right)!} \frac{\left(\sum\limits_{\ell=1}^\infty  \ell L_\ell^c\right)!}{\prod\limits_{\ell=1}^\infty L^c_\ell! \ell^{L^c_\ell}}.
    \end{align}
    Now
    \begin{align}
        \sum\limits_{\ell=1}^\infty  \ell (L_\ell^c+L_\ell^o) = r_w +t_w \quad \text{and} \quad \sum\limits_{\ell=1}^\infty L_\ell^o = r_w \label{eq:Constraints}
    \end{align}
     implies
    \begin{align}
        t_w-\sum\limits_{\ell=1}^{\infty} L^o_\ell(\ell-1) = \sum\limits_{\ell=1}^\infty  \ell L_\ell^c
    \end{align}
    and therefore
    \begin{align}
        |[L^oL^c]| = \frac{r_w!^2 t_w!}{\prod\limits_{\ell=1}^\infty L^o_\ell!L^c_\ell! \ell^{L^c_\ell}}.
    \end{align}
    This implies
    \begin{align}
        \mathcal{Z}_{a} = \sum\limits_{t,r} \prod\limits_{w=1}^\infty \left(   \sum\limits_{[L^oL^c]}  \prod\limits_{\ell=1}^{\infty} \frac{\left( \frac{\mu^{w\ell}}{w \ell} Z_c\left(\tfrac{\ell\hat\tst+\sum\limits_{i=0}^{w-1} i}{w}\right)\right)^{L_\ell^c}}{L^c_\ell!} \prod\limits_{\ell=1}^{\infty} \frac{\left( \frac{\mu^{w\ell}\rho_R^{2w}}{w}\hat Z_o\left(\tfrac{2\ell\hat\tst}{w}\right)\right)^{L_\ell^o}}{L_\ell^o!} \right),
    \end{align}
    where the $[L^oL^c]$ sum runs over the set of equivalence classes $S_{r_w+t_w}/\sim$.
    Finally, we trade the unconstrained sum over the sequences $t$ and $r$, which in turn constrains $L^o$ and $L^c$ to correspond to equivalence classes in $S_{r_w + t_w}$ for an unconstrained sum over $L^o$ and $L^c$ without $t$,$r$ sum. This is achieved by reinterpreting the constraints \eqref{eq:Constraints} as fixing $t_w$ and $r_w$ in terms of $L^o$ and $L^c$, leading to
    \begin{align}
        \mathcal{Z}_{a} = \prod\limits_{w,\ell=1}^\infty \left( \sum\limits_{L^c_\ell=0}^\infty\left(\frac{\left( \frac{\mu^{w\ell}}{w \ell} Z_c\left(\tfrac{\ell\hat\tst+\sum\limits_{i=0}^{w-1} i}{w}\right)\right)^{L_\ell^c}}{L^c_\ell!}\right) \sum\limits_{L^o_\ell=0}^\infty  \left( \frac{\left( \frac{\mu^{w\ell}\rho_R^{2w}}{w}\hat Z_o\left(\tfrac{2\ell\hat\tst}{w}\right)\right)^{L_\ell^o}}{L_\ell^o!} \right) \right).
    \end{align}
    The sums now simply give exponential functions
    \begin{align}
        \mathcal{Z}_{a}[\mu_\pm,0,\rho_R;\hat \tst] = \exp \left( \sum\limits_{w,\ell=1}^\infty  \left( \frac{\mu^{w\ell}}{w \ell} Z_c\left( \tfrac{\ell\hat\tst+\sum\limits_{i=0}^{w-1} i}{w}\right)\right) + \sum\limits_{w,\ell=1}^      \infty  \left( \frac{\mu^{w\ell}\rho_R^{2w}}{w}\hat Z_o\left(\tfrac{2\ell\hat\tst}{w}\right) \right) \right).
    \end{align}
    Replacing the summation variable $\ell$ by $k = w \ell$ finally gives us the result \eqref{eq:ZthermSC2} that we wanted to prove.

\bibliographystyle{JHEP}
\bibliography{biblio.bib}

\end{document}